\documentclass{iopart}
\usepackage{amscd,amssymb,a4}
\usepackage{graphicx,xspace}

\usepackage[left=2cm,top=2cm,right=2cm,bottom=2cm]{geometry}
\usepackage{color}

\begin{document}


\newcommand{\vv}{{\bf v}}
\newcommand{\vx}{{\bf x}}
\newcommand{\vu}{{\bf u}}
\newcommand{\vk}{{\bf k}}
\newcommand{\vp}{{\bf p}}
\newcommand{\vq}{{\bf q}}
\newcommand{\vdelta}{{\bf \delta}}
\newcommand{\vomega}{{\bf \omega}}

\def\be{\begin{equation}}
\def\ee{\end{equation}}

\def\bea{\begin{eqnarray}}
\def\eea{\end{eqnarray}}

\newcommand{\nn}{\nonumber}
\newcommand{\fig}[2]{\includegraphics[width=#1]{./#2}}
\newcommand{\Fig}[1]{\includegraphics[width=8.7cm]{./#1}}
\newlength{\bilderlength}
\newcommand{\bilderscale}{0.35}
\newcommand{\storebilderscale}{\bilderscale}
\newcommand{\bilderskip}{\hspace*{0.8ex}}
\newcommand{\textdiagram}[1]{%
\renewcommand{\bilderscale}{0.2}%
\diagram{#1}\renewcommand{\bilderscale}{\storebilderscale}}
\newcommand{\vardiagram}[2]{%
\newcommand{\bilderscale}\textrm{a}{#1}%
\diagram{#2}\renewcommand{\bilderscale}{\storebilderscale}}
\newcommand{\diagram}[1]{%
\settowidth{\bilderlength}{\bilderskip%
\includegraphics[scale=\bilderscale]{./#1}\bilderskip}%
\parbox{\bilderlength}{\bilderskip%
\includegraphics[scale=\bilderscale]{./#1}\bilderskip}}
\newcommand{\Diagram}[1]{%
\settowidth{\bilderlength}{%
\includegraphics[scale=\bilderscale]{./#1}}%
\parbox{\bilderlength}{%
\includegraphics[scale=\bilderscale]{./#1}}}
\bibliographystyle{KAY}

\title{Functional renormalization-group approach to decaying turbulence}

\author{Andrei A. Fedorenko${}^1$, Pierre Le Doussal${}^2$ and Kay J\"org Wiese${}^2$}
\address{${}^1$  CNRS-Laboratoire de Physique de l'\'Ecole Normale Sup{\'e}rieure de Lyon, 46 all\'ee d'Italie,  69007 Lyon, France.}
\address{${}^2$ CNRS-Laboratoire de Physique
Th{\'e}orique de l'\'Ecole Normale Sup{\'e}rieure, 24 rue Lhomond, 75005 Paris, France.}

\begin{abstract} 
We reconsider the functional renormalization-group (FRG) approach to decaying Burgers turbulence, and extend it to
decaying Navier-Stokes and Surface-Quasi-Geostrophic turbulence. The method is based on a renormalized small-time expansion, equivalent to a loop expansion, and naturally produces a dissipative anomaly and a cascade after a finite time. We explicitly calculate and analyze the one-loop FRG equations in the zero-viscosity limit as a function of
the dimension. For Burgers they reproduce the FRG equation obtained in the context of random manifolds, extending previous results of one of us. Breakdown of energy conservation due to shocks and the appearance  of a direct energy cascade corresponds to failure of dimensional reduction in the context of disordered systems. For Navier-Stokes in three dimensions, the velocity-velocity correlation function acquires a linear dependence on the distance, $\zeta_2=1$, in the inertial range, instead of Kolmogorov's $\zeta_2=2/3$;
however the possibility remains for corrections at two- or higher-loop order.
In two dimensions, we obtain a numerical solution which conserves energy and exhibits an inverse cascade,
with explicit analytical results both for large and small distances, in agreement with the scaling proposed by Batchelor.
In large dimensions, the one-loop FRG equation for Navier-Stokes converges to  that of Burgers. 
\end{abstract}

\maketitle

\section{Introduction}

Describing Navier Stokes (NS) turbulence with the tools of statistical physics
has remained a major challenge since Kolmogorov's dimensional arguments leading
to the $E(k) \sim \bar \epsilon^{2/3} k^{-5/3}$ energy spectrum for the 3D energy
cascade \cite{Kolmogorov1941,Kolmogorov1962,FrischBook,Oboukhov1962}. The simplest analytical method, Kraichnan's {\em direct
interaction approximation} closure scheme \cite{Kraichnan1959} (equivalent to mode
coupling)  failed to recover Kolmogorov's prediction. There were numerous attempts
to overcome these difficulties using a variety of methods, e.g.\ more refined closure
schemes \cite{LesieurBook},
large number of components \cite{MouWeichman1993,MouWeichman1995}, renormalization-group (RG) \cite{Kraichnan1968,ForsterNelsonStephen1977, DeDominicisMartin1979,Mayo2005, AdzhemyanAntonovGoldinKimKompaniets2008, AdzhemyanAntonovKompanietsVasilev2003,AdzhemyanAntonovVasilev1989}
 conjectures for short-distance expansions \cite{Polyakov1995,BoldyrevLindePolyakov2004,Polyakov1993},
 study of short time singularities \cite{frisch},
   tetrad  models \cite{ChertkovPumirShraiman1999}, and shell models
 \cite{LvovPodivilovPomyalovProcacciaVandembroucq1998}, with various degrees
 of success. At the heart of the cascade phenomenon is that
 non smooth velocity field do not conserve energy. The
 main physics challenge, i.e.\ to describe the statistics of
 the energy transfer via singular or almost singular structures, is
 only partially captured. Predicting the multi-fractal
 corrections for velocity moments
 { $S_p(u,t)=\langle \left[(\vv_{\vu t}-\vv_{0t})\cdot {\bf u}/|\vu|\right]^p \rangle%
  \simeq C_p |\vu|^{\zeta_p}$} to Kolmogorov's prediction ($\zeta^{\mathrm K}_p=p/3$)
  remains a challenge, despite the analytical progress in the simpler passive scalar
   problem 
   \cite{AdzhemyanAntonovHonkonenKim2005,ChertkovFalkovichKolokolovLebedev1995,%
GawedzkiVergassola2000,FairhallGatLvovProcaccia1996,EyinkXin2000,%
GawedzkiVergassola1998,LvovProcacciaFairhall1994,Wiese1999,AdzhemyanAntonov1998,%
BernardGawedzkiKupiainen1996,BernardGawedzkiKupiainen1998,ChenKraichnan1998,%
ChertkovFalkovich1996,FrischMazzinoNoullezVergassola1999,%
GawedzkiKupiainen1995,KraichnanYakhotChen1995,FalkovichGawedzkiVergassola2001,GatLvovPodivilovProcaccia1997,
ShraimanSiggia1996}. For the inverse cascade in 2D~\cite{Batchelor1969},
due to an infinity of conserved quantities, and a simpler numerical modeling,
more is known; the most recent  analysis unveils a tempting connection to conformal field theory and
SLE \cite{BernardBoffettaCelaniFalkovich2007}, but remains based on numerics
or speculative \cite{Flohr1996}. Also relations between Navier-Stokes and the membrane dynamics of black
holes have been discussed in the context of the AdS/CFT correspondence \cite{ElingFouxonOz2009,BredbergStrominger2011,BredbergKeelerLysovStrominger2011} 
as well as relations to the physics of graphene \cite{MullerSchmalianFritz2009}.

The problem of $N$-dimensional Burgers turbulence, i.e.\ of a potential
flow without pressure, exhibits similarities with Navier-Stokes turbulence, such as the existence of an inertial range which supports an
energy cascade and the multi-scaling of the velocity moments. Although, as NS, it lacks a small control parameter and hence is non-trivial,
it is simpler, since the Burgers equation can be integrated explicitly via the
Cole-Hopf transformation
\footnote{It is often
said \cite{BecKhanin2007} that it lacks an essential property of Navier-Stokes turbulence, namely the
sensitivity to small perturbations in the initial data, and
thus the spontaneous appearance  of randomness by the chaotic dynamics.
However, there is, in some cases (denoted SR below), decorrelation in time of
two slightly different  initial conditions, a property sometimes termed chaos in
the community of disordered systems \cite{LeDoussal2006a}.}, hence it has allowed for some progress \cite{FrischBec2001,BecKhanin2007}. A
remarkable mapping to an elastic object in a quenched random potential maps the
shocks in both a decaying or stirred Burgers velocity field to the jumps of the
equilibrium position of the pinned elastic object (which is a point for decaying Burgers or a line for
stirred Burgers) upon variation of an external field. This mapping was used to
study the large-dimension $N$ limit of {\it stirred} Burgers turbulence using
replica symmetry breaking \cite{BouchaudMezardParisi1995,Mezard1997}
and, more recently, of {\it decaying} Burgers turbulence \cite{LeDoussalMuellerWiese2010,LeDoussalMuellerWiese2007}.
The detailed statistics of shock cells which is obtained from these works is consistent with
the physical expectation, and important open questions are now (i) whether this is a good starting point to perform
an expansion towards finite $N$; (ii) whether it can inspire an approach to Navier Stokes in
large dimension, a notably difficult problem \cite{frishlarged}.
 An RG inspired method bypassing the Cole-Hopf transformation
has been proposed very recently for the KPZ equation which is closely related
to the Burgers equation~\cite{Hairer2012}.

Another powerful method able to handle singularities such as shocks and avalanches in disordered
systems, which does not rely on large $N$, is the {\em Functional Renormalization Group}  (FRG) \cite{DSFisher1986,LeDoussalWieseChauve2003} (for an introduction
and review see \cite{WieseLeDoussal2006,LeDoussal2004}). The connection between the FRG and {\it decaying}
Burgers turbulence was elucidated in \cite{LeDoussal2006b,LeDoussal2008} (see \cite{BalentsBouchaudMezard1996}
for an earlier attempt). It turns out that the force felt by an elastic manifold of internal dimension $d$ submitted to a random potential plus
a quadratic well can be seen as a generalized velocity field: it satisfies an exact evolution equation which is a functional generalization of the decaying
Burgers equation, where the role of time is played by the (inverse) curvature of the well.
For $d=0$ the manifold is a point and one recovers the standard Cole-Hopf representation of the
Burgers equation. The hierarchy of equations relating $n$-point equal-time velocity
correlation functions identifies with the (exact) hierarchy of FRG flow equations, and the loop expansion in the
field theory corresponds to the (renormalized) small time expansion in the (generalized) Burgers problem,
as will be detailed below. The amazing property is that this hierarchy {\it becomes
controlled} in an expansion in $\epsilon=4-d$ around $d=4$, which is the crucial
property of the FRG approach to disordered systems. Hence Burgers turbulence, i.e.
$d=0$, becomes accessible via this expansion. Furthermore the physics of the
generalized Burgers problem (i.e.\ of the manifold) has features which are independent
of the parameter $d$. For instance, energy conservation for smooth flows is obtained
as well as an infinite number of conserved quantities (the first property being called
"dimension reduction" in the context of manifold, and the second corresponds to the non-renormalization of the
moments of the
so-called Larkin random force). Non-conservation of energy via shocks occurs for
any $d$, and the dissipative anomaly at the heart of the energy cascade, i.e.\ the non-vanishing limit
of the energy flux $- \partial_t {\cal E} = \bar \epsilon= \nu \langle (\nabla v)^2 \rangle$ as $\nu \to 0$,
is naturally captured by the FRG \cite{LeDoussal2006b,LeDoussal2008}.
Finally the FRG allows to compute shock-size distributions in the controlled expansion around
$d=4$ \cite{LeDoussalWiese2008c,LeDoussalWiese2011b}. Most of these
studies focused on $N=1$, but recently we also investigated $N>1$ \cite{LeDoussalRossoWiese2011}.

The aim of this paper is thus to investigate whether FRG-inspired methods can be developed to
describe Navier-Stokes turbulence as well. Here, our scope is relatively modest
and it should be seen as a first exploration of the FRG method into the domain of non-linear physics.
We focus on the decaying
Burgers and Navier-Stokes equations; however, the stirred case can  also be studied
within the same framework. We derive the 1-loop FRG equations, first for Burgers in $N$-dimension
(since most explicit calculations in \cite{LeDoussal2006b,LeDoussal2008}
were for $N=1$), and then for Navier-Stokes. We discuss some features of the fixed-point solutions
which correspond to a decaying turbulent state, leaving a detailed analysis for the future.
In $N=2$ we perform a detailed analysis
of the NS fixed point.  At this stage, the method for NS is not a controlled perturbative expansion
scheme, since there is no equivalent of the Cole-Hopf mapping. The method however
does capture some of the physics of the singularities. We analyze the nature of the
singularities at small distance. While our analysis is restricted to one loop, we discuss at the end possible
extensions to higher loops.

Let us stress that in the so-called inertial range of length and time scales,
it is widely expected that the statistics of decaying turbulence
is rather similar to the forced one. Indeed, due to strong separation of the
large and small time scales, the  eddies in the
inertial range have enough time to reach an  equilibrium for the energy-flux,  before
the larger eddies will significantly decay.
The scaling behavior of decaying turbulence in the inertial range is thus indistinguishable from
the forced case,  while differences will occur at the scales of the large eddies.
This is the universality assumption entering most  theories of turbulence, see e.g. Ref. \cite{LvovPasmanterPomyalovProcaccia2003}
for a detailed discussion.

The paper is organized as follows: In section \ref{s:MOdel}, we introduce
the model and notations. In section \ref{s:Known results and phenomenology},
we review known results, both for Burgers and Navier Stokes. The FRG equations
are derived in section \ref{s:FRG equations}. We start by the general scheme,
before giving results for Burgers and Navier Stokes; finally we discuss conserved
quantities. In section \ref{s:Short-distance analysis, cusp or no cusp?}, we discuss
the short-distance singularity of the FRG equations: Are there solutions with other
power laws than a cusp?
In  section \ref{s:NS:d=2} we focus on the analysis of the FRG equations for
Navier-Stokes turbulence in two dimensions. In section \ref{sec:N-inf} we
discuss the limit of large $N$ and in section \ref{sec:SQG} we give
the FRG equations for  surface quasi geostrophic turbulence.

Finally note that this work was started a while ago. For an early exposition see \cite{LeDoussalTalk}.

\section{Model and notations}\label{s:MOdel}

We study here two models:

(i) the $N$-dimensional decaying {\em Burgers equation} for a $N$-component
velocity field $\vv_{\vu t}$ at point $\vu$ and time $t$,
\begin{eqnarray}
\partial_t \vv^\alpha_{\vu t} = \nu \nabla_{\vu}^2 \vv^\alpha_{\vu t} -
\frac{1}{2} \partial_{\vu}^\alpha (\vv_{\vu t})^{2} \ .
\label{burgers1}
\end{eqnarray}
The velocity is assumed to be vorticity-free so it can be  expressed
as gradient of a potential function,
$\vv^\alpha_{{\vu t}}=\partial^\alpha_{\vu} \hat V({\vu, t})$,
that implies $\frac{1}{2} \partial^\alpha (\vv)^2=(\vv \cdot\nabla) \vv^\alpha$.
Note that to streamline notations we attach  space and time indices to the fields,
which {\em are never to be understood as derivatives.} We use boldface to
indicate vectors, and normal font for scalars, so $k:=|\vk|$.

(ii) the incompressible $N$-dimensional {\em Navier-Stokes equation} 
\begin{eqnarray}\label{NS}
\partial_t \vv^\alpha_{\vu t} = \nu \nabla_{\vu}^2 \vv^\alpha_{\vu t} - P^{\rm T}_{\alpha \beta}(\partial_{\vu})
\partial^\gamma_{\vu} ( \vv^\gamma_{\vu t} \vv^\beta_{\vu t} ),
\ \ \ \ \ \ \ \
P^{\rm T}_{\alpha \beta}(\partial_{\vu})=
\delta^{\alpha\beta}-\frac{\partial^\alpha_{\vu}\partial^\beta_{\vu}}{\nabla^{2}}
\end{eqnarray}
with the pressure eliminated using the transverse projection operator
$P^{\rm T}_{\alpha \beta}(\partial_{\vu})$. The latter  implies
the divergence free constraint (incompressibility) $\nabla \cdot \vv =0$ at all
times. 
In Fourier space, both equations can be written as
\begin{eqnarray}
\partial_t \vv^\alpha_{\vk t} = - \nu k^2 \vv^\alpha_{\vk t} - \frac{1}{2} P_{\alpha ; \beta \gamma}(\vk) \sum_{\vp+\vq=\vk}
\vv^\beta_{\vq t} \vv^\gamma_{\vp t} \ ,
\end{eqnarray}
with
\begin{equation}\label{PNS}
P_{\alpha ; \beta \gamma}(\vk) = \left\{\begin{array}{ll}
\displaystyle  i \vk^\alpha \delta_{\beta \gamma} \quad& \mbox{(Burgers)},\\
\displaystyle  i \vk^\beta P^{\rm T}_{\alpha \gamma}(\vk) +  i \vk^\gamma P^{\rm T}_{\alpha \beta}(\vk) \quad& \mbox{(Navier Stokes)}. \\
\end{array} \right.
\end{equation}
The transverse and longitudinal projection operators written in Fourier space read
\begin{equation}\label{PBurgers}
 P^{\rm T}_{\alpha\beta}(\vk) = \delta^{\alpha\beta}-\frac{\vk^{\alpha}\vk^{\beta}}{k^{2}},
 \ \ \ \ \ \ \
 P^{\rm L}_{\alpha\beta}(\vk) = \frac{\vk^{\alpha}\vk^{\beta}}{k^{2}}.
\end{equation}
In both cases we are interested in the small-viscosity (large Reynolds number)
limit $\nu \to 0$, in which case a broad inertial range develops. In that limit the
Navier-Stokes equation formally becomes the Euler equation, and in both cases
weak solutions exist \cite{leray1933,leray1934,DuchonRobert2000}
(for review see \cite{Gawedzki1999}).

We study homogeneous turbulence  with random initial conditions, i.e.\
we chose an initial condition at time $t=0$ which is statistically translational invariant.
Everywhere we denote $\langle \dots \rangle$ the average over initial conditions.
We assume that the initial velocity field is Gaussian distributed
and that its spectral support is around  a
characteristic wave number $k_0$. The averaged squared initial velocity is
$\bar{v}_0=   \langle \vv_{00}^2 \rangle^{1/2}$. The initial Reynolds number
is $\mathcal{R}=\bar{v}_0/(\nu k_0)$ which we assume to satisfy $\mathcal{R}\gg
1$. The initial  range (where viscosity is subdominant), given by $k_0\ll k \ll \bar{v}_0/\nu$, has no
particular structure, but
will develop, as time increases, into a self-similar inertial range where the energy
cascade (in three dimensions) will take place.

To stress the similarity with the FRG, we denote the equal-time velocity two-point function as
\begin{eqnarray}
 \langle \vv^{\alpha}_{\vu t} \vv^{\beta}_{0 t} \rangle = \Delta_{t,\alpha \beta}({\vu}) \quad , \quad
\langle \vv^{\alpha}_{\vk t} \vv^{\beta}_{\vk' t} \rangle = \delta_{ \vk-\vk'} \Delta_{t,\alpha \beta}(\vk) \ .
\end{eqnarray}
We denote $\delta_{\vk}:=(2 \pi)^N \delta^{N}(\vk)$, and in Fourier space
$\Delta_{t,\alpha \beta}(\vk)=P^{\rm L,T}_{\alpha \beta}(\vk) \Delta_t(\vk)$
for Burgers and Navier-Stokes respectively.
By definition, $\Delta_{\alpha\beta}({\vu})=\Delta_{\beta\alpha}(-{\vu})$,
while symmetry is only assured for isotropic turbulence. We study a system in
a periodic cube (torus) of volume $L^N$ with mostly two distinct cases:

 (A) fixed $t$ and $L \to \infty$ in which case we further restrict to isotropic (homogeneous) turbulence where $\Delta_t(\vk)$ depends only on $|\vk|$.

 (B) fixed $L$, which becomes periodic turbulence. In case (B) we use discrete Fourier modes $\sum_\vq$, which implicitly become $ \sum_\vq \to \int_\vq \equiv \int \frac{\rmd^N \vq}{(2 \pi)^N}$ in all formula below in case (A).

The total kinetic energy per unit volume, and the kinetic-energy spectrum are denoted as in \cite{LesieurBook}\bea
{\cal E}(t) = \frac{1}{2} \langle \vv^2 \rangle = \int_0^\infty \rmd k\, {\cal E}(k,t)\ .
\eea
However for convenience we also use a non-standard normalization for energy and
energy spectrum and denote ($S_N$ is the area of the unit sphere in dimension $N$):
 \bea
&&  E(t) := \frac{(2\pi)^N }{S_N} \langle \vv^2 \rangle =
  \int_0^\infty \rmd k\, E(k,t)\ ,\label{Enonst} \\
 && E(k,t):= \int \frac{ \rmd^{N }\vk}{S_{N}} \delta(k-|\vk|)  \Delta_{t,\alpha \alpha}(\vk)\ ,
\eea
which for isotropic turbulence becomes
\begin{eqnarray} \label{energyN}
 E(k,t) = k^{N-1} \Delta_{t,\alpha \alpha}(k)\ .
\end{eqnarray}
Hence ${\cal E}(k,t)=  \frac{1}{2} \frac{S_N}{(2\pi)^N} E(k,t)$.
The  decay of turbulence depends on the initial spatial behavior of the energy spectrum at large scales, hence we
denote (in the isotropic case)
\begin{eqnarray}  \label{initial}
 E(k,t=0) \sim_{k \to 0} c \, k^{n} \ ,\  \quad \Delta_{0,\alpha \alpha}(\vk) \sim_{k \to 0} c\, k^{n - N + 1}\ ,\ \quad \Delta_{0,\alpha \alpha}(\vu) \sim_{u \to \infty} c' u^{-n - 1}
\ .\end{eqnarray}
The total (kinetic) energy (per unit volume)
$E(t)=\int_0^\infty \rmd k\, E(k,t) \sim  \Delta_{t,\alpha \alpha}(\vu=0 )$
 must be finite (it  grows as $L^{n+1}$ for $n<-1$) and the initial velocity field is usually assumed to be smooth, i.e.\  $\Delta_{0,\alpha \beta}({\vu})$ is an analytic function in each of its components $\vu$  near $|\vu|=0$.

In all cases below, when the system reaches a statistically scale invariant decaying state,
we denote by
\bea
{\ell}(t) = t^{\zeta/2}
\eea
the characteristic length scale, which usually separates the inertial range from the large-scale region (i.e.\ $\ell(t=0) \sim 1/k_0$). The notation is motivated by the relation, in the Burgers case, to the roughness exponent $\zeta$ for random manifolds. We distinguish it from the exponent $\zeta_2$, which describes the leading short-distance singular behavior
of the two-point equal-time velocity correlator in the inviscid limit. We will also assume a {\it dissipation scale} $\ell_d(t) \ll \ell(t)$ which is the lower boundary of the inertial range,
and is often set to zero in the following, corresponding to the inviscid limit.

\section{Known results and phenomenology}
\label{s:Known results and phenomenology}

\subsection{Decaying Burgers}
In the following we use the mapping of the Burgers
equation onto the problem of a particle in the $N$-dimensional potential
$W = V({\bf u}_0) + \frac{1}{2 t} ({\bf u} - {\bf u}_0)^2$, where $V({\vu})$ is the random
potential which parameterizes the initial condition, i.e.\ $\vv_{\vu t=0}=\nabla_{\vu} V(\vu)$.
Denoting by $Z$ the canonical partition function of the particle at temperature $T=2\nu$,
 the velocity at all times is
$\vv_{\vu t}=\nabla_{\vu} \hat V(\vu,t)$, where $\hat V(\vu,t)=- T \ln Z$ is the free energy.
 In the inviscid limit it becomes $\hat V(\vu,t) = \min_{\vu_0} W$. Let us summarize what is expected for Burgers (most results are shown for $N=1$, and conjectured for $N>1$),
for more details see e.g.\ \cite{BecKhanin2007} and the discussion in Section E of \protect\cite{LeDoussal2008}
including connections to the FRG. We assume a smooth Gaussian initial velocity  field with the correlator
(\ref{initial}), i.e., with a spectrum proportional to $|k|^n$ at small $k$ and
decreasing quickly at large $k$.
For isotropic turbulence (A) there are two cases:\smallskip

(1) {\it Long-range initial condition (LR)} $n<1$:
The correlator of the random potential $\langle [V({\vu})-V(0)]^2 \rangle$ grows as $u^{1-n}$ at large $u$ and the particle is always in a glass phase, i.e.\  the effective viscosity scales to zero (see below). The evolution is expected to reach an asymptotic statistically scale invariant form $\vv_{\vu t}=\ell(t) t^{-1} \tilde {\bf v}(u/\ell(t))$ (in law) with $\ell(t)=t^{\zeta/2}$, where $\zeta=4/(3+n)$ and with energy decay $E(t) \sim t^{-2 + \zeta} = t^{-2(n+1)/(n+3)}$. Shocks, i.e.\  codimension-one manifolds (together with some additional lower dimensional singularities for $N>1$), where the velocity is discontinuous (at $\nu=0^+$) or nearly discontinuous (at small $\nu>0$), form in  finite time and, convected by the flow, keep merging when they meet. The growing scale of this coarsening process (quite complicated for $N>1$) is expected to scale as $\ell(t) \sim t^{\zeta/2}$. This is clear at least for $N=1$ \footnote{At least for dilute shocks $n<-1$; shocks are dense for $n=2$ and the analysis is more difficult.}. While the width of an isolated shock grows as $L_d(t) \sim \nu t$, the width of the {\it surviving} shocks grows as $L'_d(t) \sim \nu t^{1-\zeta/2}$  \cite{LeDoussal2006b,LeDoussal2008}, hence the rescaled width $L'_d(t)/\ell(t) \sim  \nu t^{1-\zeta} \equiv \nu_{\rm eff}$ scales to zero for $n<1$. This corresponds in the FRG to an attractive zero-viscosity (i.e.\ zero-temperature) fixed point (describing a glass phase for the particle):
One can write $\nu_{\rm eff} = \nu t^{\theta/2}$, i.e.\  $T_{\rm eff}=T t^{\theta/2}$. For a  random manifold, the
glass exponent is $\theta=d-2+2\zeta>0$, with here $d=0$. This is a LR fixed point, with exponents given by their dimensional values (also  called Flory values in the context of elastic manifolds), i.e.\  they are fixed by the initial condition. This property, called {\it the persistence of large eddies} in turbulence,
means that the energy spectrum for $k \ll 1/\ell(t)$ retains its original form (\ref{initial}) with an amplitude $c$ independent of time.

\smallskip

(2) {\it Short-range initial condition (SR)} $n>n_c=1$: This is the Kida
regime \cite{BecKhanin2007,Kida} with an asymptotic statistically scale invariant
form (see however below)
with a scale $\ell(t) = [t/(\ln t)]^{1/2}$ and a decay of the energy
$E(t) \sim 1/[t (\ln t)^{1/2}]$ (for gaussian initial conditions).
The (rescaled) shock width now grows (there is no glass phase); hence
it exists only for $\nu \to 0$ before $t \to \infty$ (see \cite{BouchaudMezard1997, FyodorovLeDoussalRosso2010} for
a more refined analysis of the double limit).

There is an additional crossover region (e.g.\ for  $1<n<2$ for $N=1$) where
the persistence of large eddies (i.e.\  of the tail of the FRG function)
still holds, but the system flows to the SR (Kida) fixed point: This is known as
the Gurbatov phenomenon \cite{GurbatovSimdyankinAurellFrischToth1997}, i.e.\  the velocity statistics is not
scale invariant. The resulting energy spectrum then consists of three regions:
(i) the ``outer region", $0< k< k_s(t)\sim t^{-1/[2(2-n)]}$,
where the velocity correlations preserve its initial form (\ref{initial}); (ii)
the ``inner region", $k_s(t)<k<{\ell(t)}^{-1} \sim t^{-1/2}$ ($\zeta=1$), with spectrum $k^2$,
and (iii) the shock-dominated region with spectrum $k^{-2}$ for $k> {\ell(t)}^{-1}$.
All scales are given up to logarithmic corrections and for $N=1$. More
details can be found in \cite{GurbatovSimdyankinAurellFrischToth1997}.
Within the FRG analysis, this crossover region can be seen as
crossover from the LR to the SR FRG fixed point\footnote{
For instance the value $n = 2$ corresponds to the Flory exponent $\zeta= 4/5$, and
at short scale the SR correlator of the random potential is behaves effectively as
$\delta(x) \sim 1/x$, while at large scale it flows to the SR Kida FP.}.
In the FRG analysis of random manifolds a similar crossover was
described in \cite{BalentsDSFisher1993}.

In the marginal case $n=1,$ a LR fixed point exists where the potential retains logarithmic correlations,
with a phase transition as a function of $\nu$ \cite{FyodorovLeDoussalRosso2010}.

Note that originally Burgers \cite{Burgers74} distinguished only two cases (for $N=1$),
assuming that $J=\Delta(k=0)=\int \rmd u \Delta(u)$ exists. The case $J>0$
then corresponds to the LR case $n=0$, hence $\ell(t) \sim t^{2/3}$ ($\zeta=4/3$) and
is usually called ``random-field" fixed point in the language of random manifolds.
The case $J=0$ was solved by Kida \cite{Kida}, and corresponds here to the SR case $n>1$
(for instance for $n=2$, $\Delta(k)$ is analytic and the random potential has $\delta$-correlations).
This is usually called the ``random-bond" fixed point in the language of random manifolds.
The summary presented above contains many more cases, i.e.\ the
LR models form a line of FPs, continuously parameterized by $n$, and
the SR case can also be modified by the Gurbatov LR-SR crossover.

Finally for the periodic case, the system converges, for $N=1$, to a single
random shock per period with $E(t)\sim t^{-2}$. This corresponds to the
FRG random-periodic FP $\zeta=0$ (i.e.\  $n=\infty$), and a similar picture should hold for any $N$.

\subsection{Decaying Navier Stokes}
\label{s:Known results NS}

A similar discussion can be given for decaying NS, though on a much less firm basis, mostly phenomenology, scaling arguments,
closure calculations and some support from experiments.
Again since Von Karman and Howarth \cite{KarmanHowarth1938}
one assumes a decaying state $\vv_{\vu t} = \ell(t) t^{-1} \tilde \vv(\vu/\ell(t))$ (in law).  Then $\tilde \vv$ satisfies a equation where $\nu \to \nu t/\ell(t)^2$, which flows to zero if $\zeta>1$. In Fourier this can be written $\vv_{\vk t} = \ell(t)^2 t^{-1} \hat \vv\big(\vk \ell(t)\big)$ (in law) and the energy spectrum takes the form $E(k,t)=\ell(t)^3/t^2 \tilde E\big(k \ell(t)\big)$
with $\ell(t) = t^{\zeta/2}$ and a total kinetic energy decay $E(t) \sim t^{-2 + \zeta}$. The persistence of large eddies, i.e.\ the invariance of $E(k,t) \simeq c k^n$, implies $\zeta=4/(3+n)$. This corresponds to a LR initial condition (i.e.\ regime (1) in the previous section). The energy spectrum can then be divided into a low-wavenumber range $k \ell(t) \ll 1$ with $E(k, t) \simeq c k^n$, and the inertial range $k \ell(t) \gg 1$, with, in 3D ($N=3$), ${\cal E}(k,t) \simeq C_K \epsilon(t)^{2/3} k^{-5/3}$ (assuming Kolmogorov spectrum) with $d{\cal E}/dt =- \bar \epsilon(t)$. For more details see
\cite{LesieurBook} VII-10, Ref. \cite{Shiromizu1998} for RG arguments,
and  \cite{schaeffer2008}.

There is agreement that this LR regime cannot hold for $n \geq 4$ because of the $C(t) k^4$ backtransfer in the energy spectrum discovered by Proudman and Reid \cite{ProudmanReid1954}
for the 3D Navier-Stokes equation, and found in EDQNM closure calculations \cite{LesieurBook} (analogous to the $k^2$ backtransfer for the Burgers
dynamics). In other words, the low-$k$ energy spectrum cannot be softer than $k^4$. For $n>4$ it is argued that the small-$k$ part of the spectrum is
replaced by $E(k,t) \sim C(t) k^4$ at small $k$, with the inertial range at large $k$. Because of the Gurbatov phenomenon (analogous to the situation in Burgers discussed above)
it is then argued \cite{EyinkThomson2000} that the LR regime cannot hold for $n<n_c$ with $3<n_c<4$. In the range $n_c < n < 4$ there are three spectral regions
$E(k,t) \simeq c k^n$ for $k < 1/\ell_*(t)$ (outer region), $E(k,t) \simeq C(t) k^4$ for $1/\ell_*(t) < k < 1/\ell(t)$ (inner region)
and finally the inertial range for $k>1/\ell(t)$, leading to a breakdown of global self-similarity. This global picture
seems compatible with experiments \cite{DitlevsenJensenOlesen2004,Comte-BellotCorrsin1971}.

Note that in 3D NS there is another conserved quantity, the
helicity $h=\epsilon_{\alpha \beta \gamma} \vv_\alpha \partial_\beta \vv_\gamma$.
It is locally fluctuating even if its average is
zero. If its average is non-zero, as in MHD, then we need
to consider
$\Delta_{\alpha \beta}({\vu}) \neq \Delta_{\beta \alpha}({\vu}) = \Delta_{\alpha \beta}(-{\vu})$.
Its presence makes possible a joint
cascade with two fluxes, one of energy and one of helicity,
both to small length scales~\cite{ChenChenEyink2003}.

\section{FRG equations}\label{s:FRG equations}

\subsection{Loop expansion: General strategy}

We now write FRG-like equations able to access directly the strong-coupling regime (i.e.\ finite non-linearity) using either a graphical
method or, equivalently, starting from the exact infinite hierarchy obeyed by the equal-time  $n$-point
correlation functions denoted here $\langle \vv^{\alpha_1}_{\vu_1 t} \vv^{\alpha_2}_{\vu_2 t} ... \vv^{\alpha_n}_{\vu_n t} \rangle = C^{(n)}_{\alpha_1...\alpha_n}({\vu_1,\vu_2,...,\vu_n})$. They obey, for $n\geq 2$,
\begin{eqnarray}  \label{hierarchy1}
 \partial_t C^{(n)}_{\alpha_1...\alpha_n}({\vu_1,\vu_2,...,\vu_n}) &=& n \nu \; {\rm Sym}\left[ \nabla_{\vu_1}^2
C^{(n)}_{\alpha_1...\alpha_n}({\vu_1,\vu_2,...,\vu_n})\right] \nn\\
&& - \frac{n}{2} \; {\rm Sym}\left[ P_{\alpha_1 ; \beta \gamma }(\nabla_{\vu_1})  C^{(n+1)} _{\beta\gamma\alpha_2...\alpha_n}({\vu_1,\vu_1,
 \vu_2,...,\vu_n})\right] \ .
\end{eqnarray}
The time dependence is implicit, and $P_{\alpha ; \beta \gamma }(\nabla)$ is
given by Eqs.~(\ref{PNS}) and (\ref{PBurgers})
rewritten in real space
using $i\mathbf{k} \leftrightarrow \nabla $;  symmetrization with respect to the $n$ pairs
$(\vu_i,\alpha_i)$ for $i=1,...,n$ is  denoted by $ {\rm Sym}[...]$. We recall that $C^{(2)}_{\alpha_1\alpha_2}({\vu_1,\vu_2})= \Delta_{t,\alpha_1\alpha_2}(\vu_{1}-\vu_{2})$. In Fourier space the hierarchy reads
\begin{eqnarray}
\partial_t C^{(n)}_{\alpha_1...\alpha_n}({\vk_1,\vk_2,...,\vk_n}) &=& - n \nu\; {\rm Sym}\left[ \vk_1^2
C^{(n)}_{\alpha_1...\alpha_n}({\vk_1,\vk_2,...,\vk_n})\right] \nn\\
&& - \frac{n}{2} \; {\rm Sym} \Big[ P_{\alpha_1 ; \beta \gamma}(\vk_1) \sum_{\vp+\vq=\vk_1}
 C^{(n+1)}_{\beta\gamma\alpha_2...\alpha_n} ({\vp,\vq,\vk_2,...,\vk_n})\Big]
 \end{eqnarray}
for the correlations $\langle \vv^{\alpha_1}_{\vk_1 t} ... \vv^{\alpha_n}_{\vk_n t} \rangle =
C^{(n)}_{\alpha_1...\alpha_n}(\vk_1,...,\vk_n) = \hat C^{(n)} _{\alpha_1...\alpha_n}({\vk_1,...,\vk_n}) \delta_{\vk_1+...+\vk_n}$,
and we also define $\hat C^{\alpha_1 \alpha_2}(\vk,-\vk)=C^{\alpha_1 \alpha_2}(\vk)$.

If equation (\ref{hierarchy1}) is considered in the inertial range,
i.e.\  all $|\vu_{ij}| \gg \ell_d$, it is expected (and for Burgers in some cases
shown) that one can neglect the viscosity term in the hierarchy. To study
the inertial range it is thus tempting to consider the limit $\nu=0^+$ of these
equations. To argue that this can be done, and that the result is still
given by equation~(\ref{hierarchy1}) setting $\nu=0$, we need two conditions:
(i) the physical requirement that the $n$-point velocity correlations $C^{(n)}$
are {\it continuous} functions, i.e.\ that limits at coinciding arguments exist;
(ii) the property that the $\nu \to 0$ limit of averages such
as $\langle \Phi(\vu_i) P^{\rm T}_{\alpha \beta}(\partial_{\vu}) \partial^\gamma_{\vu} ( \vv^\gamma_{\vu t} \vv^\beta_{\vu t} ) \rangle$
where $\Phi$ is any product of velocities with $\vu_i \neq \vu$ is equal
to $P^{\rm T}_{\alpha \beta}(\partial_{\vu}) \partial^\gamma_{\vu} \langle \Phi(\vu_i) \vv^\gamma_{\vu t} \vv^\beta_{\vu t}  \rangle$.
This appears to be related to the existence  of weak solutions of the Euler equation,
which is discussed in \cite{leray1933,leray1934,DuchonRobert2000,Gawedzki1999}.
In some cases, e.g.\ for  the inviscid Burgers equation and $N=1$, it can be
justified \cite{LeDoussal2006b,LeDoussal2008} from the dilute shock picture
of \cite{BernardGawedzki1998}.

Assuming that the $\nu=0$ hierarchy holds, let us describe the strategy of the loop expansion as it was constructed in the case of
Burgers \cite{LeDoussal2006b,LeDoussal2008}. It amounts to looking for a solution of the hierarchy in the schematic form (complicated convolutions  are indicated by $*$):
\begin{eqnarray}
 \partial_t \Delta &= t \Delta * \Delta + t^3 \Delta * \Delta * \Delta + ... \label{betaschem} \\
 C^{(3)} &= t \Delta * \Delta +  t^3 \Delta * \Delta * \Delta + ...  \label{C3}
\quad , \quad C^{(4)} = \Delta * \Delta + t^2 \Delta * \Delta * \Delta + ...\\
 C^{(5)} &= t \Delta * \Delta * \Delta + ... \quad , \quad \qquad\quad ~~~\, C^{(6)} = \Delta * \Delta * \Delta + ...\ .
\end{eqnarray}
It is illustrated here to two loops, and more generally $C^{(n)}=\sum_{q \geq 0} t^{2 q+\epsilon_q} C^{(n)}_q[\Delta]$ with $\epsilon_q=0,1$ for $n \geq 3$ respectively even and odd. We impose that at $t=0$ the distribution is Gaussian, hence  the functionals $C^{(2 k)}_0[\Delta]$, $k \geq 1$, are given by the Wick decomposition. This allows to compute iteratively all the terms in the beta function (\ref{betaschem}): e.g.\ to one loop we start from $C^{4}_0=3[\Delta \Delta]$, and use (\ref{hierarchy1}) with successively $n=3$ and $n=2$, first to get $C^{3}_0$, then to get the one-loop $\Delta^2$ term in (\ref{betaschem}). Thus, the beta function appropriate to the rescaling $\Delta=t^{-2} \tilde \Delta$ is  obtained directly. Counter-terms are produced automatically by successive corrections, due to consistent evaluations of $\partial_t \Delta$ terms at each step. Higher-loop calculations will be presented elsewhere \cite{FedorenkoLeDoussalWieseInPrep}. The first corrections to each {\it cumulant}, i.e.\  the $C^{(2 k)}_1[\Delta]$ and $C^{(2 k+1)}_0[\Delta]$ are the {\it tree approximation}. For calculations per se, an equivalent procedure, which we also performed to one loop using a graphical method directly on the Burgers and Navier-Stokes equations, is to compute $\Delta_t=\Delta_{t=0} + \sum_q t^{2q} \Delta_{t=0}^{*(q+1)}$ as a direct small-time expansion, then compute $\partial_t \Delta_t$ and re-express the result in terms of $\Delta_t$ itself as given in (\ref{betaschem}) by inverting the series. The information contained in the beta function can thus be described as a {\it renormalized} small-time expansion (for a direct small-time expansions see \cite{frishlarged,frisch}).

Note that the present FRG is different from the usual RG for turbulence as developed in
 \cite{Kraichnan1968,ForsterNelsonStephen1977, DeDominicisMartin1979,Mayo2005, AdzhemyanAntonovGoldinKimKompaniets2008, AdzhemyanAntonovKompanietsVasilev2003,AdzhemyanAntonovVasilev1989}. Here we keep the complete crucial information about $\Delta(k)$ (in Fourier) while the usual RG integrates out shells in $k$. In particular, the information in the small-$u$ behavior of $\Delta({\vu}) \sim u^{\zeta_2}$ determines the exponent $\zeta_2$.

\subsection{FRG equation for Burgers}

We now display the resulting one-loop $\beta$-function for Burgers.  Calculations  in real space are given in  \ref{realspaceRG}, and  in Fourier space in \ref{frgfourier}. A graphical derivation can be found in \ref{a6}.  The result reads \footnote{Note that we have assumed that $\Delta_{ab;b}(\vdelta)-\Delta_{ba;b}(\vdelta)$ is continuous and vanishes at $\vdelta=0$.}
\begin{eqnarray} \label{flow1}
&& \partial_t \Delta_{\alpha \beta}({\vu})= - t \partial_{\alpha} \partial_{\beta} \Big[ \Delta_{\gamma \delta}({\vu}) - \Delta_{\gamma \delta}(0)\Big]^{2}\ .
\end{eqnarray}
It has the usual form of the $N$-component one-loop FRG equation \cite{BalentsDSFisher1993,LeDoussalWiese2005a}.
To study the evolution from the initial condition (\ref{initial}), it is more convenient to
introduce the rescaled velocity correlation $\tilde \Delta_{t,\alpha \beta}({\vu})$ defined through
\begin{eqnarray} \label{scaling1}
&& \Delta_{t,\alpha \beta}({\vu}) = \frac{1}{4} t^{\zeta-\frac{\epsilon}{2}} \tilde \Delta_{t,\alpha \beta}(\vu/t^{\zeta/2})\ .
\end{eqnarray}
everywhere in this section we introduce:
\bea
\epsilon = 4
\eea
for Burgers, while $\epsilon=4-d$ for the manifold allows perturbative control. The rescaled velocity correlation satisfies
\begin{eqnarray} \label{FRGBurgers}
&& t \partial_t \tilde \Delta_{\alpha \beta}({\vu}) = \left(\frac{\epsilon}{2} -\zeta + \frac{\zeta}{2} \vu \cdot \nabla_{\vu}\right) \tilde \Delta_{\alpha \beta}({\vu}) - \frac{1}{4} \partial_{\alpha} \partial_{\beta} \Big[ \tilde \Delta_{\gamma \delta}({\vu})  - \tilde  \Delta_{\gamma \delta}(0)\Big]^2\ .
\end{eqnarray}
Using $\tilde \Delta_{\alpha \beta}({\vu})=- \partial_\alpha \partial_\beta \tilde R({\vu})=: - \tilde R''_{\alpha \beta}({\vu})$, it can be recast as an equation for the correlator of the random potential of the particle problem
\begin{eqnarray} \label{FRGBurgersR}
&& t \partial_t \tilde R(\vu)= \left(\frac{\epsilon}{2} - 2 \zeta + \frac{\zeta}{2} \vu \cdot \nabla_{\vu}\right) \tilde R(\vu)+   \frac{1}{4} \left[ \tilde R''_{\alpha \beta}(\vu)^2 - 2 \tilde R''_{\alpha \beta }(0) R''_{\alpha \beta }(\vu) \right]\ .
\end{eqnarray}
One recovers the familiar zero-temperature FRG equation  for manifolds, derived here directly for the inviscid Burgers problem in $N$ dimension, i.e.\ for $d=0$,
by identifying $t \partial_t = - \frac{1}{2} m \partial_m$ (for $N=1$ it was
obtained to 4 loops in \cite{LeDoussal2006b,LeDoussal2008}).

Let us focus on isotropic turbulence and denote $R(\vu) = h(u)$, the general case being very similar. The first property of the above FRG equations is that as long as the velocity flow is
smooth, $\Delta_{\gamma \delta}({\vu}) - \Delta_{\gamma \delta}(0) \simeq - \frac{1}{3!} h''''(0) (u^2 \delta_{\gamma \delta} + 2 u_\gamma u_\delta),$
hence from equation~(\ref{flow1}) the energy is conserved,
\bea
\partial_t { \mathcal{E}}(t) = { \frac12} \partial_t \Delta_{\alpha \alpha}(0)= 0\ ,
\eea
in agreement with standard knowledge for Burgers flows. However it is known (since Larkin, for review see e.g. \cite{BlatterFeigelmanGeshkenbeinLarkinVinokur1994}
in the context of elastic manifolds) that $h''''(0)$ diverges in a finite time $t_c$. Furthermore it is known since \cite{DSFisher1986,BalentsDSFisher1993,GiamarchiLeDoussal1995}
that the solution of the one-loop equation $\Delta_{t,\alpha \beta}({\vu}) $ develops a cusp at the origin,
more precisely $h(u) = h(0) + h''(0) \frac{u^2}{2}  + h'''(0) \frac{u^3}{6} +...$, a property which was found to hold also to next order (two loop) \cite{LeDoussalWiese2005a} and, from the physics
of shocks, is believed to hold to any order. Hence in the present context of
Burgers turbulence it implies non-conservation, and decay, of the kinetic energy,
\bea \label{decay1}
\partial_t { \mathcal{E}}(t) = { \frac12} \partial_t \Delta_{\alpha \alpha}(0) =%
 - t \frac{N(N+3)}{ 4} h'''(0)^2 \simeq  - \frac{N(N+3)}{ 4^3} %
 \tilde h'''(0)^2 t^{\zeta-3},
\eea
using that now $\Delta_{\gamma \delta}({\vu}) - \Delta_{\gamma \delta}(0) \simeq  - \frac{1}{2} h'''(0) u (\delta_{\gamma \delta} + \hat u_\gamma \hat u_\delta)$
where in the last equation we have substituted the scaling form (\ref{scaling1}) valid at large $t$.
In the large-time regime, $\tilde \Delta_{t,\alpha \beta}(u)$ flows to a fixed-point function $\tilde \Delta^*_{\alpha \beta}(u)$, which represents an asymptotic self-similar decaying solution. The energy decay can also be written as
\bea \label{Eburgers}
{ \mathcal{E}}(t) = { \frac{1}{8}}  \tilde \Delta^*_{\alpha \alpha}(0) t^{\zeta-2} \ ,
\eea
an exact relation for the amplitude if one knows the fixed point to all orders. To one-loop accuracy this is equivalent
to (\ref{decay1}). Equation~(\ref{Eburgers}) generalizes the result (324), or (325), in \cite{LeDoussal2008} to any dimension $N$.

As is well known from studies of the FRG equations \cite{DSFisher1986,BalentsDSFisher1993,GiamarchiLeDoussal1995},
 there are two types of fixed points. First a family of LR fixed points, such
that at large $u$, $\Delta_{\alpha \beta}(u) \sim 1/u^{n+1}$, and $R(u) \sim u^{1-n}$ can be obtained  by neglecting the non-linear terms
(which are subdominant at large $u$) and considering only the linear part (rescaling) of the FRG equation (\ref{FRGBurgers}).
This easily recovers $\zeta = \zeta_{\rm LR}(n)=\epsilon/(3+n)$, the result discussed in Section \ref{s:Known results and phenomenology}, and the
so-called persistence of large eddies. Second, the SR fixed point, for which only one value of $\zeta$
is possible, and which is obtained by shooting in the fixed-point equation (\ref{FRGBurgers}), (\ref{FRGBurgersR}) from $u=0$, asking for a fast decay of
$R(u)$ at infinity. This gives a non trivial $\zeta = \zeta_{\rm SR}= \frac{\epsilon}{4 + N} + \delta_N$, where $\delta_N$ decreases
exponentially at large $N$ \cite{BalentsDSFisher1993}. The LR behavior holds up to $\zeta(n_c) = \zeta_{\rm\ SR}$, hence
for small $\epsilon$ and large $N$ it suggests $n_c \approx N+1$.

On the other hand, we know that for $d=0$ and any $N$, the analog of the SR fixed point should be the one given by Kida, i.e.
$\zeta=1$, and (up to the Gurbatov LR-SR crossover) that $n_c=1$ separates LR from SR. Hence, we see
that while the LR regime is well captured by the FRG, i.e.\ the loop expansion in powers of $\epsilon$, the
SR exponent $\zeta_{\rm\ SR}$ and the SR FP of decaying Burgers is not well approximated. One reason seems to be that Kida physics is
controlled by rare events and extremal statistics, and seems to be
better captured by the replica-symmetry breaking (RSB) method, which even leads to some exact
results for the Kida FP \cite{BouchaudMezard1997, FyodorovLeDoussalRosso2010} (these can
be extended to any $N$). In fact even the $n=1$ marginal case also involves some replica-symmetry
breaking physics (as well as a connection to random matrix theory) \cite{FyodorovLeDoussalRosso2010}.

Of course the above discussion concerned scales larger than $\ell(t)$. For $u < t^{\zeta/2}$, i.e.\ in the
inertial range, the FRG gives the correct physics of shocks and energy transfer, with a cusp in $\Delta$. To which
extent the agreement (shock size distributions, etc.) can be made quantitative remains to be worked out in detail.

\subsection{FRG equation for Navier-Stokes in momentum space}

For Navier-Stokes the one-loop beta function is non-local in real space and thus
easier to display in Fourier space (for a real-space expression see \ref{realspaceRG}).
While the general case is displayed in   \ref{frgfourier}, we give here an
expression valid for the subspace of flows such that
 $\Delta_{\alpha \beta}(\vk)=P^{\rm T}_{\alpha \beta}(\vk) \Delta(\vk)$.
 For later use we introduce the potential $R(\vk)$ such that
 $\Delta(\vk)= k^2 R(\vk)$
 In real space we can also write
$\Delta_{\alpha \beta}(\vu)=P^{\rm T}_{\alpha \beta}(\partial_{\vu}) \Delta(\vu) = -(\delta_{\alpha \beta} \nabla_{\vu}^2 - \partial^\alpha_{\vu} \partial^\beta_{\vu}) R(\vu)$.
The function $\Delta(\vk)$ is then the Fourier transform of the trace $\Delta_{\alpha \alpha}(\vu)/(N-1)=\Delta(\vu)$, and the
potential function $R(\vk)$ is the Fourier transform of $R(\vu)$.

For $N=2$ all incompressible tensors can be written in this form,  and this is not a restriction; we can even use discrete Fourier sums.
For $N>2,$ this requires $\Delta(\vk)=\Delta(k)$ i.e.\  isotropic turbulence, and the $\vk$-continuum limit, i.e.\  an infinite box; the sums below  thus become momentum integrals, as explained in   \ref{frgfourier}. The FRG equations are
\begin{eqnarray} \label{NSFRG}
 \partial_t \Delta(\vk)& =&   \frac{2 t}{N-1}
\sum_\vq \tilde b_{\vk,\vk-\vq,\vq} \left[ \Delta(\vq) \Delta(\vk-\vq) -
 \Delta(\vq) \Delta(\vk) \right], \\
  \tilde b_{\vk,\vk-\vq,\vq} &=&  \frac{k^2 q^2 - (\vk \cdot \vq)^2}{k^2 q^2 (\vk-\vq)^2} \Big\{(k^2-q^2) \left[(\vk-\vq)^2 - q^2\right] + (N-2)
k^2 (\vk-\vq)^2 \Big\}\ .
\end{eqnarray}
Note that $\tilde b_{\vk,\vk-\vq,\vq}:= - P_{cjm}(\vk) P_{jbc}(\vk-\vq) P^{\rm T}_{mb}(\vq)$ is not invariant under $\vq \to \vk-\vq$, hence the first term can also be written in a symmetrized form, given in   \ref{frgfourier}.
We now use the rescaled variables
\begin{eqnarray}
&& \Delta_{t,\alpha \beta}(\vk) = t^{\zeta-2 + N \frac{\zeta}{2}} \tilde \Delta_{t,\alpha \beta}(\vk t^{\zeta/2})
\end{eqnarray}
to obtain
\begin{eqnarray}
&&  \! \! \! \! \! \! \! \! \! \! \! \! \! \! \! \! \! \! \!
\! \! \! \! \!\! \! \! \! \! \! \! \! t
\partial_t \tilde \Delta(\vk) =  \left(2-\zeta - N \frac{\zeta}{2}
- \frac{\zeta}{2} \vk \cdot \nabla_\vk\right)
\tilde \Delta(\vk) + \frac{2}{N-1}
\sum_\vq \tilde b_{\vk,\vk-\vq,\vq} \left[ \tilde \Delta(\vq) \tilde \Delta(\vk-\vq) -
 \tilde \Delta(\vq) \tilde \Delta(\vk)\right] \label{NSFRGresc}\ .
 \end{eqnarray}

Let us  point out that this one-loop FRG equation, i.e.\ the unrescaled form (\ref{NSFRG}), is very similar to the so-called Quasi-Normal approximation (QN). For $N=3$, one can
check that one recovers here the direct $\nu=0$ limit of equation~(VII-2-9) of Ref.\ \cite{LesieurBook}. Let us however point out  that the spirit here is a bit
different. First, we are looking at $\nu=0$ directly. Second, we are searching for a fixed point valid for all $k$, with the appropriate choice of
$\zeta$. Third, we use these equations as a first step in a systematic renormalized small-time (i.e.\ loop) expansion, which must be analyzed before
carrying out the loop-expansion program. In addition we have kept  $N$, the dimension of space, arbitrary.

In the turbulence ``closure" literature one often sees quoted the EDQNM, the Eddy damped quasi normal
approximation, which is believed to improve on the QN. It amounts to replacing in the above FRG equation $\tilde b_{\vk,\vk-\vq,\vq}  \to \tilde b_{\vk,\vk-\vq,\vq} \theta_{k,p,q}$.
where $\theta_{k,p,q}=1/\mu_{k,p,q}$ are the ``eddy damping rates", phenomenological parameters of the theory,
a standard choice being $\mu_{k,p,q}=\mu_k+\mu_p+\mu_q$;  there are two choices for $\mu_k$ either \cite{LesieurBook} $ \mu_k^2 = k^3 E(k)$ or $\mu_k^2 = a_1 (\int_0^k \rmd p\, p^2 E(p))^{1/2}$.
An interesting question, left for the future, is to understand how the
next order in the systematic (renormalized) small-time expansion
would compare with these phenomenological extensions.

\subsection{FRG equation for Navier-Stokes in real space: isotropic turbulence}
\label{ss:FRG-NS-real}
The flow equation (\ref{NSFRGresc}) for an
isotropic solution can be rewritten in real space.
As for Burgers we introduce the rescaled correlators via
\bea
R({\bf u}) = t^{2 \zeta-2} \tilde R({\bf u} t^{-\zeta/2})  \quad , \quad \Delta({\bf u}) = t^{\zeta-2} \tilde \Delta({\bf u} t^{-\zeta/2}) \ .
\eea
For isotropic turbulence, $R({\bf u})=R(u)$ and $\Delta({\bf u})=\Delta(u)$ (no numerical factor  is introduced for NS).
The flow equation  is  simpler in terms of the  function $\tilde {R}(\vu)$,
since a large part of equation~(\ref{NSFRGresc})  is {\em almost} local
as a function  of $\tilde R(\vu)$. We parameterize the
 solution as
\begin{eqnarray}
\tilde R(\vu) = r(y=u^2/2) \quad , \quad  \tilde \Delta(\vu) =
- \nabla^2_{\vu} \tilde R(\vu) = -  \big[N r'(y)+ 2y r''(y)\big]\ . \label{220}
\end{eqnarray}
Then equation~(\ref{NSFRGresc}) turns into
\begin{eqnarray}\label{FRG-real-space}
&& t \partial_{t} \tilde \Delta(\vu) =  (2 - \zeta) \tilde \Delta(\vu)
+ \frac \zeta2 \vu \cdot \partial_\vu \tilde \Delta(\vu) +
\delta \tilde\Delta_{\rm L} (\vu) + \delta \tilde\Delta_{\rm NL} (\vu).
\end{eqnarray}
The Laplacian of the first nonlinear term in equation (\ref{NSFRGresc}) is
local in real space and reads
\begin{eqnarray}
&& \fl   \nabla^2_{u} \delta \tilde\Delta_{\rm L} (\vu)  =
   2\left( 3 - N \right) \,N\,{\left( 2 + N \right) }^2\,
     {r''(y)}^2 - 8\,y^2\,r^{(3)}(y)\,
    \left[ \left(   5N^2+3N -40   \right) \,
        r^{(3)}(y) + \left( 8N-20  \right) y\,r^{(4)}(y)
       \right]  \nonumber  \\
 && \fl\hphantom{  \nabla^2_{u} \delta \tilde\Delta_{\rm L} (u)  = }  - 4yr''(y)\,
     \left[ \left( 2 + N \right) \,
        \left( 7N^2+ 3N -44   \right) \,
        r^{(3)}(y)
         + 4\,y\,
        \left( \left(  4N^2+ 6N   -26   \right)
             \,r^{(4)}(y) \right.
 \right.
  \left. \left.    +
          \left( 3 N -5  \right) y\, r^{(5)}(y) \right)  \right]
\nonumber \\
&& \fl \hphantom{  \nabla^2_{u} \delta \tilde\Delta_{\rm L} (u)  = }
        - 2\left( N -1\right) \,r'(y)\,
     \left[ \left( 4 + N \right) \,
        \left( \left( 2 + N \right) \,
           \left( N\,r^{(3)}(y) + 6\,y\,r^{(4)}(y) \right)  +
          12\,y^2\,r^{(5)}(y) \right)  + 8\,y^3\,r^{(6)}(y) \right].
        \ \ \ \ \label{99-2}
\end{eqnarray}
The second nonlinear term in the sum in equation (\ref{NSFRGresc})
is strongly non-local
in real space. Performing the angular average we obtain
\begin{eqnarray}
  &&  \delta\tilde \Delta_{\rm NL} (\vu)  =  -  \sum_\vk e^{i\vk\cdot\vu}
  \tilde R(k)
\sum_{\vp} A(k,p)p^2 \tilde R(p), \label{53-2}
\end{eqnarray}
where
\begin{eqnarray}\label{a44-2}
&&\fl  A(k,p) = \frac{k^4 (4 + N (4 N - 9) ) - 4 k^2 (N-1) p^2 + N p^4
}{2 N(N-1)} + \frac{(k^2-p^2)^3}{2 (N-1) (k^2+p^2)}\,
{}_{2}F_{1}\left(\frac{1}{2},1,\frac{N}{2},\frac{4 k^2 p^2}{(k^2+p^2)^2}\right)
\ .\end{eqnarray}
This expression considerably simplifies for $N=3$, with
\bea
{}_{2}F_{1}\left(\frac{1}{2},1,\frac{3}{2},\frac{4 k^2 p^2}{(k^2+p^2)^2}\right)=\frac{k^2+p^2}{2 k p}\mbox{atanh} \left(\frac{2 k p}{k^2+p^2}\right)\ ,
\eea
and especially for $N=2$ with
\bea
_{2}F_{1}\left(\frac{1}{2},1,1,\frac{4 k^2 p^2}{(k^2+p^2)^2}\right)=\frac{k^2+p^2}{k^2-p^2}\Theta(p<k) \ ,
 \eea
 see  equation~(\ref{53}) below.

\subsection{Energy conservation and energy anomaly.}

Let us note some properties of the FRG equation for NS. First, as long as $\Delta_{\alpha \beta}(\vu)$ is analytic at $\vu=0$ one has $\partial_t \Delta_{\alpha \beta}(0)=0$, which implies energy conservation. It can be seen by integrating equation~(\ref{NSFRG}) over $\vk$, and relabeling $\vp=\vk-\vq$ in the first integral and $
\vp=\vk$ in the second (also changing $\vq \to -\vq$ there), that
\begin{eqnarray} \label{Delta0}
 \partial_t \sum_\vk \Delta(\vk) =   \frac{2 t}{N-1} \sum_{p,q} \frac{\tilde b_{\vp+\vq,\vp,\vq} - \tilde b_{\vp,\vp+\vq,-\vq}}{p^2 q^2 (p+q)^2}
\Delta(\vq) \Delta(\vp) = 0\ ,
\end{eqnarray}
since the integrand vanishes by symmetrization of $\vp,\vq$. Since each integral contains terms of the form $\sum_\vq \vq^{\alpha_{1}}...\vq^{\alpha_{i}} \Delta(\vq)$, with $i=0,...,4$
(\ref{Delta0}) holds if $\Delta_{\alpha \beta}(\vu)$ is smooth enough.

Let us now recall where the energy anomaly comes from. The exact equation for $n=2$ in the NS hierarchy implies the exact relation
\begin{eqnarray} \label{exact1}
&& \frac{1}{2} \partial_t \langle \vv_{\vu't} \cdot \vv_{\vu t} \rangle =  \nu
\nabla_{\vu}^2 \langle \vv_{\vu't} \cdot \vv_{\vu t} \rangle + \frac{1}{4}
\nabla_{\vu'}^\alpha   \langle (\vv^\alpha_{\vu't} - \vv^\alpha_{\vu t})
(\vv_{\vu't} - \vv_{\vu t})^2 \rangle \ .
\end{eqnarray}
In the limit $\nu \to 0$ it can be read either in the dissipative region $\vu'=\vu$ (where the velocity is sufficiently smooth and the cubic term can be set to zero); or in the inertial range with $\vu' \to \vu$, where the first term is negligible.  This expresses the energy decay rate as
\begin{eqnarray}\label{25}
&& \partial_t {\cal E} = - \bar \epsilon = \lim_{\nu \to 0} \nu \langle \vv_{\vu t} \cdot \nabla_{\vu}^2 \vv_{\vu t} \rangle =
\lim_{\vu' \to \vu} \frac{1}{4}
\nabla_{\vu'}^\alpha   \left< (\vv^\alpha_{\vu't} - \vv^\alpha_{\vu t})
(\vv_{\vu't} - \vv_{\vu t})^2 \right>\ ,
\end{eqnarray}
with ${\cal E}=\frac{1}{2} \Delta_{\alpha \alpha}(u=0)$.
The relation ({\ref{exact1}) can be derived from equation~(\ref{c33}) in the Appendix
which leads to
\begin{eqnarray}
&& \partial_t \Delta_{\alpha \alpha}(\vu) = - 2 \nabla_{u}^\gamma C^{(3)}_{\gamma \beta \beta}(\vu,\vu,0)
= - 2 \nabla_{\vu}^\gamma \langle \vv^\gamma_{\vu t} \vv_{\vu t} \cdot \vv_{0t} \rangle\ ,
\end{eqnarray}
noting that $\nabla^\alpha_{u'} \langle \vv^\alpha_{\vu t} \vv_{\vu't} \cdot \vv_{\vu't} \rangle=0$ using translational invariance and incompressibility and that $C^{(3)}$ vanishes at coinciding points.

Hence if there is an energy anomaly $\bar \epsilon>0$, the above implies
$C^{(3)}_{\gamma \beta \beta}(\vu,\vu,0) \simeq \frac{\bar \epsilon}{N} \vu_\gamma$ at small $u$.
In the case of isotropic turbulence, using incompressibility, the third-order tensor can be parameterized by a single function of the distance $h_3(u)
$, with \(h_3(0)=0\) as
\cite{KarmanHowarth1938,landauFluid}
\bea \label{kh}
\fl \langle \vv^\alpha_{\vu} \vv^\beta_{0} \vv^\gamma_{0} \rangle = \frac{h_3(u)}{u} \vu_\alpha \delta_{\beta \gamma}
- \frac{1}{2} \Big[(N-1)\frac{h_3(u)}{u} +  h_3'(u)\Big]  ( \vu_\beta \delta_{\alpha \gamma}  + \vu_\gamma \delta_{\alpha \beta} )
+ \vu_\alpha  \vu_\beta  \vu_\gamma \frac{u h_3'(u)-h_3(u)}{u^3}\ .
\eea
In the small-distance limit,
\bea
&& C^{(3)}_{\alpha \beta \gamma}(\vu,0,0) \simeq h_3'(0) \vu_\alpha \delta_{\beta \gamma} - \frac{N}{2} h_3'(0) ( \vu_\beta \delta_{\alpha \gamma}  + \vu_\gamma \delta_{\alpha \beta} )
\quad , \quad h_3'(0) = \frac{2 \bar \epsilon}{N(N+2)(N-1)}\ .
\eea
This is often expressed as\bea
&& \left< (\vv^\alpha_{\vu}-\vv^\alpha_0) (\vv^\beta_{\vu}-\vv^\beta_0) (\vv^\gamma_{\vu}-\vv^\gamma_0) \right>
= \frac{- 4 \bar \epsilon}{N(N+2)} (\delta_{\alpha \beta} \vu_\gamma + \delta_{\alpha \gamma} \vu_\beta + \delta_{\beta \gamma} \vu_\alpha)\ ,
\eea
and in particular leads to Kolmogorov's $4/5$ law (for $N=3$),\bea
\left< \Big[(\vv^\alpha_{\vu}-\vv^\alpha_0) \cdot \frac{\vu}{u} \Big]^3 \right> \simeq \frac{- 12 \bar \epsilon}{N(N+2)} u\ .
\eea
Note that for isotropic turbulence the two-point velocity-correlation can be
written \cite{KarmanHowarth1938} as
\bea
\Delta_{\alpha \beta}(\vu) = \Big[f(u) + \frac{u f'(u)}{N-1} \Big] \delta_{\alpha \beta} - \frac{u f'(u)}{N-1} \hat u_\alpha \hat u_\beta\ ,
\eea
where the function $f(u)$ defined in \cite{KarmanHowarth1938} is related to the
potential function $R(\vu)=h(u)$,
as $f(u) = - (N-1) h'(u)/u$.
Using equation~(\ref{c3}), one obtains the exact relation between
the flow of $f(u),$
and the third-moment function $h_3(u)$ \cite{KarmanHowarth1938},
\bea
\partial_t f(u) = - \left[({N-1}) h_3'(u) + \frac{N^2-1}{u} h_3(u)\right]\ ,
\eea
generalized to any $N$, which recovers
the above value for $\bar \epsilon$ using
  $\partial_t {\cal E} =  \frac{N}{2} \partial_t f(0) =%
 - \frac12 N (N-1)(N+2) h'_3(0) = - \bar \epsilon$.
Now, in principle, from equation~(\ref{a13}) we have an expression for the third-order tensor (\ref{kh}), to lowest order
in the (renormalized) small-time expansion, hence we can in principle relate the function $h_3(u)$ to $\Delta(u)$
(within our lowest-order FRG). The expression however is highly non-local and the dissipation rate
$\sim h_3'(0)$ is not easy to calculate in general (while in Burgers it is, to one loop, simply proportional
to $\Delta'(0)^2$).

\protect\subsection{Enstrophy conservation and enstrophy anomaly in dimension 2.}
As is well known in dimension 2, $N=2$, the velocity field is sufficiently regular so that the energy is conserved (no dissipative anomaly), and the energy flows to large scales
(inverse cascade). There is no energy cascade towards small scale. We will show below that the solutions of the
FRG equation satisfy these properties.

In dimension $N=2$ one also considers the vorticity field $\omega_{\vu t}= \epsilon_{\alpha \beta} \partial_\alpha \vv^\beta_{\vu t}$.
Taking the curl of the unforced Navier-Stokes equation (\ref{NS}), one gets
\begin{eqnarray}
&& \partial_t \omega_{\vu t} + \vv_{\vu t} \cdot \nabla_{\vu}\, \omega_{\vu t} = \nu \nabla^2_{\vu} \omega_{\vu t} - \mu \,\omega_{\vu t}\ .
\end{eqnarray}
We have temporarily added a friction term $\mu$, which is often used to model dissipation at large scales. This implies
that
\begin{eqnarray} \label{ff}
&& \partial_t \int_{\vu} f(\omega_{\vu t}) = \int_{\vu} f'(\omega_{\vu t}) ( \nu \nabla^2_{\vu} \omega_{\vu t} - \mu \omega_{\vu t} )\ ,
\end{eqnarray}
since the convection term integrates to a surface term, using compressibility, which is discarded. Hence there is
conservation of any power, or function, of the local vorticity in the limit
$\nu \to 0$, provided the right-hand-side has a vanishing limit. In particular
\begin{eqnarray} \label{aw}
&& (\partial_t  + 2 \mu) \langle \omega_{\vu t}^2 \rangle = 2 \nu \langle \omega_{\vu t} \nabla^2_{\vu} \omega_{\vu t} \rangle\ ,
\end{eqnarray}
hence under regularity conditions in the inviscid limit (existence of $R^{(6)}(0)$ is sufficient) one
finds that the enstrophy $D(t)=\frac{1}{2} \langle \omega_{0t}^2 \rangle $ is conserved (setting
friction to zero),
\bea
\partial_t D(t) = - \frac{1}{2} \partial_t \nabla^2_{\vu} \Delta(0) = 0\ .
\eea
Note that (\ref{aw}) is the limit $u' \to u$ of the more general relation for the decaying Navier-Stokes equation
\begin{eqnarray}
(\partial_t + 2 \mu) \langle \omega_{\vu t} \omega_{\vu't} \rangle = 2 \nu \nabla^2_{\vu} \langle \omega_{\vu t} \omega_{\vu't} \rangle + \frac{1}{2}
\nabla^\alpha_{\vu}   \langle (\vv^\alpha_{\vu t} - \vv^\alpha_{\vu't})
(\omega_{\vu't} - \omega_{\vu t})^2 \rangle\ ,
\end{eqnarray}
which allows to relate enstrophy non-conservation (the enstrophy anomaly) to the non-smoothness of the flow.
The argument leading to (\ref{ff}) can be generalized to show that
\begin{eqnarray}
&& \partial_t \langle e^{\lambda \omega_{\vu t}} \rangle = - \mu \lambda  \partial_\lambda \langle e^{\lambda \omega_{\vu t}} \rangle +
\nu \langle \nabla_{\vu}^2 \omega_{\vu t} e^{\lambda \omega_{\vu t}} \rangle \ .
\end{eqnarray}
In the absence of an enstrophy anomaly the limit $\nu \to 0$ of the last term vanishes, and the solution is
\begin{eqnarray}
&& Z_t(\lambda) = Z_{t=0}(\lambda e^{- \mu t})\ ;
\end{eqnarray}
hence for $\mu=0$ the full probability distribution of local vorticities is conserved.

Until now these relations  were exact. One now checks that the FRG equation
does conserve enstrophy; indeed for a sufficiently smooth $\Delta(\vec u)$ one has
\begin{eqnarray} \label{vort0}
\partial_t \sum_\vk k^2 \Delta(\vk) = 0\ .
\end{eqnarray}
This is shown as above from ${\rm Sym}_{\vp,\vq} \big[ (\vp+\vq)^2 \tilde b_{\vp+\vq,\vp,\vq} - p^2 \tilde b_{\vp,\vp+\vq,-\vq} \big] = 0$, and  is valid  for $N=2$ only.

\section{Short-distance analysis, cusp or no cusp?}

\label{s:Short-distance analysis, cusp or no cusp?}

Here we study the behaviour of the velocity correlation $\Delta_{\alpha \beta}(u)-\Delta_{\alpha \beta}(0) \sim |\vu|^{\zeta_2}$ at small $u$, i.e.\  in the inertial range predicted by the FRG equation. To prepare for Navier Stokes, we first analyze Burgers, where we already know that the singularity is a linear cusp, i.e.\  $\zeta_2=1$, and we study it in Fourier space, since Navier Stokes is easier to express in Fourier space. Note that the analysis below can only {\it exclude} a range of values of $\zeta_2$, but to confirm that the selected values do  occur in the solution, one must solve the fixed-point equation.

\subsection{Burgers}
\label{s:FP:Burgers}

For decaying Burgers, the FRG equation reads, in Fourier space
\begin{equation}
  \! \! \! \! \! \! \! \! \! \! \! \! \! \! \! \! \! \! \! \! \! \! \! \! \! \! \! \! \! \! \! \!  \! \! \! \! \! \! \! \! \! \! \!  t \partial_t \tilde \Delta(\vk) =  \Big(2-\zeta - N \frac{\zeta}{2} - \frac{\zeta}{2} \vk \cdot \nabla_\vk\Big)
\tilde \Delta(\vk) + \sum_{\vq \neq 0,\vk-\vq\neq 0} \frac{k^2[\vq \cdot (\vk-\vq)]^2 }{2q^2 (\vk-\vq)^2} \tilde \Delta(\vq) \tilde \Delta(\vk-\vq)
- \sum_{\vq \neq 0}  \frac{(\vq \cdot \vk)^2}{q^2} \tilde \Delta(\vq) \tilde \Delta(\vk) \ .
\label{frgburgersf}
\end{equation}
It is true for any $N$ and any symmetry (periodic sums, etc...). Note that while
this equation seems to be non-local in real space, if one expresses it first
using $\Delta_{\alpha \beta}(\vk)=P_{\alpha \beta}^{\rm L}(\vk) \Delta(\vk) $,
then performs the Fourier transform, it  becomes local as a function of $\Delta_{\alpha \beta}(\vu)$.

We want to understand why there is a linear cusp, and why there can be nothing  but a linear cusp.
For that we start with the isotropic case and look for a solution which at large $k$ takes the form
\begin{eqnarray}
&& \label{asymptG}
\tilde \Delta(\vk) \sim A\, G(k) \quad , \quad G(k) = k^{-N - \zeta_2} g(m^2/k^2)
\end{eqnarray}
with $g(0)=1$, and in \ref{s:amplitudes} we note $ b =N+  \zeta_2$. A further assumption is that $g(p)$ admits an expansion of the form $1 - \frac{N + \zeta_2}{2} p^2 + ... $; the mass $m^2$  parameterizes the amplitude of the leading subdominant term, and its value is unimportant for the following \footnote{Functions $g(p)=1+p^a$ can also be tried but they lead to additional conditions hence do not change the main point of the discussion}. A convenient heuristic form in that case is $G(k)=(k^2+m^2)^{-(N+\zeta_{2})/2}$. One first notes that (\ref{asymptG}) implies that for  $\zeta_2>0$, both integrals in (\ref{frgburgersf}) are UV convergent, thus $\Delta(\vu=0)$ exists. The cusp corresponds to $\zeta_{2}=1$. In principle we can start by restricting our search to $\zeta_{2}<2$, i.e $\Delta(\vu)$ has no second derivative. If one tries to expand the first part of the non-linear term in equation~(\ref{frgburgersf}) for $q \ll k$, one finds that it cancels half of the second term, and that corrections are of the order of $\sum_q q^2 \Delta(q)$, which is a divergent series. The second half of the second term comes from the region $q-k \ll k$. Hence the integral in (\ref{frgburgersf}) is dominated by large $q$. One can thus insert the ansatz~(\ref{asymptG}) into equation~(\ref{frgburgersf}), which shows that,  up to the global factor of $A^2$,  the non-linear term behaves as
\begin{eqnarray}
 \fl   \int_\vq  \frac{1}{2} k^2 \frac{[\vq \cdot (\vk-\vq)]^2 }{q^2 (\vk-\vq)^2} G(q) G(|\vk-\vq|) - \frac{(\vq \cdot \vk)^2}{q^2} G(q) G(k)
=  B k^{2-\zeta _2-N} + C k^{2-2 \zeta _2-N} +D k^{-\zeta
   _2-N}   + ... \label{expansion}
\end{eqnarray}
It is easy to see that $B$ cancels between the two non-linear terms, see  \ref{s:Ints-Burgers}, and that for the ansatz
(\ref{asymptG}) to be a solution for  $\zeta_2<2$,  we need that the amplitude $C=C(N,\zeta_2)$ vanishes. The latter is computed in  \ref{s:Ints-Burgers}, equation~(\ref{Capp}) as
\begin{eqnarray}
&& C = 2^{-1 - N + \zeta_2} \pi^{1/2 - N/2} \frac{\left[N + \zeta_2 (2 + \zeta_2)\right] \Gamma\big(-\frac{\zeta_2}2\big) \Gamma\big(\frac N 2 + \zeta_2\big)}{
\Gamma\big(\frac{1 - \zeta_2}2\big) \Gamma\big(\frac{2 + N + \zeta_2}2\big)^{\!2} }= 0\ .
\end{eqnarray}
It has a unique solution $\zeta_2=1$ in the interval $\zeta_2 \in [0,2]$, and
thus there  only is a solution with a cusp. \footnote{Note however that there are other roots, i.e.\  $\zeta_2=2 p+1$, $p=1,2,...$ which are
potentially possible behavior. For $N=1$ such solutions are $\tilde \Delta(u)-\tilde \Delta(0) = \sum_{m=1}^\infty a_{2m} u^{2m} + a_{2p+1} u^{2p+1} + ...$, and
are formally possible solutions at small $u$, but do not seem to correspond to globally physical fixed points.}

\subsection{Navier-Stokes}
\label{s:FP:NS}
Inserting the ansatz (\ref{asymptG}) in the Navier-Stokes FRG equation (\ref{NSFRGresc}),
since $
{ \tilde b}_{\vk,\vk-\vq,\vq} \sim k \cdot q$ at large $q$, the second integral
is UV convergent for { $\zeta_2>0$.}
An analysis similar to the previous section yields
\begin{eqnarray}
\fl
\frac{ 2}{N-1}  \sum_\vq { \tilde b}_{\vk,\vk-\vq,\vq} \Big[ G(q) G(|\vk-\vq|) -
 G(q) G(k) \Big] =  B k^{2-\zeta _2-N} + C k^{2-2 \zeta _2-N} +D k^{-\zeta
   _2-N}   + ... \label{expansion2}
\end{eqnarray}
The contributions to the amplitude $B$ again cancel between the two integrals,
hence $B=0$.
The amplitude $C=C(N,\zeta_{2})$ is computed in  \ref{s:Ints-NS}:
\begin{equation}
\fl
C = \frac{- \sqrt{\pi}}{4 (4 \pi)^{N/2}} \left[
\frac{2^{\zeta_2} [(N-2) \zeta_2 - (N-1)] (N + \zeta_2) \Gamma(-\frac{\zeta_2}{2}) \Gamma(\frac{N}{2} + \zeta_2)}{
\Gamma(\frac{3 - \zeta_2}{2}) \Gamma(\frac{2 + N + \zeta_2}{2})^2}
- \frac{ 4 \sqrt{\pi} N \Gamma(\frac{N}{2})}{\sin(\frac{\pi \zeta_2}{2}) \Gamma(\frac{4+N-\zeta_2}{2}) \Gamma(\frac{N+\zeta_2}{2})} \right]= 0  \ .\label{Ctext}
\end{equation}
Solutions of this equation are plotted on figure \ref{CofN}. First, one can check that for any $N$, the function $C(N,\zeta_{2})$ vanishes
for $\zeta_2=1$: $C(N,1)=0$. Next one finds that for $N>N_2=2.1155$, the root $\zeta_2=1$ remains  unique  in $[0,2]$
and  {\it the linear cusp is the only possible solution}.
For $N<N_2$, an additional pair of solutions appears on both sides
of $\zeta_2=0.87...$. The largest one reaches $1$ at $N_3=2.1145$,
while the smallest root is $\zeta_2\approx 0.8$. For $N<N_3$ the
two additional roots are on both sides of $1$ and, as $N \to 2^+$,
one reaches 0 while the other reaches $2$. For $N =2$ (and $N<2$)
$C(N,\zeta_{2})$ is decreasing as a function of $\zeta_{2}$ for
$\zeta_{2}\in[0,2]$ and the cusp is again the unique root \footnote{For $N<0$, there are again additional solutions for
$\zeta_{2}\in [0,2]$, but for NS these roots are not  physically interesting.}.
\begin{figure}
\centerline{\Fig{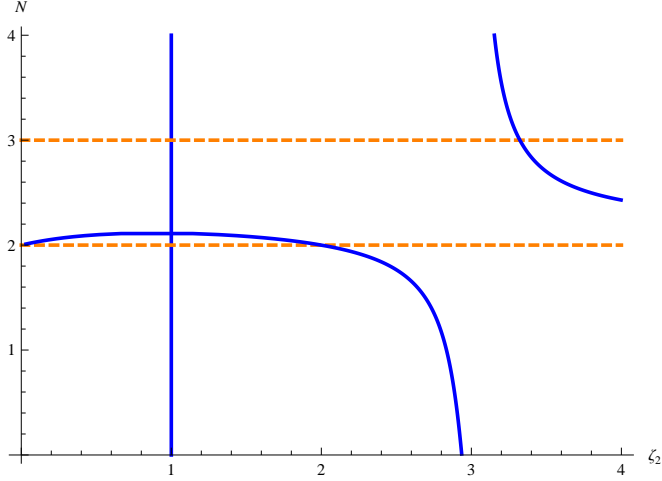}}
\caption{Blue solid curves: Locations in the $(\zeta_{2},N)$ plane, where $C$ given in equation~(\ref{Ctext}) vanishes. Orange dashed lines: $N=2$ and $N=3$. For $N=3$ the solutions in $[0,4]$ are $\zeta_{2}=1$ and $\zeta_{2}=3.32358$. For $N=2$ they are $\zeta_2=0, 1,2$. Note that the maximum of the lowest curve is at
$N=N_2=2.1155$, $\zeta_2=0.87...$ hence in the (very small) interval $2 \leq N  < N_2$ there are two roots on both sides of this maximum (in addition to the
root at $\zeta_2=1$). At $N=N_3=2.1145$ the
upper root reaches $\zeta_2=1$.}
\label{CofN}
\end{figure}

In addition to the cusp, there are other roots with $\zeta_2>2$. For $N > 2.43...$ one finds that $C$
vanishes exactly once in each interval $\zeta_2 \in [2p,2p+2]$, $p=0,1,2,...$ and diverges to $\pm \infty$ for $\zeta_2=(2 p)^{\mp}$. In the other intervals, $p \geq 1$, the root tends to $\zeta_2=2p+1$ for large $N$. For $N=3$ the other roots are at $\zeta_2=3.32358, 4.98205, 7.,...$. For $2<N < N_4 = 2.43...$ the root for $\zeta_{2}\in[2,4]$ may not exist in which case there is a double root in the interval $[4,6]$. One root crosses $\zeta_2=4$ at $N=N_4$.
For $N=2$ the roots are $\zeta_2=5.02421,7.0006,...$.

A peculiar result is that as $N \to 2^+$ one root tends to $2^{-}$, but then seemingly disappears for $N=2$,
suggesting that this case has to be treated with more care. Indeed, we will see in section \ref{s:N=2:largeK} that $N=2$ and $\zeta_{2}=2$ is indeed a solution. The calculations are rather non-trivial, since the integrals are not defined without proper regularization.

 To conclude on an optimistic note, although the cusp seems to be the only solution for $N=3$, we did find some
non-trivial values for $\zeta_2$ for $N$ slightly larger than 2. One possible scenario may be that these become valid in a
larger domain in $N$ when higher loop corrections (higher powers of time) are
included.

\section{Two-dimensional decaying turbulence ($N=2$)}
\label{s:NS:d=2}
\subsection{Basic properties}
In dimension $N=2$, since Kraichnan and Batchelor it is believed that
\cite{Batchelor1969,LesieurBook}:

(i) the energy flow is to small $k$ while the enstrophy flow is to large $k$;

(ii) since there is no direct energy cascade there is no
energy anomaly, i.e.\ $\lim_{\nu \to 0^+} \nu \Delta''(0)=0$, and energy is conserved
i.e.\  $ \partial_t \Delta(0)= \frac{1}{2 \pi} \partial_t  \int \rmd k E(k,t)=0$,
once one neglects dissipation on large scales due to e.g. friction;

(iii) there is an enstrophy cascade, i.e.\  there
is an enstrophy anomaly $\lim_{\nu \to 0^+} \nu \Delta''''(0) \neq 0$
and enstrophy is not conserved 
$ - \partial_t \Delta''(0)= \frac1{2\pi} \partial_t \int \mathrm{d}k\, k^2 E(k,t)<0$.

Let us see how these feature arise from the FRG equation. To facilitate the calculations, we introduce the stream function $\psi$ and the vorticity $\omega$ such that $\vv_{\vu t}^\alpha=\epsilon_{\alpha \beta} \partial_\beta \psi_{\vu t}$ and $\omega_{\vu t}= \epsilon_{\alpha \beta} \partial_\alpha \vv^\beta_{\vu t} = - \nabla^2 \psi_{\vu t}$. In terms of the stream function, $R$ and $\Delta_{\alpha\beta}$ are
\begin{eqnarray}
&& R(\vu-\vu') =\langle \psi_{\vu t} \psi_{\vu't} \rangle \ ,\\
&& \Delta_{\alpha \beta}(\vu) =  -  (\delta_{\alpha \beta}{ \nabla^{2}} - \nabla_\alpha \nabla_\beta) R(\vu)\ .
\end{eqnarray}
For isotropic turbulence
$\Delta(\vu)=\Delta_{\alpha \alpha}(\vu)/(N-1)=\Delta_{\alpha \alpha}(\vu)$   and
$\Delta(k)=k^2 R(k) =  E(k)/k$.

\subsection{Isotropic turbulence}

\subsubsection{FRG equations.}

In section \ref{ss:FRG-NS-real} the  FRG equation has been given in real space. We remind that one parameterizes the isotropic solution as
\begin{eqnarray}
\tilde R(u) = r(y=u^2/2) \quad , \quad  \tilde \Delta(u) = - \nabla^2_{\vu} \tilde R(u) = - 2 \big[r'(y)+ y r''(y)\big]\ . \label{220bis}
\end{eqnarray}
 For isotropic turbulence and $N=2$, equations (\ref{99-2})--(\ref{a44-2}) simplify to
\begin{eqnarray}\label{51}
&&\fl t \partial_{t} \tilde \Delta(u) =  (2 - \zeta) \tilde \Delta(u) + \zeta y \partial_y \tilde \Delta(u) + \delta \tilde\Delta_{\rm L} (u) + \delta \tilde\Delta_{\rm NL} (u),
  \\
&&\fl\delta \tilde \Delta_{\rm L} (u)=  8 \left[  3\,\int _{\infty}^{y}{r''(x)}^2\,\rmd x - 2 r'(y)\,r''(y) +
  3 y\,{r''(y)}^2 - 4 y\,r'(y)\,r^{(3)}(y) +
  y^2\,r''(y)\,r^{(3)}(y) - y^2\,r'(y)\,r^{(4)}(y)\right],     \label{52}
\\
  &&\fl  \delta\tilde \Delta_{\rm NL} (u)  =  -  \int_\vk e^{i\vk\cdot\vu}   \tilde R(k)
\int_{|{\vq}|<k} (k^2-q^2)^2 q^2 \tilde R(q), \label{53}
\end{eqnarray}
where $\int_{|{\vq}|<k} :=\int_0^k \frac{q \rmd q}{2 \pi}$.
The integration constant has been fixed assuming that $\tilde \Delta(u)$, and thus the above correction, vanishes at $|\vu|=\infty$.

The above can be turned into an equation for $r'(y)$:\
\begin{eqnarray}
\fl t \partial_{t} r'(y) =  (2 - \zeta) r'(y) + \zeta y \partial_y r'(y) + 4 \bigg[
2 r'(y) r''(y) - y r''(y)^2 + y r'(y) r'''(y)  - 3
\int_{\infty }^{y} \rmd x\,  r''(x)^{2} \bigg] \nn\\
\fl ~~~~~~~~~~~~~~+ \frac{4}{\pi}
\frac{\rmd}{\rmd y} \int_{0}^{1} {\rmd \lambda} \, (\lambda-1)\,
\int_{0}^\infty {\rmd z}\, \int_{0}^{2\pi } \rmd \theta \, \left[ f
(y+\lambda z+2 \sqrt{\lambda yz}\cos\theta)-f (y)\right] f'
(z), \\
\fl f (x) = [xr' (x)]'= r' (x)+x r''(x)\ .
\end{eqnarray}In Fourier space, the FRG equation reads
\begin{eqnarray}\label{F.13}
t \partial_t \tilde \Delta(k) =  (2- 2 \zeta)  \tilde \Delta(k) - \frac{\zeta}{2} k \tilde \Delta'(k) + \delta \tilde \Delta(k)\ .
\eea
Using the distance-geometry representation of  \ref{distance geometry}, the sum of the  two non-linear terms $\delta \tilde \Delta(k) = \delta \tilde \Delta_{\rm L} (k) + \delta \tilde \Delta_{\rm NL} (k) $ can be rewritten in Fourier space  as
\begin{eqnarray}\label{F.13bis}
\fl  \delta \tilde \Delta(k):=
\frac{k^4}{4 \pi^2} \int_{1}^{\infty}\rmd s
\int_{-1}^{1}\rmd t
\left[ (s-t)^2 - 4\right] \frac{s t}{s^2-t^2} \sqrt{(s^2-1)(1-t^2)}\, \tilde \Delta\Big(\frac{k}{2} (s-t)\Big) \left[ \tilde \Delta(k)-\tilde \Delta\Big(\frac{k}{2} (s+t)\Big)\right]\ .
 \end{eqnarray}
 We now study the scaling form and the asymptotic behavior
of the fixed-point solution ${\Delta}^*(k)$ for large and small~$k$.

\subsubsection{Searching for an isotropic fixed point}
\label{2d-isotrop-fp-search}
We  look for a fixed point of the form
\begin{eqnarray}\label{55}
&& \Delta(u) = t^{\zeta-2} \tilde \Delta(u t^{-\zeta/2}) \quad , \quad \Delta(k) = t^{2 \zeta-2} \tilde \Delta(k t^{\zeta/2}) \ .
\eea
The asymptotic behaviors at small and large distances are
\bea
&& \tilde \Delta(0)-\tilde \Delta(u) \sim u^{\zeta_2}\quad~~~\mbox{\quad for } u \ll 1 \qquad \Leftrightarrow \qquad \tilde \Delta(k) \approx A  k^{- (2+\zeta_2)} \mbox{\quad for }  k \gg 1 \label{2dlarge} \\
&& \tilde \Delta(u) \sim u^{-(1+n)} \mbox{\quad ~~~~~~~~~~for }  u \gg 1 \qquad \Leftrightarrow \qquad \tilde \Delta(k) \approx A_2 k^{n-1} \mbox{~~~\quad for } k \ll 1 \quad . \quad  \label{2dsmall}
\eea
valid only up to logs (and for $\tilde \Delta(u)$ at large $u$ only an upper bound since for integer $n$ a faster decay is possible from analyticity
in Fourier space).

Consider now the mean kinetic energy  $E(t) ={ 2\pi}\langle {{\bf v}^2}\rangle $ which
is given in the non-standard units introduced in equation (\ref{Enonst}).
If we assume scaling, then
\begin{eqnarray}
&& E(t) =  \int_0^\infty  \Delta(k)  {k\, \rmd k} = t^{2- \zeta} \int_0^\infty  \tilde \Delta(p) {p\, \rmd p}\ .
\end{eqnarray}
If $\zeta_2$ is large enough, the energy should be conserved (the cusp seems necessary for the violation so let us consider $\zeta_2>1$). Then the value naturally compatible with energy conservation is $\zeta=2$. In the context  of disordered system this is  called the Larkin exponent, i.e.\ the dimensional reduction exponent $\zeta=\epsilon/2$ with $\epsilon=4$.

Furthermore Batchelor \cite{Batchelor1969} proposed that $E(k,t)=v^3 t f(k v t)$ which implies conservation of energy, if $\int \rmd z f(z)$ converges. This is again $\zeta=2$.
Let us recall that
$E(k,t)= k \Delta_t(k)$. It also implies a decay of the total enstrophy, i.e.\ of $\int \rmd k\, k^3 \Delta_t(k)$, proportional to $t^{-2}$ if $\int \rmd z\, z^2 f(z)$ converges.
Assuming that $E(k)$ is independent of $v$ at large $k$ implies $f(x) \sim x^{-3}$ and $E(k,t) \sim t^{-2} k^{-3}$ at large $k$. This leads to
\begin{eqnarray}
&&\zeta_{2}=2\ .
\end{eqnarray}
This is indeed the only solution we found to be consistent with our analysis
of the flow equations \footnote{from the FRG equation in Fourier at large $k$ one sees that
   $\delta \tilde \Delta^*(k)/\tilde \Delta^*(k)$ should go to a constant at large $k$ equal to $  - (2 - \zeta + \frac{1}{2} \zeta \zeta_2)$.
   This can be checked numerically and we found it holds only for $\zeta_2=1,2$ and that $\zeta_2=1$ can be excluded
   as it would be compatible only with $\zeta>4$ a value much too large}
as we discuss below. At small $k$ it behaves as $\Delta_t(k) \sim t^4 k^2$.
Note that it {\it is not} a long-range fixed
point with $n=-1$, which would also give $\zeta=2$ according to the general
discussion of section~\ref{s:Known results NS}. The reason is that the
amplitude of the $k^2$ depends on time (in addition the energy would be infinite). Rather it corresponds to $n=3$, but a short-range fixed point.}

We now consider the general properties of the small- and large-$k$ expansions of
the fixed-point solution $\tilde{\Delta}^*(k)$. Since the fixed-point equation
is neither local  in real space nor in Fourier space, the expansions of
interest are expected to contain unavoidably {\em global} properties of the
fixed point
$\tilde{\Delta}^*(k)$.

\subsubsection{Small-$k$ expansion.}
The expansion of the nonlinear term (\ref{F.13bis}) in the distance geometry
representation  is given  in~\ref{sec:app-small-k}. To lowest orders
it reads
\begin{eqnarray}\label{F.18bis-A}
\delta \tilde \Delta(k) &=& \frac{k^2}{4 \pi}  \int_{0}^\infty \rmd q\,  q
  \tilde \Delta(q)^2 -\frac{k^4}{16 \pi}  \int_0^\infty\rmd q\, q \tilde \Delta'(q)^2
  + O(k^{6}),
\end{eqnarray}
where we used the small-$k$ behavior (\ref{2dsmall}) that implies, for $n>2$,
$\tilde{\Delta}(k=0)=\tilde{\Delta}'(k=0)=0$. Assuming that
$\tilde{\Delta}^*(k)=A_2k^2+A_4k^4+...$ we find from the FRG equations in Fourier space:
\begin{eqnarray}
&& (3 \zeta - 2) A_2 = 4 A_2 = \frac{1}{4 \pi}  \int_0^\infty \rmd k\, k
\tilde \Delta(k)^2,   \label{eqA2}\\
&& (4 \zeta - 2) A_4 = 6 A_4 = -\frac{1}{16 \pi}  \int_0^\infty\rmd k\, k
\tilde \Delta'(k)^2.
\label{eqA4}
\end{eqnarray}

\subsubsection{Large-$k$ expansion.} \label{s:N=2:largeK}
As shown in Section~\ref{2d-isotrop-fp-search}, the large-$k,$ i.e.\ small-$u$
asymptotics of the fixed-point solution is given by equation~(\ref{2dlarge}). Assuming that
$\Delta(k)=A/{k^4} + O(1/k^6)$ we find in \ref{s:NS-D=2-large-k-Andrei}, that
the nonlinearity gives
\begin{eqnarray}\label{76}
\delta \tilde \Delta(k) = -\frac{A^2}{8\pi k^4} + O\left(\frac{\ln k}{k^6}\right).
\end{eqnarray}
Inserting equation~(\ref{76}) into the flow equation with $\zeta=2$ and
asking that it be at a fixed point yields
\begin{eqnarray}
A=16\pi  \quad  \leftrightarrow   \quad \tilde \Delta(k) \simeq \frac{16 \pi}{k^4}
\end{eqnarray}
at large $k$, which implies $\Delta(k) \simeq \frac{16 \pi}{t^2 k^4}$,
noting that this power of $k$ is preserved by rescaling by $t^{\zeta/2}$
in (\ref{55}). This means that in real space, using the definition (\ref{220}) to pass to the second line:
\begin{eqnarray}
\tilde\Delta(u) - \tilde\Delta(0) &=&  2 u^{2}\ln u +...  \\
~~~~~~~~~~ \tilde R(u) &=& - \tilde \Delta(0) \frac{u^2}{4}  - \frac{1}{8}
u^{4}\ln u +...
\end{eqnarray}

\subsubsection{Numerical solution}
\begin{figure}
\centerline{\fig{10cm}{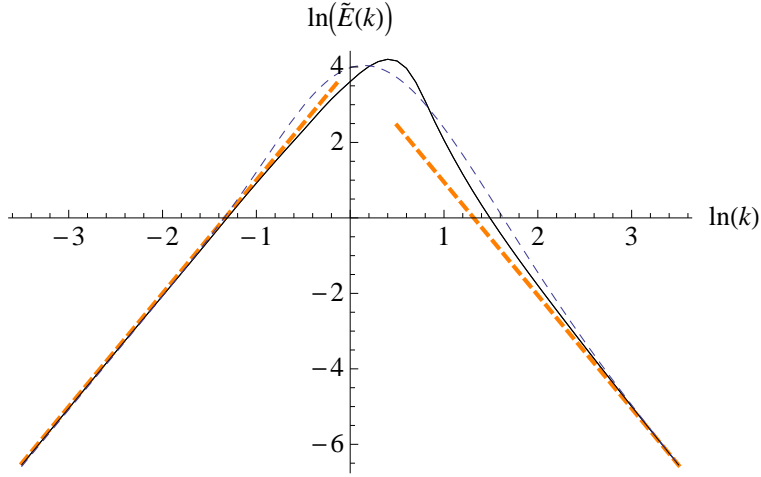}}
\caption{Double logarithmic plot for the rescaled energy \(\tilde E(k):=k \tilde \Delta (k) \) as a function of
$k$. The solid black line is our solution of equation~(\ref{51}); the thick dashed lines show the asymptotic slopes $\pm3$. The thin grey-blue dashed line is  $\tilde E_{\rm guess}(k) := k \tilde \Delta_{\rm guess}(k)$, see equation~(\ref{Deltaguess}).}
\label{f:2D-E(k)}
\end{figure}
To find numerically a fixed-point solution $\tilde \Delta^{*}(u)$ of equation~(\ref{51}) is highly non-trivial. Since equation~(\ref{51}) is an integral-equation, none of the standard techniques, such as Taylor-expansion, or solution as an eigenvalue problem are
available. From a decent physical fixed point, we expect that it is attractive w.r.t.\ all (sensible and small) perturbations. Thus, if we propose a   guess   $\tilde \Delta_{\mathrm{guess}}(u)$, which satisfies the above mentioned asymptotic behaviors and constraints, and is close to the true solution, it should converge against the fixed point. This is what we  succeeded in doing, starting from \(\Delta_{\mathrm{cor}} (k)=0\):
\begin{eqnarray}\label{70}
\tilde \Delta (k) = \tilde \Delta_{\mathrm{guess}} (k) + \tilde
\Delta_{\mathrm{cor}} (k),\\
\label{Deltaguess}
\tilde \Delta_{\mathrm{guess}}(k) = 16 \pi  \left(\frac{k^2}{\left(k^2+1\right)^3}+\frac{54.7237   k^6}{\left(k^2+1\right)^6}+\frac{2.65177 k^4}{\left(k^2+1\right)^5}\right)\ .
\end{eqnarray}
The main problem then was that numerically the flow-equation (\ref{51}) is rather unstable. The technique we finally succeeded in getting to work was:
Starting with $\tilde \Delta_{\rm corr}(k)=0$, we recursively inject $\tilde \Delta_{\rm corr}(k)$ into the flow equation (\ref{51}), and use the latter to evolve $\tilde \Delta(k)$ during a small time-step, giving us an improved approximation for  $
\tilde \Delta (k)$, calculated for approximately 100 $k$-points.  The latter is then  projected onto an optimal spline
with only 20 supporting points, or more precisely a non-linear transformation thereof. This procedure is numerically much more stable than using a best polynomial fit, a Fourier representation, or any of the other known sets of orthogonal functions we tried. The projection effectively smoothes the function, while still   capturing the necessary details. The complete technical details can be found in  \ref{a6c}, most importantly a check of the convergence of the function, see figure \ref{fig:precision}, as well as its tabulated values.
Here we  illustrate the result in the form of the classical double-logarithmic plot for the energy as a function of
$k$, see figure \ref{f:2D-E(k)}. The small-\(k\) asymptotics is \(E(k)\sim k^3\), and the large-\(k\) asymptotics is  \(E(k)\sim k^{-3}\). We remark that the solution remains below the asymptotic small-$k$ behavior, but then converges from above towards the asymptotic large-$k$ behavior.

\subsubsection{Physical interpretation of the solution}
We have found a fixed point  \(\tilde \Delta(k)\) for 2D decaying turbulence. Having in mind equation~(\ref{55}), the velocity-velocity correlation is \(\Delta(k) = t^2 \tilde \Delta( t k) \)
where $\tilde \Delta$ is time independent. Thus \begin{equation}
t \partial_{t} \Delta(k) = 2 \Delta(k) + k \Delta'(k).
\end{equation}
This implies similar relations for the time evolution of energy $E(k)=k \Delta(k)$ and enstrophy $\Omega(k)=k^3 \Delta(k) $
\begin{equation}
t \partial_{t} E(k) = 2k\Delta(k) + k^2 \Delta'(k)\  , \qquad t \partial_{t} \Omega(k) = 2k^3 \Delta(k) + k^4 \Delta'(k)\
.
\end{equation}
We define scaled momentum-dependent energy and enstrophy decay rates, written in terms of the scaled momentum $\tilde k=k t^{\zeta/2}=k t$ and
correlator as:
\bea
&& \tilde{\dot E}(\tilde k) := \partial_{t} E(k) = 2 \tilde k \tilde \Delta(\tilde k) + \tilde k^2 \Delta'(\tilde k) = - \partial_{\tilde k} \tilde j_E(\tilde k)  \\
&& \tilde{\dot \Omega}(\tilde k) := t^2 \partial_{t} \Omega(k) = 2 \tilde k^3 \tilde \Delta(\tilde k) + \tilde k^4 \tilde \Delta'(\tilde k) =  - \partial_{\tilde k} \tilde j_\Omega(\tilde k)
\eea 
where we have defined the scaled energy and enstrophy fluxes:
\bea \label{76-3}
\tilde j_E(\tilde k) = - \tilde k^2 \tilde \Delta(\tilde k) \quad , \quad \tilde j_\Omega(\tilde k) = -\tilde k^4 \tilde \Delta(\tilde k)+2\int_0^{\tilde k} \rmd p\, p^3 \,\tilde \Delta(p)
\eea
The scaled energy and enstrophy decay rates, and the scaled fluxes, all as functions of $\tilde k$
are plotted on figure~\ref{f:4}. (For convenience we revert to the notation of the rest of the paper
for the argument of $\tilde \Delta$, i.e. the $x$-axis is called $k$ but it is more properly $\tilde k$). 
We see that the energy flux \(\tilde j_E(k)\) is negative, thus to small momentum scales, and moreover energy is conserved, since  $\lim_{k\to \infty}\tilde j_E(k)=0$. 
Hence the total energy decay rate vanishes, $\bar \epsilon = - \partial_t {\cal E} = \frac{-1}{4 \pi t} \int \rmd \tilde k \tilde{\dot E}(\tilde k)=0$. The enstrophy flux is mostly positive, thus to   large momentum scales, and  enstrophy is {\em not} conserved. However,  note that for $\Delta(k)\sim 1/k^4$, the integral in (\ref{76-3}) grows as $\ln k$, thus the  enstrophy conservation is only {\em marginally violated}. For  $\Delta(k)\sim 1/[k^4 (\ln k)^c]$, and $c>1$, the enstrophy would be conserved, since the flux at large $k$ would vanish. Thus a small modification of the asymptotic behavior, which might not be given correctly by our leading-order fixed point,  would be enough to ensure enstrophy conservation.

\begin{figure}
\centerline{\Fig{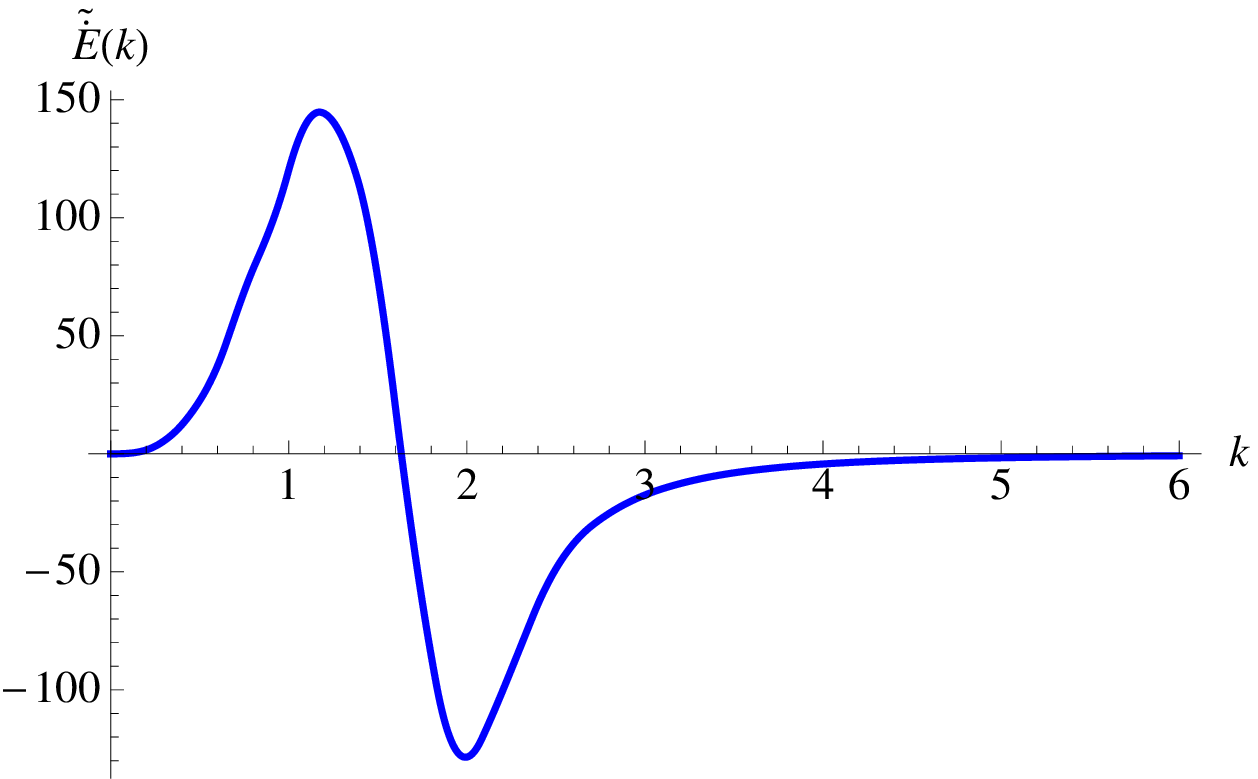}\Fig{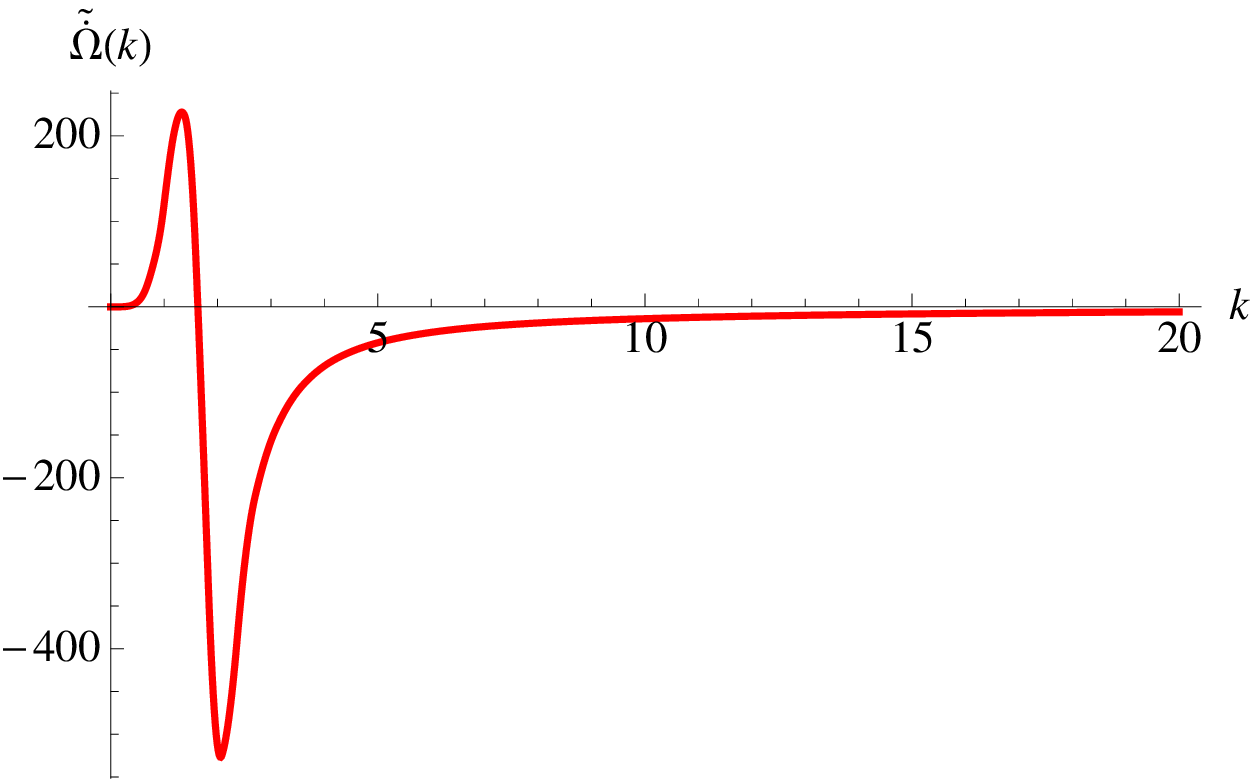}}
\centerline{\Fig{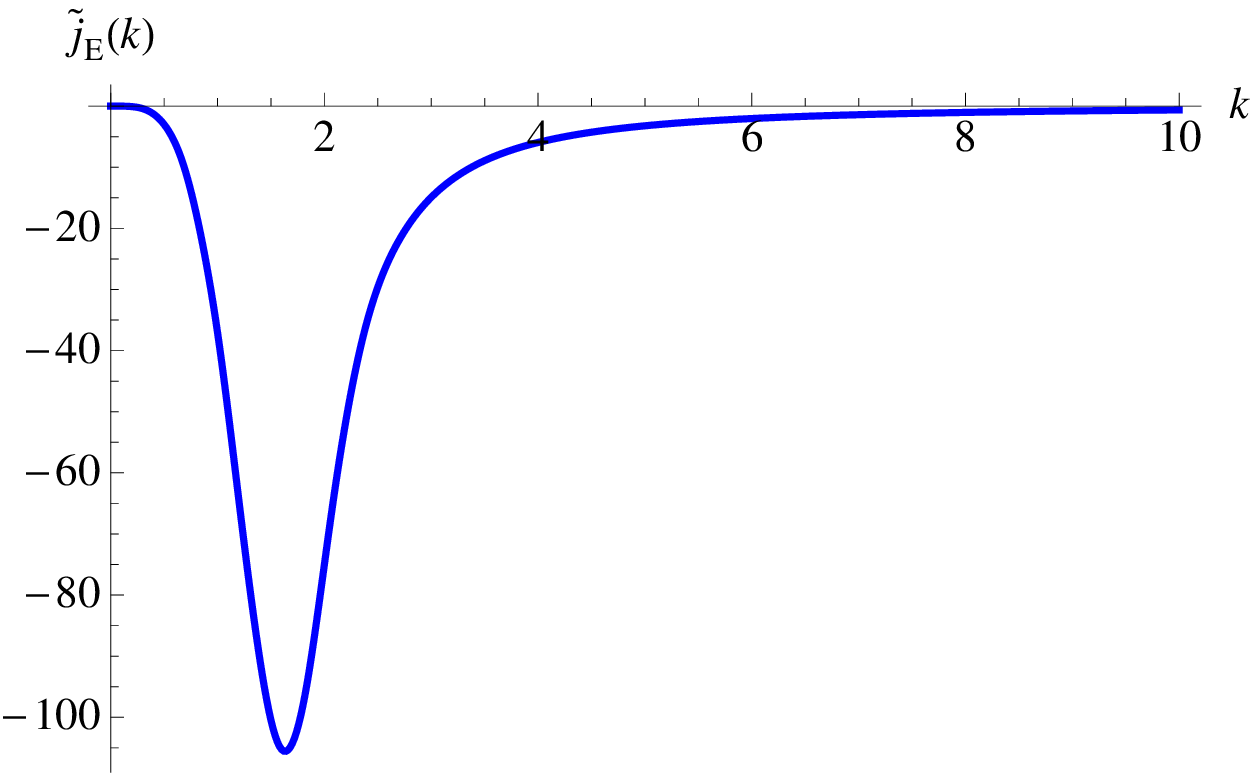}\Fig{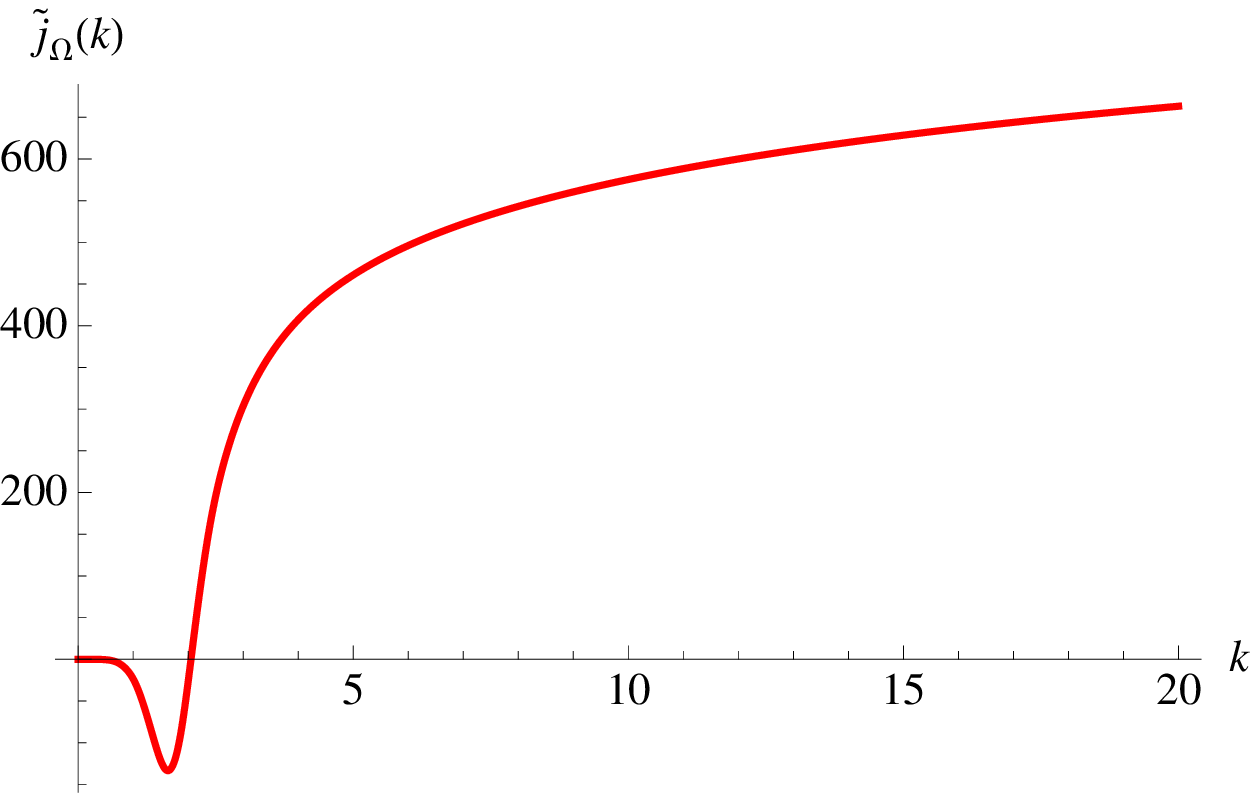}}
\caption{The scaled energy decay rate $\tilde{\dot E}$ and enstrophy $\tilde{\dot \Omega}$ (top) as well as their fluxes \(\tilde  j_E\) and \(\tilde j_\Omega \) (bottom), defined in the text,
as a function of the rescaled momentum (noted $k$ here).}
\label{f:4}
\end{figure}

To summarize, it seems that the FRG fixed point is compatible with the Batchelor-Kraichnan scenario \cite{Batchelor1969,Kraichnan1971}
with an enstrophy anomaly $- \partial_t \frac{1}{2} \langle{\omega_{\vu t}^2}\rangle =
\lim_{\nu \to 0} \nu \langle{ (\nabla \omega)^2}\rangle = \bar \epsilon_\omega$. Since
$\bar \epsilon_\omega$ has dimension $({\rm time})^{-3}$, the energy spectrum within the
Batchelor-Kraichnan 2D enstrophy cascade is $E(k,t) \sim \bar \epsilon_\omega^{2/3} k^{-3}$, with, in decaying turbulence
$\bar \epsilon_\omega \sim 1/t^3$. However, it is also known that this is not the
end of the story, and that more detailed arguments and assumptions lead to additional logarithmic corrections,
e.g. $E(k,t)\sim \bar \epsilon_\omega^{2/3}/(k^3 (\ln k)^{c})$ with $c=1/3$ \cite{Kraichnan1971,LesieurBook}.
It was also argued that in real space the vorticity correlations
are {\em\ not} $\langle(\omega_{\vu} - \omega_0)^2 \rangle \sim \ln u$ at small scale, as would be
the case if $E(k) \sim 1/k^3$. Instead
they have fractional  powers of \(\ln u \),  claimed to extend to all moments of the
vorticity field, as\ $\langle (\omega_{\vu} -\omega_0)^n \rangle \sim (\ln u)^{n/3}$
\cite{FalkovichLebedev1994}, as a consequence of the infinite number of
conservation laws; each conservation law is violated and leads to a flux of the corresponding (almost conserved) quantity. (Equivalently one can write
that $\langle v \cdot \nabla \omega^{n} \omega^{n} \rangle$ is a constant \cite{FalkovichLebedev1994}).
Most of these issues were discussed for forced turbulence, but
remain relevant for  the decaying case. Note that in forced 2D turbulence there
is an additional regime with an  inverse energy cascade $E(k) \sim k^{-5/3}$
(see e.g.\ \cite{LesieurBook}). The  energy flows to large scales,
until the largest scale is reached, where coherent structures form. (These lowest $k$ modes may be called
a condensate). At small
scales however, the behavior should not be qualitatively different from decaying turbulence.
 For numerical and experimental tests of the enstrophy cascade see \cite{ParetTabeling1998,ParetJullienTabeling,PasqueroFalkovich2002}.
As  discussed in \cite{Bernard1999a,Bernard1999b},  friction is a marginal perturbation, hence
should change the logarithms of the distance in the vorticity correlations into power laws.
Finally, for a more mathematical discussion of the enstrophy anomaly
see \cite{Eyink2001,Constantin2007}.
In particular, the anomaly was proven to vanish in the forced 2D Euler equation
with friction \cite{Constantin2006}.

A challenging question is whether some of this physics can be captured in higher-loop extensions of the present approach.

\subsection{Periodic 2D-turbulence}

Let us consider the NS equation in a square box of size $L=2 \pi$ with periodic boundary conditions.
We study the FRG equation (\ref{NSFRG}) for $N=2$, using   integer Fourier
modes $(k_x,k_y) \in \mathbb Z^2$, which are summed over.
It is easy to analyze numerically the FRG equation projected onto a grid $[-Q,Q]^2$,
setting
$\Delta(k)=0$ outside.
Equation (\ref{NSFRG}) can be written schematically
as $\partial_\tau R(k) = (R*R)(k)$ with $\tau=t^2/2$ and $R(k)=\Delta(k)/k^2$.
Rescaling
 is not crucial here, since
periodic turbulence corresponds to $\zeta=0$.
One finds, for any $Q$, that the flow asymptotically behaves as
\begin{eqnarray} \label{asympt}
R(k) = r_1 \delta_{k=(\pm 1,0),(0,\pm1)} + \tilde r_k e^{-  \lambda \tau}
\end{eqnarray}
and that the energy becomes entirely concentrated in the lowest modes $k^2=1$.
The Fourier coefficient  $r_1$ of these modes tends to a constant at large times,
while all other modes decay.
There is a transient regime where the other modes first increase before decaying,
following (\ref{asympt}).
The $\tilde r_k$ are obtained by diagonalizing
$\partial_\tau \tilde r_k = 2 (R_1*\tilde r)(k)=- \lambda \tilde r_k$,
where $R_1=r_1 \delta_{k=(\pm 1,0),(0,\pm1)}$ is  a real  non-symmetric matrix. We  truncated numerically on a grid $k \in [-Q,Q]^2$ and found, apart from one trivial eigenvalue $\lambda=0,$ corresponding to $\tilde R=R_1$, that all eigenvalues are  negative.
Only the leading one corresponds to a vector with all positive entries for $k^2>1$, which is requested from (\ref{asympt}).
A plot of $\lambda$ versus $1/Q^2$ is approximately linear and  the numerical
solution suggests $\lambda= - 0.6 r_1$.
 The leading eigenvector  is very well fitted by
$\tilde r_k/\tilde r_1=1.57 k^{-a}$ with $a=5.8 \pm 1$, which we checked up to $Q=16$.
Since $a>4$, this is
consistent with the absence of an  energy anomaly. It is interesting that
this value of  $a$ seems to be indeed near  the Batchelor value  of $a=6,$
which corresponds to $\Delta(k)\sim k^{-4}$ and $E(k)\sim k^{-3}$
\cite{Batchelor1969}, consistent with our analysis in the last section.

In conclusion,   we want to note that $R(k)$ written in real space tends
to a fixed point which corresponds to  the average over the set of exact,
time-independent solutions of the Euler equation,
\begin{eqnarray}
&& \psi_{\vu t} = w_{1x} \cos(u_x) + w_{2x} \sin(u_x) + w_{1y} \cos(u_y) + w_{2y} \sin(u_y),  \\
&& \vv_{\vu t} = \big( - w_{1y} \sin(u_y) + w_{2y} \cos(u_y) , w_{1x} \sin(u_x) - w_{2x} \cos(u_x) \big).
\end{eqnarray}
This is easily checked by inserting into Euler's equation.
The four parameters  $w_{ix,y}$ are  independent
Gaussian random variables with zero mean and variance $r_1/2$. The FRG suggests that the way it tends to this fixed point
 is non-trivial (with a non-analytic correlator). It would be of great interest to study that question
 in  detail.

\section{Analysis of the FRG equation in three dimensions ($N=3$), and in large dimensions (\(N\to \infty\))}
\label{sec:N-inf}

Obtaining a numerical solution for the fixed point of the FRG equation for $N>2$ is difficult.
In three dimension $N=3$ we have studied the FRG equation (\ref{NSFRG})
for a periodic flow, in Fourier
space, very much as we did in the previous Section for $N=2$. We have
found that on a
grid in Fourier space $[-Q,Q]^3$ equation (\ref{NSFRGresc}) does flow to a
fixed point with $\zeta=0$. This fixed
point, to our numerical accuracy, was compatible with a cusp behavior at large
$k$, i.e.\ $\zeta_2=1$.
Since the result is not too surprising, and consistent with the analysis of
Section \ref{s:Short-distance analysis, cusp or no cusp?},
we will not reproduce the details here. Instead we now turn to the large dimension
limit (large $N$ limit) analysis of the FRG equation.

Large dimensions are often a means of controlling an expansion.
The aim would be to sum contributions from all loops at \(N\to \infty\),
thus avoiding any artifact from a closure scheme, as has been done for
random manifolds, i.e., Burgers,
\cite{LeDoussalMuellerWiese2007,LeDoussalWiese2004a,LeDoussalWiese2003b,LeDoussalWiese2001}.
While this remains a project for future research, we have analyzed the one-loop
equations for \(N\to \infty\). In terms of the function \(r(y)\)
introduced in equation~(\ref{220}), the RG equations at large \(N\) are derived
in  \ref{a:K} and reads
\begin{eqnarray}
&& \fl N\left( 2 + N \right) t\partial_{t} {r}''({y}) +
  4\,{y}\left[ \left( 2 + N \right) t\partial_{t} {r}'''({y}) +
     {y}t\partial_{t} {r}''''({y}) \right] = 2 r''({y}) +  \zeta_0 y
r''' ({y})+ {\ 2} r''({y})^2 + {\ 2} r'''({y})\big[r'({y})-r'(0)\big]\nn \\
&& + \frac{1}{N} \Big[ 4 r''({y}) +y (8+6\zeta_0+\zeta_{-1}) r'''  ({y})+ 4 \zeta_0
 y^2 r''''({y}) + {\ 2} r''({y})^2 + {\ 10} r''' ({y})\big[r'({y})-r'(0)\big]\nn \\
&&  ~~~~~~~+ 12 y r'''' ({y})\big[r'({y})-r'(0)\big] +  28 y r''({y})
r'''({y})  - 4 \int_0^\infty \rmd t\, r'''(y + t) r''(t) \Big]  + O(1/N^2),
\label{K8-bis}
\end{eqnarray}
where $\zeta= \zeta_0 + \zeta_{-1}/N+ ...$
We now look for a fixed point.
To leading  order we have
\begin{eqnarray}
&& 0 = 2 r''(y)   + \zeta_0 y r''' (y)+ {\ 2\big[} r''(y)\big]^2
+ {\ 2} r'''(y)\big[r'(y)-r'(0)\big], \label{K8-bis-2}
\end{eqnarray}
which coincides, up to a numerical prefactor, with the large-$N$ limit of
the  1-loop Burgers equation, see e.g.\ equation~(7.7) of Ref.~\cite{LeDoussalWiese2003b}.
This confirms that at least to one loop the infinite-$N$ limit of the  decaying
Navier Stokes equation  reproduces that of the Burgers equation.
Equation (\ref{K8-bis-2}) has an exponentially decaying solution
only for $\zeta_0=0$; an analytic solution for the inverse function  can be written as
\begin{eqnarray}
&& z := - r'(y), \\
&& y =  z - z_0 - z_0 \ln(z/z_0). \label{eq-N}
\end{eqnarray}
The asymptotic behavior is $z=z_0e^{-1-y/z_0}$ for large $y$.
However, we cannot neglect the terms of order $1/N$
for $y/N\gg 1$, and the above solution is valid only for $y\ll N$,
the {\em primary} region. On the other hand, for $y\gg z_0$, one can neglect the nonlinear
terms in equation (\ref{K8-bis-2}) due to the
exponential decay
of $r(y)$. Both solutions are expected to match  in the {\em inner}
region $z_0\ll y\ll N$,
which becomes quite wide for $N\to\infty$ \cite{BalentsDSFisher1993,LeDoussalWiese2005a}.
Presumably in the inner
region both solutions have a simple
exponential behavior, to order $1/N$. The linearized equation to order
\(1/N\) reads  
\begin{eqnarray}
\fl && 0 = 2r''(y) - {\ 2}r'''(y) r'(0) + \frac{1}{N} \Big[ 4 r''(y) + 8 y r'''(y)
+ \zeta_{-1} y r'''(y)
 - {\ 10} r'''(y) r'(0)  \nn \\
 && \mbox{}\hspace{20mm} - {\ 12} y r''''(y) r'(0)   - {\ 4} \int_0^\infty \rmd t\, r'''(y + t) r''(t)\Big ].
\end{eqnarray}
We assume that this equation has a solution which behaves as $e^{-y/z_0}$
for  $y\gg z_0$.
We are free to fix $z_0={\ 1}$ for the sake of simplicity. Substituting
$r(y)=C e^{-y}$  into the linearized equation, and collecting all
terms proportional to $e^{-y}$ and $ye^{-y}$, we obtain
\begin{eqnarray}
&& e^{-y}\ \ \ \  ~\,: \ \ \ \ 0= 2+{\ 2}r'(0)+  \frac{1}{N}\Big[4+{\ 10}r'(0)+2 \int_0^\infty \rmd t\, e^{-t} r''(t)\Big],
 \nonumber \\
&& y e^{-y}\ \ \ \ : \ \ \ \ 0= \frac{1}{N}\Big[-8 -\zeta_{-1} - {\ 12}r'(0)\Big].
\end{eqnarray}
The first line gives  $r'(0)=-{\ 1}+O(1/N)$ and the second $\zeta_{-1}=4$,
so that
\begin{equation}
   \zeta=\frac4{N}+O\left(\frac1{N^2}\right).
\end{equation}
Close to  \(z_0\) the solution of  equation (\ref{eq-N}) has the form
\(z-z_0= \sqrt{2z_0 y} \sim u\). This implies that the cusp persists,
i.e.\ $\zeta_2=1$.

\section{Decaying surface quasi-geostrophic turbulence}
\label{sec:SQG}

An interesting and still much studied generalization of 2D NS is the
Surface Quasi Geostrophic (SQG) equation. It is defined in dimension $N=2$, and
depends on a continuously varying parameter $a$. In real space it reads
\begin{equation}\label{212}
  \partial_{t}T_{\vu t} + \mathbf{v}_{\vu t} \cdot \nabla T_{\vu t} =
\nu \nabla^2 T_{\vu t} \ , \quad
\mathbf{v}_{\vu t} = \hat{z} \times \nabla \psi_{\vu t} \  , \quad
(-\nabla^2)^{a/2} \psi_{\vu t}= T_{\vu t}.
\end{equation}
It describes the convection of a quantity
$T_{\vu t}$ by the velocity field \(\vv_{\vu t}\), which  in turn is related
to the velocity. For $a=2$ one recognizes that the quantity $T_{\vu t}$ is
precisely the vorticity $\omega_{\vu t}$, and one recovers the usual
2D NS equation. For $a=1$ the field $T_{\vu t}$ represents the temperature
in the ``true" SQG turbulence, which is used to model the 2D atmospheric flow
on the surface of the Earth. Finally, for $a=-2$
the model  was obtained by Charney and Oboukhov for waves in rotating fluids, and
by Hasegawa and Mima for drift waves in a magnetized
plasma in the limit of a vanishing Rossby radius. It is thus called the
Charney-Hasegawa-Mima equation \cite{BoffetaDeLilloMusacchio2002}.
The naive scaling dimension of the field $T_{\vu t}$  is $\delta_u T\sim u^H$ with
$H=(2-2a)/3$. Recently, it has been conjectured that isolines of $T_{\vu t}$ in
the inverse-cascade regime of the
forced Charney-Hasegawa-Mima equation were SLE lines with
$\kappa=\frac{4}{3}(1+2 a)$, with some numerical
evidence \cite{BernardBoffettaCelaniFalkovich2007,FalkovichMusacchio2010}.

The SQG equation for arbitrary $a$ shares some properties with
the 2D NS equation, in the sense that in the inviscid limit both
the enstrophy $D=\frac{1}{2} \langle T^2 \rangle$ (and all powers of $T$)
and the energy $E= \frac{1}{2} \langle T \psi \rangle$ are conserved for
flows smooth enough. To show the latter one goes to Fourier space, where
the relation between $\psi_{\vu t}$ and $\vv_{\vu t}$ is
$\psi_{\vq t}= q ^{-a} T_{\vq t}$. This yields
\begin{equation}\label{212b}
  \partial_{t} E = \sum_\vq T_{-\vq t} q^{-a} \partial_t T_{\vq t} =
  \sum_{\vk,\vq} \frac{\epsilon_{\alpha \beta} \vk_\alpha \vq_\beta}{q^a k^a}
  T_{\vk t} T_{\vq-\vk t} T_{-\vq t} = 0\ ,
  \end{equation}
due to the  symmetry $\vq \leftrightarrow - \vk$.

Here we consider the decaying inviscid SQG equation. We  display  the FRG equation
to one loop,  leaving its analysis
for the future. One defines the 2-point correlation in Fourier space,
\begin{eqnarray}
&& \Delta_T(\vk) = \langle{T_{-\vk t} T_{\vk t}}\rangle \ .
\end{eqnarray}
The FRG equation for $\Delta_T(\vk)$ is derived in~\ref{a6}, and reads
\begin{eqnarray}\label{229}
  \partial_{t} \Delta_{ T}(\vk)&=&  t
\sum_\vq\Delta_{ T}(\vq)\Delta_{ T}(\vq+\vk)[\vq\times \vk]^2\left
\{q^{-2a}-2q^{-a}|\vq+\vk|^{-a}+|\vq+\vk|^{-2a} \right \}
\nonumber \\
&&- 2t \sum_\vq
\Delta_{ T}(\vq)\Delta_{ T}(\vk)[\vq\times \vk]^2\left\{q^{-2a}-q^{-a}
k^{-a}-q^{-a}|\vq+\vk|^{-a} +k^{-a}|\vq+\vk|^{-a} \right \}.
\end{eqnarray}
Note that the correlator associated to the velocity is
$\Delta(\vk)=k^{2-2a} \Delta_T(\vk)$.
Again, it is convenient to introduce the rescaled correlators via
\begin{eqnarray}
&&  \Delta_T(\vk) = t^{-2 + \frac{\zeta}{2} (6-2 a)} \tilde  \Delta_T(\vk t^{\zeta/2}), \\
&& \Delta(\vk) = t^{-2 + 2 \zeta} \tilde  \Delta(\vk t^{\zeta/2}).
\end{eqnarray}
For the rescaled correlator, the FRG equation can be written as\begin{eqnarray}\label{229b}
&&\!\!\!\!\!  t \partial_{t} \tilde \Delta_{T}(\vk)=[2 - (3-a) \zeta]
 \tilde  \Delta_{T}(\vk) - \frac{\zeta}{2} \vk \partial_{\vk}
 \tilde \Delta_{T} (\vk)\nonumber \\
&&~~~~~~~~~~~~~ + \sum_{\vq,\vp=\vk-\vq} [\vq\times \vk]^2 (q^{-a}-p^{-a}) \tilde \Delta_{T}(\vq)
\Big[ (q^{-a}-p^{-a}) \tilde \Delta_{T}(\vp)  - 2 (q^{-a}-k^{-a})
\tilde \Delta_{T}(\vk)  \Big].
\end{eqnarray}
One can  use $\vq\times \vp=\vq\times \vk$.
The equation for the rescaled velocity correlator reads
\begin{eqnarray}\label{229c}
&&  t \partial_{t} \tilde \Delta(\vk)=(2 - 2 \zeta)
 \tilde  \Delta(\vk) - \frac{\zeta}{2} \vk \partial_{\vk} \tilde \Delta(\vk)\nonumber \\
&& ~~~~~~~~~~~~~+ \sum_{\vq,\vp=\vk-\vq} \frac{[\vq\times \vk]^2}{q^2} \tilde \Delta(\vq)
\bigg[ \frac{(p^a-q^a)^2}{k^{2 a-2} p^2} \tilde \Delta(\vp) -
2 \frac{(p^a-q^a)(k^a-q^a)}{k^{a} p^a} \tilde \Delta(\vk) \bigg]\ .
\end{eqnarray}
In this form it is easy to check that for $a=2$ one recovers the FRG equation
for the NS equation in $N=2$.

The FRG equation (\ref{229c}) written in real space for general value of $a$ is
a nonlocal integro-differential equation. It is interesting that there are
some values of $a$ for which the FRG equation
becomes quasi-local, i.e.\ involving  only derivatives of finite
order at the point $\vu$ and  at the origin $\vu=0$. For instance that happens
in the case of the Charney-Hasegawa-Mima turbulence corresponding to $a=-2$.

In  \ref{sec-SQG-integral} we have studied, as we did for Burgers and NS, the
possible values for the exponent $\zeta_2$, defined from 
$\tilde \Delta({\bf k}) \sim k^{- (2+\zeta_2)}$ at large $k$, and isotropic turbulence. More work is necessary
to study the fixed points of the FRG equation as a function of the parameter $a$.

\section{Conclusions and Perspectives}
In this article, we have applied functional-renormalization-group methods to decaying turbulence. In contrast to standard perturbative RG, the functional RG approach takes into account a coupling function i.e.\ an infinity of couplings rather than one or few. It naturally leads to a non-analytic 2-point function. While the method is in principle exact, as any RG treatment, in practice the flow has to be projected onto a lower-dimensional subspace, here the equal-time 2-point velocity correlation function. With this projection in mind, the FRG equations are organized in an expansion in powers of the 2-point velocity correlation
function
itself, equivalent to a loop expansion. As we have discussed, they correspond to a small-time {\it renormalized} perturbation theory.
Here, we  studied the 1-loop equations. For Burgers, they reproduce the FRG equations derived in the context of random manifolds, and correctly describe the singular structure of the flow, made out of shocks. While this had been worked out in details before for $N=1$, here we  extended it to any dimension $N$.

Let us stress  that the method works at least qualitatively  for Burgers;  that it correctly   accounts  for shocks, and that the
distribution of velocities is not close to a Gaussian. The reason is that the extension to a manifold provides a model which can
be  controlled perturbatively (in $d=4-\epsilon$), while at the same time exhibiting shock singularities, non-conservation of energy (called
failure of dimensional reduction in  the context of disordered systems) and energy cascades.  This is because shock sizes and the magnitude of the
energy decay rate are $O(\epsilon)$ in that expansion. That in itself is remarkable
in the turbulence context, and motivated us to consider Navier-Stokes with this method.

For Navier Stokes, the fixed point depends on the dimension. For \(N\to \infty\), the FRG  equations converge (at leading 1-loop order) to those of the decaying Burgers equation. Thus the 2-point velocity correlation
function
should grow linearly with distance, i.e.\ have a cusp. This cusp is also the only possible solution for the 3-dimensional FRG equation, at 1-loop order, in contradiction to  experimental evidence.  It is possible, that at second (2-loop) or higher order, new non-trivial fixed points emerge. If this is not the case, one would have to understand why the method seemingly does not admit the correct singularities. Since for large $N$ we find that the FRG equation reduces to the one of Burgers, hence has shock singularities,
one possible way to understand that may be via a large-dimension expansion combined with a loop expansion.

Finally, in two dimensions the FRG equations allow for a  fixed point which is consistent with Batchelor's scaling. While the equations are similar to the quasi-normal Markovian approximation, we give here an explicit solution. Again it seems a good starting point to include higher-loop corrections, one challenge being to confirm, or infirm, the conjectured
logarithmic corrections. We have also written the flow equations corresponding to  SQG turbulence, which await a more detailed analysis.

We hope that this work helps to bring a new perspective in a long-debated subject.

\bigskip

{\it Acknowlegments}: We thank  D.~Bernard, B.~Birnir, and 
I.\ Procaccia for useful discussions. We are especially indebted to  B.\ Shraiman and G.~Falkovich for many enlightening remarks. This work was supported by
ANR under two successive programs, 05-BLAN-0099-01 and 09-BLAN-0097-01/2, and in part through NSF grants PHY05-51164  and~PHY11-25915 during
 several stays at KITP. We also thank the KITPC for its hospitality.

\appendix
\section{1-loop FRG equation in real space}
\label{realspaceRG}

The two equations from (\ref{hierarchy1}) needed to one loop in the inviscid limit are
\begin{eqnarray}
\qquad\quad \ \  \partial_t \Delta_{\alpha_1 \alpha_2}(\vu) &=&{\rm Sym}   \Big[ P_{\alpha_1 ; \beta \gamma}(\partial_{\vu}) C^{(3)}_{\beta \gamma \alpha_2}(0,0,\vu)\Big] = -{\rm Sym}  \Big[ P_{\alpha_1 ; \beta \gamma}(\partial_{\vu}) C^{(3)}_{\beta \gamma \alpha_2}(\vu,\vu,0)\Big]\\
& =& 2\,{\rm Sym} \Big[ P^{\rm T}_{\alpha_1 \beta} (\partial_{\vu}) \partial_{\vu}^\gamma C^{(3)}_{\beta \gamma \alpha_2}(0,0,\vu)\Big] \qquad\quad ~~~({\rm Euler}) \label{c33} \\
& =& {\rm Sym}\Big[\partial_{\vu}^{\alpha_1} C^{(3)}_{\beta \beta \alpha_2}(0,0,\vu)\Big] \qquad\qquad\qquad \quad  ({\rm Burgers})   \\
 \partial_t C^{(3)}_{\alpha_1 \alpha_2 \alpha_3}(\vu_1,\vu_2,\vu_3) &=&  - \frac{3}{2} {\rm Sym}\Big[ P_{\alpha_1 \beta \gamma}(\partial_{\vu_1}) C^{(4)}_{\beta \gamma \alpha_2 \alpha_3}(\vu_1,\vu_1,\vu_2,\vu_3)\Big]  \label{c3}
\end{eqnarray}
with $\vu=\vu_{12}$ and where $P_{\alpha ; \beta \gamma}(\partial)=P^{\rm T}_{\alpha \beta}(\partial) \partial_\gamma + P^{\rm T}_{\alpha \gamma}(\partial) \partial_\beta$ (Euler), $P_{\alpha ; \beta \gamma}(\partial)=\delta_{\beta \gamma} \partial_\alpha$ (Burgers). In the first three lines ${\rm Sym}[...]$ means symmetrization over $\alpha_1,\alpha_2$ and we have used that
$\Delta_{\alpha_1 \alpha_2}(\vu)$ is even in $\vu$ (no average helicity). In the last line, {\it and everywhere below} ${\rm Sym}[...]$ means symmetrization over $\vu_1^{\alpha_1},\vu_2^{\alpha_2},\vu_3^{\alpha_3}$, i.e simultaneous exchange of the points in space and the indices.

To lowest order we replace (denoting $\vu_{ij}:=\vu_i - \vu_j$) \begin{eqnarray}
\fl  C^{(4)}_{\beta \gamma \alpha_2 \alpha_3}(\vu_1,\vu_1,\vu_2,\vu_3) =
\Delta_{\beta \gamma}(0) \Delta_{\alpha_2 \alpha_3}(\vu_{23}) + \Delta_{\beta \alpha_2}(\vu_{12}) \Delta_{\gamma \alpha_3}(\vu_{13})
 + \Delta_{\beta \alpha_3}(\vu_{13}) \Delta_{\gamma \alpha_2}(\vu_{12})\ .
\end{eqnarray}
Hence integrating equation~(\ref{c3}) one gets
\begin{eqnarray}
&& \! \! \! \! \! \! \! \! \! \! \! \! \! \! \! \! \! \! \! \! \! \! \! \!\! \! \! \! \! \! \! \! \! \! \! \! \! \! \! \!\!\!\!
C^{(3)}_{\alpha_1 \alpha_2 \alpha_3}(\vu_1,\vu_2,\vu_3) = - \frac{3}{2} t\,{\rm Sym}\Big [  P_{\alpha_1 \beta \gamma}(\partial_{\vu_1})\Big ( \Delta_{\beta \alpha_2}(\vu_{12}) \Delta_{\gamma \alpha_3}(\vu_{13})\Big ) + P_{\alpha_1 \beta \gamma}(\partial_{\vu_1})\Big (\Delta_{\beta \alpha_3}(\vu_{13}) \Delta_{\gamma \alpha_2}(\vu_{12})\Big)\Big  ] \nonumber \\
&& =  - 3 t\,{\rm Sym}\Big [  P^{\rm T}_{\alpha_1 \gamma}(\partial_{\vu_1})\Big( \Delta_{\beta \alpha_3}(\vu_{13}) \partial_{\vu_1}^\beta \Delta_{\gamma \alpha_2}(\vu_{12}) + \Delta_{\beta \alpha_2}(\vu_{12}) \partial_{\vu_1}^\beta \Delta_{\gamma \alpha_3}(\vu_{13}) \Big) \Big] \quad  ~~~({\rm Euler}) \\
&& = - 3 t\,{\rm Sym}\Big [\partial_{\vu_1}^{\alpha_1}  \big(  \Delta_{\beta \alpha_2}(\vu_{12}) \Delta_{\beta \alpha_3}(\vu_{13}) \big) \Big] \quad  ~~~~~~~~~~~~~~~~~~~~~~~~~~~~~~~~~~~~~~~~~~~~~~~~~~~~({\rm Burgers})
\end{eqnarray}
where for Euler we used the transversality of $\Delta_{\alpha \beta}$. Expanding, one finds for Burgers\begin{eqnarray}
 \fl
- t^{-1} C^{(3)}_{\alpha_1 \alpha_2 \alpha_3}(\vu_1,\vu_2,\vu_3) =  \partial_{\vu_1}^{\alpha_1}  \Big(  \Delta_{\beta \alpha_2}(\vu_{12}) \Delta_{\beta \alpha_3}(\vu_{13}) \Big) + \partial_{\vu_2}^{\alpha_2}  \Big(  \Delta_{\beta \alpha_1}(\vu_{21}) \Delta_{\beta \alpha_3}(\vu_{23}) \Big)  + \partial_{\vu_3}^{\alpha_3}  \Big(  \Delta_{\beta \alpha_2}(\vu_{32}) \Delta_{\beta \alpha_1}(\vu_{31}) \Big). \nn\\
\end{eqnarray}
This expression  is  symmetric and does not need to be symmetrized.
Taking the limit of $\vu_2 \to \vu_1$ one finds
\begin{eqnarray}
&& \fl
- t^{-1} C^{(3)}_{\alpha_1 \alpha_2 \alpha_3}(\vu_1,\vu_1,\vu_3) =  \partial_{\vu_3}^{\alpha_3}  \Big(  \Delta_{\beta \alpha_2}(\vu_{31}) \Delta_{\beta \alpha_1}(\vu_{31}) \Big)  + \Delta_{\beta \alpha_2}(0) \partial_{\vu_1}^{\alpha_1}  \Delta_{\beta \alpha_3}(\vu_{13})
+  \Delta_{\beta \alpha_1}(0) \partial_{\vu_2}^{\alpha_2}  \Delta_{\beta \alpha_3}(\vu_{23})\ .
\end{eqnarray}
We have used that
\begin{eqnarray}
&&  \lim_{\vu_2 \to \vu_1}
\big( \partial_{\vu_1}^{\alpha_1}   \Delta_{\beta \alpha_2}(\vu_{12}) \big) \Delta_{\beta \alpha_3}(\vu_{13})
+\big(  \partial_{\vu_2}^{\alpha_2}  \big(  \Delta_{\beta \alpha_1}(\vu_{21}) \big) \Delta_{\beta \alpha_3}(\vu_{23}) = 0\ ,
\end{eqnarray}
which comes from $R'''_{\alpha_1 \beta \alpha_2}(\vu)$ being odd. One then gets
\begin{eqnarray}
&& \partial_t \Delta_{\alpha_1 \alpha_2}(\vu) = - t \partial_{\vu}^{\alpha_1} \partial_{\vu}^{\alpha_2} \Delta_{\beta \beta'}(\vu)^2
+ 2 t \Delta_{\beta \beta'}(0) \Delta_{\beta \alpha_2;\beta' \alpha_1}(\vu)\ ,
\end{eqnarray}
where we have used that $\Delta_{\beta \alpha_2;\beta'}(\vu)$ is odd and that $\Delta_{\beta \alpha_2;\beta' \alpha_1}(\vu)=-R''''_{\beta \beta' \alpha_1 \alpha_2}(\vu)$ is symmetric in $\alpha_1,\alpha_2$. This gives the equation in the text.

For Euler one finds by expanding
\begin{eqnarray}
\fl
- t^{-1} C^{(3)}_{\alpha_1 \alpha_2 \alpha_3}(\vu_1,\vu_2,\vu_3) &=& P^{\rm T}_{\alpha_1 \gamma}(\partial_{\vu_1}) \big( \Delta_{\beta \alpha_3}(\vu_{13}) \partial_{\vu_1}^\beta \Delta_{\gamma \alpha_2}(\vu_{12}) + \Delta_{\beta \alpha_2}(\vu_{12}) \partial_{\vu_1}^\beta \Delta_{\gamma \alpha_3}(\vu_{13}) \big) \nonumber \\
&& + P^{\rm T}_{\alpha_2 \gamma}(\partial_{\vu_2}) \big( \Delta_{\beta \alpha_3}(\vu_{23}) \partial_{\vu_2}^\beta \Delta_{\gamma \alpha_1}(\vu_{21}) + \Delta_{\beta \alpha_1}(\vu_{21}) \partial_{\vu_2}^\beta \Delta_{\gamma \alpha_3}(\vu_{23}) \big) \nn \\
&& + P^{\rm T}_{\alpha_3 \gamma}(\partial_{\vu_3}) \big( \Delta_{\beta \alpha_1}(\vu_{31}) \partial_{\vu_3}^\beta \Delta_{\gamma \alpha_2}(\vu_{32}) + \Delta_{\beta \alpha_2}(\vu_{32}) \partial_{\vu_3}^\beta \Delta_{\gamma \alpha_1}(\vu_{31}) \big)\ .
\end{eqnarray}
This expression  is  symmetric and does not need to be symmetrized.
We now  take the limit $\vu_2 \to \vu_1$:
\begin{eqnarray}
\fl
- t^{-1} C^{(3)}_{\beta \gamma \alpha_2}(\vu_1,\vu_1,\vu_3) &=&  P^{\rm T}_{\alpha_2 \gamma'}(\partial_{\vu_3}) \big( \Delta_{\beta' \beta}(\vu_{31}) \partial_{\vu_3}^{\beta'} \Delta_{\gamma' \gamma}(\vu_{31}) + \Delta_{\beta' \gamma}(\vu_{31}) \partial_{\vu_3}^{\beta'} \Delta_{\gamma' \beta}(\vu_{31}) \big) \nonumber \\
&& + \lim_{\vu_2 \to \vu_1} \bigg( P^{\rm T}_{\beta \gamma'}(\partial_{\vu_1})\Big ( \Delta_{\beta' \alpha_2}(\vu_{13}) \partial_{\vu_1}^{\beta'} \Delta_{\gamma' \gamma}(\vu_{12}) + \Delta_{\beta' \gamma}(\vu_{12}) \partial_{\vu_1}^{\beta'} \Delta_{\gamma' \alpha_2}(\vu_{13})\Big ) \nonumber \\
&& \qquad ~~~~~~+ P^{\rm T}_{\gamma \gamma'}(\partial_{\vu_2})\Big ( \Delta_{\beta' \alpha_2}(\vu_{23}) \partial_{\vu_2}^{\beta'} \Delta_{\gamma' \beta}(\vu_{21}) + \Delta_{\beta' \beta}(\vu_{21}) \partial_{\vu_2}^{\beta'} \Delta_{\gamma' \alpha_2}(\vu_{13})\Big) \bigg) \label{a13}
\ .\end{eqnarray}
The term $\lim_{\vu_2 \to \vu_1}  \big( ...  \big) $ involves a non-trivial coinciding-point limit. One may naively equate it with
\begin{eqnarray}
&&  P^{\rm T}_{\beta \gamma'}(\partial_{\vu_1})  \Delta_{\beta' \gamma}(0) \partial_{\vu_1}^{\beta'} \Delta_{\gamma' \alpha_2}(\vu_{13})
+ P^{\rm T}_{\gamma \gamma'}(\partial_{\vu_1})  \Delta_{\beta' \beta}(0) \partial_{\vu_1}^{\beta'} \Delta_{\gamma' \alpha_2}(\vu_{13})\ ,
\end{eqnarray}
but this is actually incorrect. It would lead to a term $+ 2 t \Delta_{\beta' \gamma}(0) \partial_{\vu}^{\beta'} \partial_{\vu}^\gamma \Delta_{\alpha_1 \alpha_2}(\vu)$ in the beta function. The correct beta function must retain the non-trivial limit, for which we obtain, with $\vu_{31}=\vu$:
\begin{eqnarray}
\fl&& \partial_t \Delta_{\alpha_1 \alpha_2}(\vu) = 2  [ P^{\rm T}_{\alpha_1 \beta} (\partial_{\vu}) \partial_{\vu}^\gamma C^{(3)}_{\beta \gamma \alpha_2}(0,0,\vu)] \nn\\
\fl&& = - 2 t
P^{\rm T}_{\alpha_1 \beta} (\partial_{\vu})  P^{\rm T}_{\alpha_2 \gamma'}(\partial_{\vu}) \big( \partial_{\vu}^\gamma \Delta_{\beta' \beta}(\vu) \partial_{\vu}^{\beta'} \Delta_{\gamma' \gamma}(\vu) + \Delta_{\beta' \gamma}(\vu) \partial_{\vu}^\gamma \partial_{\vu}^{\beta'} \Delta_{\beta \gamma'}(\vu) \big) \nonumber \\
\fl&&~~~~ +
2 t \,{\rm Sym}\Big[ P^{\rm T}_{\alpha_1 \beta} (\partial_{\vu}) \partial_{\vu}^\gamma
P^{\rm T}_{\alpha_2 \gamma'}(\partial_{\vu_3})  \lim_{\vu_2 \to \vu_1} \Big( P^{\rm T}_{\beta \gamma'}(\partial_{\vu_1}) \big( \Delta_{\beta' \alpha_2}(\vu_{13}) \partial_{\vu_1}^{\beta'} \Delta_{\gamma' \gamma}(\vu_{12}) + \Delta_{\beta' \gamma}(\vu_{12}) \partial_{\vu_1}^{\beta'} \Delta_{\gamma' \alpha_2}(\vu_{13}) \big) \nonumber \\
\fl&& \qquad ~~~~~~~~~~~+ P^{\rm T}_{\gamma \gamma'}(\partial_{\vu_2}) ( \Delta_{\beta' \alpha_2}(\vu_{23}) \partial_{\vu_2}^{\beta'} \Delta_{\gamma' \beta}(\vu_{21}) + \Delta_{\beta' \beta}(\vu_{21}) \partial_{\vu_2}^{\beta'} \Delta_{\gamma' \alpha_2}(\vu_{13})\Big) \Big]\ .
\end{eqnarray}
We have used that $\partial_{\vu}^{\beta'} \Delta_{\gamma' \alpha_2}(\vu)$ is odd, and
several times transversality i.e.\  $\Delta_{\alpha \beta}(\vu)=P^{\rm T}_{\alpha \beta}(\partial_{\vu}) R(u)$ (this does not assume any symmetry) hence $P^{\rm T}_{\alpha_1 \gamma'}(\partial_{\vu}) \Delta_{\gamma' \alpha_2}(u) = \Delta_{\alpha_1 \alpha_2}(u)$.

\section{1-loop FRG equation in Fourier space}
\label{frgfourier}
To one loop one must first solve
\begin{eqnarray}
\partial_t C^{(3)}_{\alpha_1 \alpha_2 \alpha_3}(\vk_1,\vk_2,\vk_3) =
 - \frac{3 }{2}  {\rm Sym}\Big[ P_{\alpha_1 ; \beta \gamma}(\vk_1) \sum_{\vp+\vq=\vk_1} C^{(4)}_{\beta \gamma \alpha_2 \alpha_3}(\vp,\vq,\vk_2,\vk_3)\Big]\ ,
\end{eqnarray}
where again, here and below ${\rm Sym}[...]$ means symmetrization w.r.t $\vk_1^{\alpha_1} ,\vk_2^{\alpha_2} ,\vk_2^{\alpha_2}$
(simultaneous permutations of points and indices). Here $P_{\alpha ; \beta \gamma}(\vk) = i \vk^\beta P^{\rm T}_{\alpha \gamma}(\vk)
+  i \vk^\gamma P^{\rm T}_{\alpha \beta}(\vk)$ for NS and $P_{\alpha ;
\beta \gamma}(\vk) = i \vk^\alpha \delta_{\beta \gamma}$ for Burgers,
and one  uses the Gaussian approximation
\begin{eqnarray}
 C^{(4)}_{\beta \gamma \alpha_2 \alpha_3}(\vp,\vq,\vk_2,\vk_3)
&=&  \delta_{\vp+\vq} \delta_{\vk_2+\vk_3} \Delta_{\beta \gamma}(\vp) \Delta_{\alpha_2 \alpha_3}(\vk_2) \nn \\
&& +
\delta_{\vp+\vk_2} \delta_{\vq+\vk_3} \Delta_{\beta \alpha_2}(\vp) \Delta_{\gamma \alpha_3}(\vq)
+ \delta_{\vp+\vk_3} \delta_{\vq+\vk_2} \Delta_{\beta \alpha_3}(\vp) \Delta_{\gamma \alpha_2}(\vq)\ .
\end{eqnarray}
We assume that $\Delta_{\alpha \beta}(-\vk)=\Delta_{\alpha \beta}(\vk)$.  The first term vanishes when multiplied by $P_{\alpha_1 ; \beta \gamma}(\vk_1)$.
One finds, using symmetries\begin{eqnarray}
 C^{(3)}_{\alpha_1 \alpha_2 \alpha_3}(\vk_1,\vk_2,\vk_3)
&=& \delta_{\vk_1+\vk_2+\vk_3}  \hat C^{(3)}_{\alpha_1 \alpha_2 \alpha_3}(\vk_1,\vk_2,\vk_3) \\
 \hat C^{(3)}_{\alpha_1 \alpha_2 \alpha_3}(\vk_1,\vk_2,\vk_3)  &=&
- t\Big(  P_{\alpha_1 \beta \gamma}(\vk_1) \Delta_{\beta \alpha_2}(\vk_2) \Delta_{\gamma \alpha_3}(\vk_3)
\nn\\
&& \ \ \ \ \ \ + P_{\alpha_2 \beta \gamma}(\vk_2)  \Delta_{\beta \alpha_1}(\vk_1) \Delta_{\gamma \alpha_3}(\vk_3)
+ P_{\alpha_3 \beta \gamma}(\vk_3)
\Delta_{\beta \alpha_1}(\vk_1) \Delta_{\gamma \alpha_2}(\vk_2)\Big)
\ .\end{eqnarray}
This expression  is already symmetric and does not need symmetrization anymore.

The 1-loop equation is  obtained by inserting this result into
\begin{eqnarray}
\fl \partial_t \Delta_{\alpha_1 \alpha_2}(\vk) &=& -{\rm Sym}  \Big[ P_{\alpha_1 \beta' \gamma'}(\vk) \sum_{\vp+\vq=\vk}
\hat C^{(3)}_{\beta' \gamma' \alpha_2}(\vp,\vq,-\vk) \Big ] \\
\fl& =& t\,{\rm Sym}  \Big[ P_{\alpha_1 \beta' \gamma'}(\vk)  \Delta_{\gamma \alpha_2}(\vk) \sum_{\vp+\vq=\vk} P_{\beta' \beta \gamma}(\vp) \Delta_{\beta \gamma'}(\vq)
+ P_{\alpha_1 \beta' \gamma'}(\vk) \Delta_{\gamma \alpha_2}(\vk) \sum_{\vp+\vq=\vk} P_{\gamma' \beta \gamma}(\vq)  \Delta_{\beta \beta'}(\vp) \nonumber
\\
\fl&& ~~~~~~~~~~- P_{\alpha_1 \beta' \gamma'}(\vk) P_{\alpha_2 \beta \gamma}(\vk) \sum_{\vp+\vq=\vk}
\Delta_{\beta \beta'}(\vp) \Delta_{\gamma \gamma'}(\vq)\Big  ]\ , \nonumber
\end{eqnarray}
where here ${\rm Sym}[...]$ means symmetrization w.r.t. $\alpha_1,\alpha_2$. Symmetrization finally yields the general 1-loop equation
\begin{eqnarray}
\fl \partial_t \Delta_{\alpha_1 \alpha_2}(\vk)
& =& t  \bigg(\Big( P_{\alpha_1 \beta' \gamma'}(\vk)  \Delta_{\gamma \alpha_2}(\vk) + P_{\alpha_2 \beta' \gamma'}(\vk) \Delta_{\gamma \alpha_1}(\vk)\Big )
\sum_{\vp+\vq=\vk} P_{\beta' \beta \gamma}(\vp) \Delta_{\beta \gamma'}(\vq)
 \nonumber
\\
\fl&& ~~- P_{\alpha_1 \beta' \gamma'}(\vk) P_{\alpha_2 \beta \gamma}(\vk) \sum_{\vp+\vq=\vk}
\Delta_{\beta \beta'}(\vp) \Delta_{\gamma \gamma'}(\vq)  \bigg)\ ,  \label{frggen2}
\end{eqnarray}
from which the  1-loop FRG equation for NS and Burgers  can be retrieved.

For NS one  has $\vk_{\alpha_i} \Delta_{\alpha_1,\alpha_2}( \vk)=0$ for $i=1,2$. One checks on (\ref{frggen2}) that
if $\Delta$ is transverse at a given time $t$, it remains so, i.e.\  the r.h.s.\ is automatically transverse. For $N=2$ this implies  that $\Delta_{\alpha \beta}(\vk)=P^{\rm T}_{\alpha \beta}(\vk) \Delta(\vk)$, but this is not true for $N>2$. The general form is $\Delta_{\alpha \beta}(\vk)= \sum_{i,j=1}^{N-1} e_\alpha^i(\vk) e_\beta^j(\vk) \Delta_{ij}(\vk)$
where the $e^i(\vk)$ span a basis orthogonal to $\vk$ and  $\Delta_{ij}(\vk)$ is a symmetric matrix.

For simplicity we consider here the subspace $\Delta_{\alpha \beta}( \vk)=P^{\rm T}_{\alpha \beta}(\vk) \Delta(\vk)$. One finds, using Mathematica
\begin{eqnarray}
\fl
\sum_{\vp+\vq=\vk} P_{\alpha_1 \beta' \gamma'}(\vk) P_{\alpha_2 \beta \gamma}(\vk)
\Delta_{\beta \beta'}(\vp) \Delta_{\gamma \gamma'}(\vq)
= \frac{- 2}{N-1} P^{\rm T}_{\alpha_1 \alpha_2}(\vk)
\sum_q \frac{[k^2 q^2 - (\vk \cdot \vq)^2]}{k^2 q^2 (\vk-\vq)^2} \nonumber \\
 \times  \Big[\frac{1}{2} \Big((\vk-\vq)^2 - q^2\Big)^2  + \frac{1}{2} (N-2)
k^2 \Big(q^2 + (\vk-\vq)^2\Big) \Big] \Delta(\vq) \Delta(\vk-\vq)\ .
\end{eqnarray}
At this stage this was obtained by: (i) symmetrizing w.r.t.\ $q \to k-q,$ and (ii) {\it assuming} that the result was proportional to $P^{\rm T}_{\alpha_1 \alpha_2}(\vk),$
and then contracting with $P^{\rm T}_{\alpha_1 \alpha_2}(\vk)$ (or $\delta_{\alpha_1,\alpha_2}$). For $N=2$
there is no loss of generality, and the sum over momenta can be discrete, while for $N>2$ this holds only for  isotropic turbulence $\Delta(\vk)=\Delta(k)$ and in the limit of an infinite box where the sums become integrals.

Next one finds by the same method
\begin{eqnarray}
\fl \Big( P_{\alpha_1 \beta' \gamma'}(\vk) \Delta_{\gamma \alpha_2}(\vk)
 +
P_{\alpha_2 \beta' \gamma'}(\vk) \Delta_{\gamma \alpha_1}(\vk)\Big)
\sum_{\vp+\vq=\vk}  P_{\beta' \beta \gamma}(\vp) \Delta_{\beta \gamma'}(\vq) \\
\fl = \frac{- 2}{N-1} P^{\rm T}_{\alpha_1 \alpha_2}(\vk)
\sum_q \frac{k^2 q^2 - (\vk \cdot \vq)^2}{k^2 q^2 (\vk-\vq)^2} \Big[(k^2-q^2)\Big ((\vk-\vq)^2 - q^2\Big) + (N-2)
k^2 (\vk-\vq)^2 \Big] \Delta(\vq) \Delta(\vk)\ . \nonumber
\end{eqnarray}
This yields the 1-loop FRG equation
\begin{eqnarray}
\fl \partial_t \Delta(\vk) = \frac{2 t}{N-1}
\sum_q  \frac{k^2 q^2 - (\vk \cdot \vq)^2}{k^2 q^2 (\vk-\vq)^2} \Bigg\{ \Big[\frac{1}{2} \Big((\vk-\vq)^2 - q^2\Big)^2  + \frac{1}{2} (N-2)
k^2 \Big(q^2 + (\vk-\vq)^2\Big) \Big] \Delta(\vq) \Delta(\vk-\vq) \nonumber  \\
\lo - \Big[(k^2-q^2) \Big((\vk-\vq)^2 - q^2\Big) + (N-2)
k^2 (\vk-\vq)^2 \Big]  \Delta(\vq) \Delta(\vk) \Bigg\}\ .
\end{eqnarray}
We note Kraichnan's conventions,
\begin{eqnarray}
&& 4 k^2 a_{\vk,\vp,\vq} = P_{\alpha_1 \beta' \gamma'}(\vk) P_{\alpha_1 \beta \gamma}(\vk)
P^{\rm T}_{\beta \beta'}(\vp) P^{\rm T}_{\gamma \gamma'}(\vq)\ ,  \\
&& 2 k^2 b_{\vk,\vp,\vq} = P_{cjm}(\vk) P_{jbc}(\vp) P^{\rm T}_{mb}(\vq)\ .
\end{eqnarray}
This corrects a misprint in  equation~(VII-2-7) of \cite{LesieurBook}. The various symbols satisfy
\begin{eqnarray}
\fl 2 \tilde a_{\vk,\vk-\vq,\vq} = - 4 k^2 a_{\vk,\vk-\vq,\vq} =   \frac{[k^2 q^2 - (\vk \cdot \vq)^2]}{k^2 q^2 (\vk-\vq)^2}  \Big[ \Big((\vk-\vq)^2 - q^2\Big)^2  + (N-2)
k^2\Big (q^2 + (\vk-\vq)^2 \Big)\Big] \label{aformula-1} \\
\fl  \tilde b_{\vk,\vk-\vq,\vq} =  - 2 k^2 b_{\vk,\vk-\vq,\vq} =  \frac{[k^2 q^2 - (\vk \cdot \vq)^2]}{k^2 q^2 (\vk-\vq)^2} \Big[(k^2-q^2)\Big ((\vk-\vq)^2 - q^2\Big) + (N-2)
k^2 (\vk-\vq)^2 \Big] \label{bformula-1} \\
\fl \tilde a_{\vk,\vk-\vq,\vq} = \frac{1}{2} \left[\tilde b_{\vk,\vk-\vq,\vq} + \tilde b_{\vk,\vq,\vk-\vq}\right]\ . \label{aformula}
\end{eqnarray}
The FRG equation can thus be written in various forms,
\begin{eqnarray}
 \partial_t \Delta(\vk) &=&  \frac{2 t}{N-1}
\sum_q \tilde a_{\vk,\vk-\vq,\vq}  \Delta(\vq) \Delta(\vk-\vq) - \tilde b_{\vk,\vk-\vq,\vq}
 \Delta(\vq) \Delta(\vk)\ ,
 \eea
 as well as the form given in the text.

Note that $\tilde b_{\vk,\vk-\vq,\vq}$ is not invariant under $\vq \to \vk-\vq$;  $\tilde a_{\vk,\vk,0}=\tilde a_{\vk,0,\vk}=\frac12 \tilde b_{\vk,\vk,0}$ while $\tilde b_{\vk,0,\vk}=0$. These properties imply that the coefficient $B$ in the expansions in the text is zero.
Further  $k^2 q^2 - (\vk \cdot \vq)^2=k^2 p^2 - (\vk \cdot \vp)^2=p^2 q^2 - (\vp \cdot \vq)^2$ if $\vk=\vp+\vq$, hence this term is already symmetric under $\vq \to \vk-\vq$.

\section{FRG equation for $N=3$ periodic flows}

A convenient parameterization of a general $N=3$ divergence-less velocity correlation matrix, i.e.\  such that $\vk_\alpha \Delta_{\alpha \beta}(\vk) = 0$, (with mirror symmetry) is \begin{eqnarray}
 \Delta_{\alpha \beta}(\vk) &=& \sum_{i=x,y,z} \epsilon_{\alpha i \gamma} \hat \vk_\gamma \epsilon_{\beta i \delta} \hat \vk_\delta \Delta_i(\vk)\nn \\
&& = \delta_{\alpha \beta} \left(\sum_i \Delta_i (\vk)(1- \hat \vk_i^2) -\Delta_\alpha(\vk)\right) - \hat \vk_\alpha \hat \vk_\beta \left(\sum_i \Delta_i (\vk)- \Delta_\alpha (\vk)- \Delta_\beta(\vk)\right)
\ .\end{eqnarray}
In coordinates this is\begin{eqnarray}
&& \Delta_{xx}(\vk) = \hat \vk_y^2 \Delta_z(\vk) + \hat \vk_z^2 \Delta_y(\vk) \quad , \quad \Delta_{xy}(\vk) = \Delta_{yx}(\vk) = - \hat \vk_x \hat \vk_y \Delta_z(\vk)\ ,
\end{eqnarray}
and similar for circular permutations. The semi-isotropic case $\Delta_{\alpha \beta}(\vk)=P^{\rm T}_{\alpha \beta}(\vk) \Delta(\vk)$ corresponds to $\Delta_{i}(\vk) = \Delta(\vk),$ and is fully isotropic when $\Delta_{i}(\vk) = \Delta(|\vk|)$. For a periodic flow with a cubic lattice symmetry we expect  that
\begin{eqnarray}
&& \Delta_x(\vk)=\Delta(\vk_y,\vk_z;\vk_x) \quad , \quad \Delta_y(\vk)=\Delta(\vk_x,\vk_z;\vk_y) \quad , \quad \Delta_z(\vk)=\Delta(\vk_x,\vk_y;\vk_z)
\end{eqnarray}
where $\Delta(\vk_1,\vk_2;\vk_3)$ is a symmetric function of its first two arguments.

We have derived FRG equations for the $\Delta_i$. They are of the form $\partial \Delta_z = \sum_{i,j=x,y,z} \sum_q \Delta_i(\vq) \Delta_j(\vk-\vq) f_{ij}(\vk,\vq)$ where the
$f_{ij}(\vk,\vq)$ are quite complicated functions of $k_x,k_y,k_z,q_x,q_y,q_z$;  we have not tried to solve them.

\section{Generating functional approach and  diagrammatics}\label{a6}

In this appendix we explain how the FRG equations can be derived for
the decaying Burgers, NS and SQG equations within the Martin-Siggia-Rose
formalism, using a small-time expansion.
We start with the Burgers equation, with an initial condition
$\vv_{\vu 0}={\bf w}_\vu$ at $t=0^+$. This is equivalent to
\bea
\partial_t \vv_{\vu t} + \frac{1}{2} \nabla_{\vu} \vv_{\vu t}^2 = \nu \nabla^2_{\vu} \vv_{\vu t} + \delta(t) {\bf w}_\vu\ ,
\eea
with $\vv_{\vu t}=0$ for $t<0$, i.e.\ a forcing which acts only at time zero. We then introduce the generating functional
$e^{-S[\vv_{\vu t},\tilde{\vv}_{\vu t}]}$ for the velocity correlators, in the
usual way, which leads to the dynamic action
\begin{eqnarray}
 S &=& \int_{\vu,t \geq 0} \Big[ \tilde \vv_{\vu t} \partial_t \vv_{\vu t} -
  \nu \tilde \vv_{\vu t} \nabla_{\vu}^2 \vv_{\vu t}
-\ \frac{1}{2} (\nabla_{\vu} \cdot \tilde \vv_{\vu t})
(\vv_{\vu t})^2 \Big]  - \int_\vu \tilde \vv_{\vu 0} {\bf w}_\vu\ .  \label{lf20}
\end{eqnarray}
 $\tilde \vv_{\vu t}$ is the response field, and the path integral should be evaluated with
$\vv_{\vu t=0^-}=0$ at the boundary. Since normalization of the path integral is one, the generating function for
averages over the initial conditions can be computed from the dynamical path integral
with action:
 \bea \label{initc2}
S' &=& \int_{\vu,t \geq 0} \Big[ \tilde \vv_{\vu t} \partial_t \vv_{\vu t} -
  \nu \tilde \vv_{\vu t} \nabla_{\vu}^2 \vv_{\vu t}
-\ \frac{1}{2} (\nabla_{\vu} \cdot \tilde \vv_{\vu t})
(\vv_{\vu t})^2 \Big] -
 \frac{1}{2} \int_{\vu,\vu'} \tilde{\vv}_{\vu 0}^{\alpha}
 \Delta^{0}_{\alpha \beta}(\vu-\vu')  \tilde{\vv}_{\vu' 0}^{\beta},
\eea
where $\langle{{\bf w}_\vu^\alpha {\bf w}_{\vu'}^{\beta}}\rangle=
\Delta^{0}_{\alpha \beta}(\vu-\vu')$.
We  use the following graphical representation.
The vertex corresponding to the cubic nonlinearity is depicted by
\begin{eqnarray} \label{lf13}
- (\partial_{\vu}^{\beta}  \tilde \vv_{\vu t}^{\beta})
 (\vv_{\vu t}^{\alpha} \vv_{\vu t}^{\alpha})
= \diagram{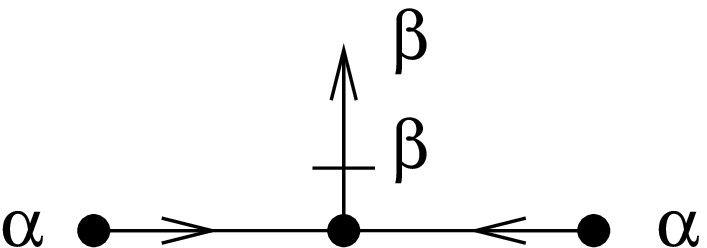}\ .
\end{eqnarray}
The response function in the limit $\nu \to 0$ reads
\begin{equation}
\Theta(t_2-t_1)\delta_{\alpha\beta} = \diagram{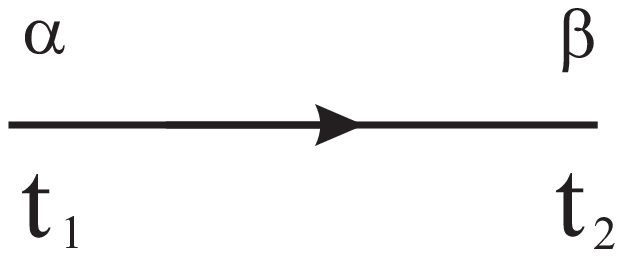}
\ .\end{equation}
The dashed line denotes the 2-point velocity correlator at $t=0$,
\begin{equation}
\Delta^{0}_{\alpha \beta}(\vu-\vu')  = \diagram{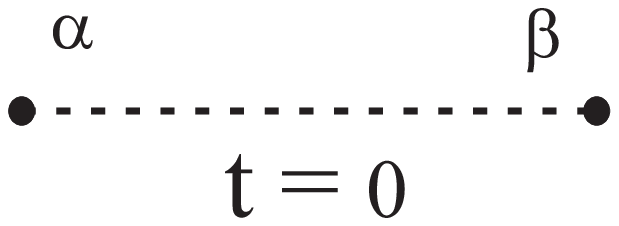}.
\end{equation}
We now switch to the Fourier representation. The 2-point velocity correlator written
in Fourier space
$\Delta_{\alpha\beta}^{t}(\vk)=\langle{\vv_{\vk t}^{\alpha} \vv_{-\vk t}^{\beta} }\rangle$
to 1-loop order is given by the  diagrams
\begin{figure}[t]\begin{center}
\includegraphics[clip,width=4.5in]{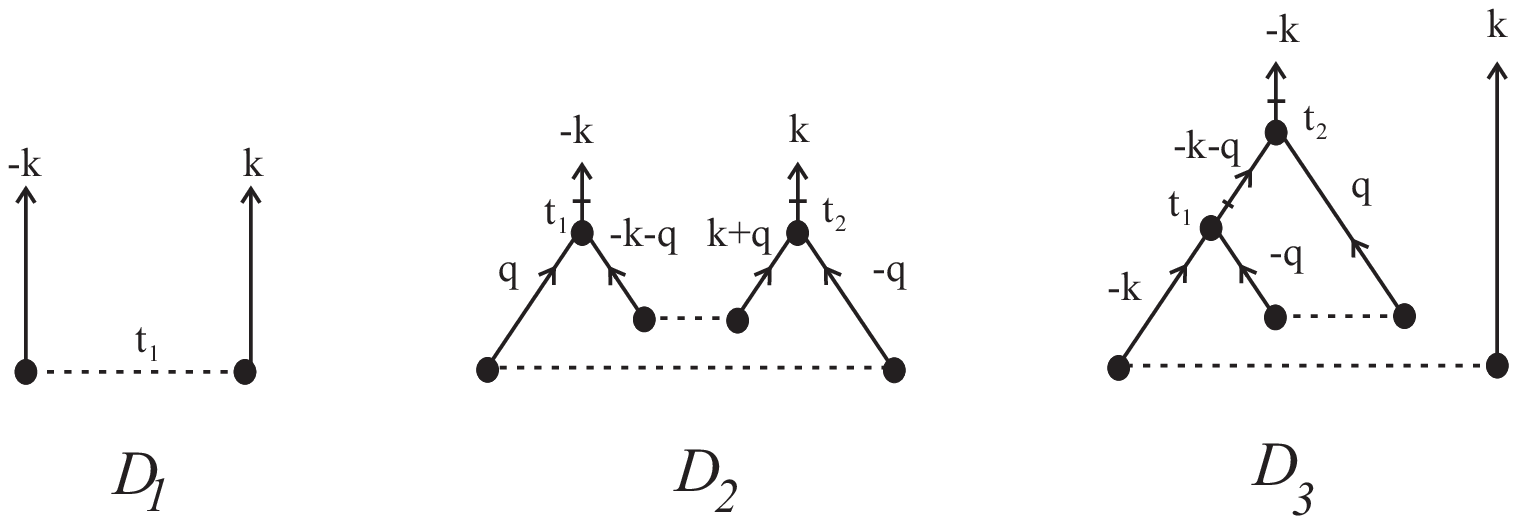}
\end{center}
\caption{The diagrams given in Eqs.~(\ref{D7}) to (\ref{D9}).}
\label{D0}
\end{figure}
 given on figure \ref{D0}. The corresponding expressions with combinatorial factors are\begin{equation}\label{D7}
  D_1 =  \Delta^0_{\alpha\beta}(\vk),
\end{equation}
\begin{equation}\label{D8}
  D_2=\frac{t^2}{2} \vk_{\alpha} \vk_{\beta} \sum_\vq \Delta^0_{\gamma\delta}(\vq)
   \Delta^0_{\gamma\delta}(\vk+\vq) ,
\end{equation}
\begin{equation}\label{D9}
  D_3=-t^2  \vk_{\alpha} \vk_{\beta} \sum_\vq \Delta^0_{\gamma\delta}(\vq)
   \Delta^0_{\gamma\delta}(\vk).
\end{equation}
In real space, the sum of the diagrams can be written as
\begin{eqnarray}
  \Delta_{\alpha\beta}^{t}(\vu) &=& \Delta_{\alpha\beta}^0(\vu) -
  \frac{t^2}2 \partial_{\alpha}\partial_{\beta}
  [\Delta_{\gamma\delta}^0(\vu)-\Delta_{\gamma\delta}^0(0)]^2.
\end{eqnarray}
To compute the $\beta$-function we take the derivative with respect to $t$
\begin{eqnarray}
 \partial_t \Delta_{\alpha\beta}^{t}(\vu) &=&  -
  t \partial_{\alpha}\partial_{\beta}
  [\Delta_{\gamma\delta}^0(\vu)-\Delta_{\gamma\delta}^0(0)]^2\ ,
\end{eqnarray}
and substitute to one loop $\Delta_{\alpha\beta}^{0}=\Delta_{\alpha\beta}^{t}$.
As a result we obtain  the FRG equation (\ref{flow1}).

We now generalize the method developed above to the Navier-Stokes equation.
It is convenient to put all derivatives in the cubic vertex on the
response field and write the dynamical action as
\begin{equation}\label{a5}
 S = \int_{\vu,t\ge0} \Big[ \tilde \vv_{\vu t} \partial_t \vv_{\vu t} - \nu \tilde \vv_{\vu t}
\nabla_{\vu}^2 \vv_{\vu t} - \lambda \vv_{\vu t}^\beta \vv_{\vu t}^\gamma (P^{\rm T}_{\alpha
\beta}(\partial_{\vu}) \partial^\gamma_{\vu} \tilde \vv_{\vu t}^{\alpha})
  \Big]
     -  \frac{1}{2} \int_{\vu,\vu'} \tilde{\vv}_{\vu 0}^{\alpha}
 \Delta^{0}_{\alpha \beta}(u-u')  \tilde{\vv}_{\vu' 0}^{\beta}\ ,\ \ \ \ \
\end{equation}
where the last term imposes the initial condition similar to that in
action (\ref{initc2}).
The nonlinear cubic vertex can then be written as
\begin{equation}\label{a6b}
\vv_{\vu t}^{\beta} \vv_{\vu t}^\alpha
(P^{\rm T}_{ \beta\beta' }(\partial_{\vu}) \partial^\alpha_{\vu} \tilde \vv_{\vu t}^{\beta'})
  = \diagram{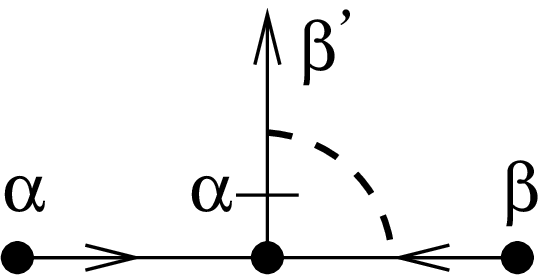}=- \diagram{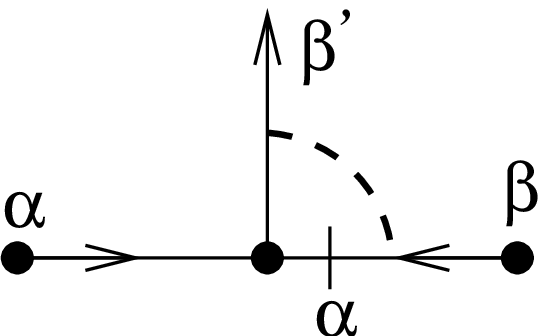}
\ .\end{equation}
The two-point function
$\Delta_{\alpha\beta}^{t}(\vk)=\langle{\vv_{\vk t}^{\alpha} \vv_{-\vk t}^{\beta} }\rangle$
to second order in $\Delta^0(\vk)$ is given by the diagrams in figure~\ref{fig4s}.
The  tree-level diagram $S_1$ reads
\begin{equation}
  S_1=  \Delta^0_{\alpha\beta}(\vk).
\end{equation}
There are six 1PI one-loop diagrams which can be split into two groups.
The first group gives
\begin{eqnarray}
 S_2 + S_3&=& t^2
     P^{\mathrm{T}}_{\alpha\gamma}(-\vk) P^{\mathrm{T}}_{\beta\sigma}(\vk)
     \vk_{\delta} \vk_{\rho} \sum_{\vq}
     \Big[ \Delta^0_{\gamma\sigma}(\vq)\Delta^0_{\delta\rho}(\vq+\vk)+
     \Delta^0_{\gamma\rho}(\vq)\Delta^0_{\delta\sigma}(\vq+\vk)\Big].
 \end{eqnarray}
Expanding the projection operators we find
\[P^{\mathrm{T}}_{\alpha\beta}(\vk) [S_2+S_3] = t^2 \sum_\vq
 \tilde{a}_{\vk,\vk-\vq,\vq}\Delta^0(\vq)\Delta^0(\vk+\vq),
\]
where we used that
$\Delta^0_{\alpha \beta} (\vq) =P^{\mathrm{T}}_{\alpha \beta} (\vq) \Delta^0 (\vq)$
and $\tilde{a}$ is given by equation (\ref{aformula-1}).
Analogously we obtain
\begin{eqnarray}
 S_4+S_5+S_6+S_7 = -  t^2 I_{\alpha\tau}(\vk)\Delta^0_{\tau\beta}(\vk),
\end{eqnarray}
where we have introduced
\begin{eqnarray}\label{a13-2}
I_{\alpha\tau}(k)& =& \sum_\vq \left[
 P^{\mathrm{T}}_{\alpha \gamma}(\vk)
 P^{\mathrm{T}}_{\gamma \tau} (\vk+\vq)
 P^{\mathrm{T}}_{\mu \sigma} (q) \Delta(\vq)
 \vk_{\sigma} (\vk_{\mu }+\vq_{\mu}) \right. \nonumber \\
&& \left. ~~~~~+
 P^{\mathrm{T}}_{\alpha \gamma}(\vk)
 P^{\mathrm{T}}_{\gamma \mu} (\vk+\vq)
  P^{\mathrm{T}}_{\mu \sigma} (\vq) \Delta(\vq)
 \vk_{\sigma} (\vk_{\tau }+\vq_{\tau}) \right. \nonumber \\
&& \left.
 ~~~~~+ P^{\mathrm{T}}_{\alpha \sigma}(\vk)
 P^{\mathrm{T}}_{\gamma \tau} (\vk+\vq)
  P^{\mathrm{T}}_{\mu \sigma} (\vq) \Delta(\vq)
 \vk_{\gamma} (\vk_{\mu }+\vq_{\mu})\right. \nonumber \\
&& \left. ~~~~~+
 P^{\mathrm{T}}_{\alpha \sigma}(\vk)
 P^{\mathrm{T}}_{\gamma \mu} (\vk+\vq)
  P^{\mathrm{T}}_{\mu \sigma} (\vq) \Delta(\vq)
 \vk_{\gamma} (\vk_{\tau }+\vq_{\tau}) \right].
\end{eqnarray}
Expanding the projection operators one  sees that
\[P^{\mathrm{T}}_{\alpha\beta}(\vk) [S_4+S_5+S_6+S_7 ] = -t^2 \sum_\vq
 \tilde{b}_{\vk,\vk-\vq,\vq}\Delta^0(\vq)\Delta^0(\vk),
\]
where  $\tilde{b}$ is defined in equation~(\ref{bformula-1}).
Taking the derivative with respect to $t$ we obtain the FRG equation (\ref{NSFRG}).

The FRG equation for the SQG equation can be obtained in the same way.
The  dynamic  action corresponding to  equation (\ref{212}) with the
initial condition imposed at $t=0$  is given by
\begin{equation}\label{action-SQG}
 S = \int_{\vu,t\ge 0} \Big\{ \tilde T_{\vu t} \partial_t T_{\vu t} - \nu \tilde T_{\vu t}
\nabla^2 T_{\vu t} + \tilde{T}_{\vu t}\hat{z} \cdot [ \nabla \psi_{\vu t}\times
\nabla T_{\vu t}]\Big \}
-\frac{1}{2} \int_{\vu,\vu'} \tilde T_{\vu 0}
 \Delta^{0}_{ T}(\vu-\vu')   \tilde T_{\vu' 0}. \ \ \ \
\end{equation}
The SQG vertex can be written in Fourier space as 
\begin{eqnarray}
&&\int_{\vu,t}  \tilde{T}_{\vu t}\hat{z} \cdot [ \nabla \psi_{\vu t}\times
\nabla T_{\vu t}] =
  \int_{\vu, \vu', t} \tilde{T}_{\vu t} \epsilon_{\alpha \beta} \partial_{\vu}^{\alpha}
   K_{\vu-\vu'}  {T}_{\vu' t}
\partial_{\vu}^{\beta} {T}_{{\vu}t}  \nonumber \\
&& =- \int_{\vu, \vu', t} K_{\vu-\vu'} {T}_{\vu' t} \epsilon_{\alpha\beta}
\partial_{\vu}^{\alpha}[ \tilde{T}_{\vu t} \partial_{\vu}^{\beta} {T}_{\vu t}]
= -\int_t \sum_{\vk ,\vq} \tilde T _{-\vk-\vq t}
\epsilon_{\alpha\beta} \vq_{\alpha} \vk_{\beta} q^{-a} T_{\vq t} T_{\vk t},
\label{222}
\end{eqnarray}
where we have introduced the inverse Fourier transform
\be  K_{\vu} :=\mbox{FT}^{-1}_{\vu\leftarrow \vq} \frac{1}{q^{a}}\ . \ee
The 2-point function $\Delta^t_T(\vk)=\langle{T_{\vk t} T_{-\vk t} }\rangle$ to second
order in $\Delta^0_T$ is given by the diagrams in figure~\ref{fig4s}.
The corresponding expressions read
\begin{figure}[t]
\begin{center}
\fig{15cm}{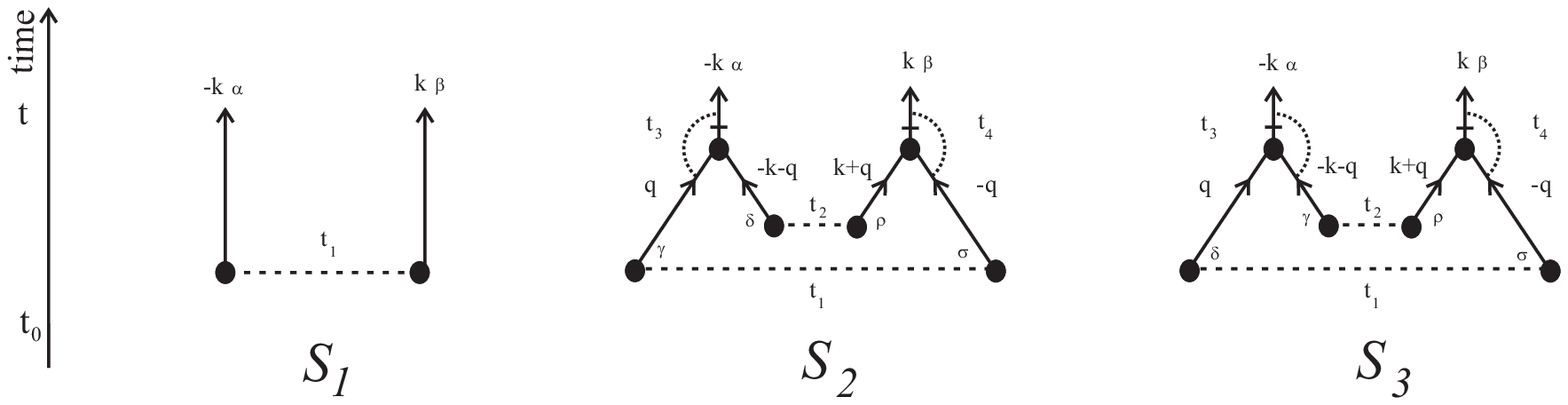}\\
\fig{15cm}{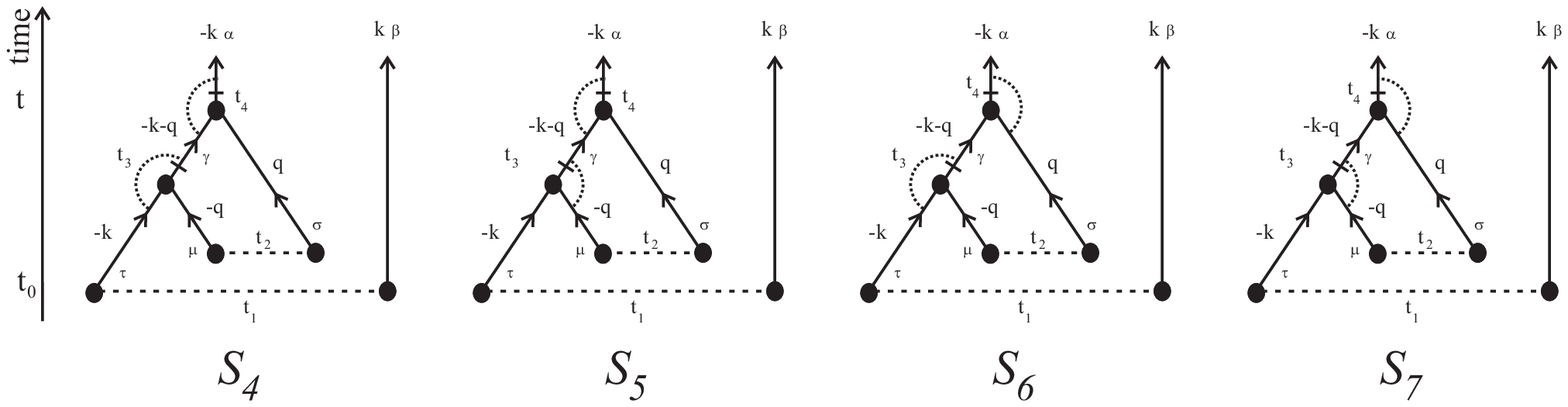}
\end{center}
\caption{Diagrams contributing to the FRG for the NS and SQG equations.}
\label{fig4s}
\end{figure}
\begin{eqnarray}\label{225}
\fl ~~~~~~~~~~~~~~~~~~~~S_{1} &=&  \Delta_{ T}^0 (\vk),\\
\label{226}
\fl ~~~~~~~~~~~~~~S_2 + S_3&=& \frac1{2}t^2 \sum_\vq
\Delta^0_{T}(\vq)\Delta^0_{ T}(\vq+\vk)[\vq\times \vk]^2
\left\{q^{-2a}-2q^{-a}|\vq+\vk|^{-a}+|\vq+\vk|^{-2a} \right \},
\\
\fl S_4 + S_5+S_6 + S_7&=& - t^2 \sum_\vq
\Delta^0_{ T}(\vq)\Delta^0_{ T}(\vk)[\vq\times \vk]^2
\left\{q^{-2a}-q^{-a}k^{-a}-q^{-a}|\vq+\vk|^{-a} +k^{-a}|\vq+\vk|^{-a} \right
\}.
\label{227}
\end{eqnarray}
Taking the derivative with respect to $t$ and reexpressing the bare  disorder$\Delta^0_{ T}$
in terms of the renormalized one $\Delta_T$  we obtain the FRG equation (\ref{229}).

\section{Distance geometry for FRG equation}\label{distance geometry}

\subsection{Navier-Stokes}

We derive the measure used in \cite{LesieurBook}, e.g.\
equation~(VII-2-9). Following \cite{DDG2}, appendix A, the integral of a function $f
(x_{1}, \dots ,x_{n})$ which depends only on $u_{ij}:=x_{i}\cdot x_{j}$ can be written as
\begin{equation}\label{a2}
\int_{\mathbb{R}^N} \prod_{i=1}^n \rmd^Nx_i\, f(u_{ij}) = {S_N\over 2}\ldots{S_{N-n+1}\over 2}
\int \prod_{i\le j} \rmd u_{ij} (\det[u_{ij}])^{N-n-1\over 2}\  f([u_{ij}]),
\end{equation}
where the domain of integration is such that the scalar-products
can be realized in $N$-dimensional space, and $S_N=2 \pi^{N/2}/\Gamma(N/2)$ is the
area of the unit sphere. Here we need $n=2$ with $\vk=\vx_{1}$,
$\vp=\vx_{2}$, $\vk+\vp+\vq=0$, $k=|\vk|$,
$q=|\vq|$, $ p=|\vp|$. One finds, for arbitrary $N$,
\begin{eqnarray}\label{a4}
\fl\int \rmd^Nk\, \rmd^Np\,  f(u_{ij}) &=&\int \rmd k^{2}\rmd p^{2}\rmd(\vk\vp)
(k^{2}p^{2}-(\vk\vp)^2)^{\frac{N-3}2}\frac{S_{N}S_{N-1}}{4} f (k^{2},p^{2},(\vk-\vp)^{2}) \nonumber \\ \label{F.4bis}
&=&S_{N}S_{N-1}\int_{\Delta} \rmd k\, \rmd p \,\rmd  q\, kpq \, \left[\frac4{ (p{+}q{-}k) (k{+}p{-}q) (k{+}q{-}p) (k{+}p{+}q)} \right]^{\frac{3-N}2}\, f (k^{2},p^{2},q^{2}), 
\end{eqnarray}
where the  triangle symbolizes the realization of the triangle
inequality. We have used that $\vk \cdot \vp=\frac{1}{2}(p^2+k^2-q^2)$ hence $\rmd k^{2}\rmd p^{2}\rmd(\vk\vp)
= \frac{1}{2} \rmd k^2 \rmd p^2 \rmd q^2$, as well as,
\begin{equation}
k^2p^2 -(\vk\vp)^2 = \frac{1}{4} (k+p+q) (p+q-k) (k+p-q) (k-p+q).
\end{equation}
The domain of integration is plotted in
figure \ref{f:domaine}. A non-trivial check is to suppose that $f$ is independent of
$p$, and to do the $p$ integration. With the domain in figure
\ref{f:domaine}, the two cases $q<k$ and $q>k$ have to be
distinguished. For $N=3$, both (!) give $\int p\, \rmd p = 2kq$, which result in
two independent 3-dimensional integrals (with correct normalization)
for $q$ and $k$.

Since this is valid for any function of $k$, we may rewrite (\ref{NSFRG}) for the isotropic case and any $N$, replacing $\sum_{q}\to
\int_{q} = \int \frac{\rmd^Nq}{(2 \pi)^N}$, using
$\delta \Delta(K)=\int \rmd^N \vk \frac{\delta(k-K)}{S_N k^{N-1}} \delta \Delta(k)$,
which yields
\begin{eqnarray} \label{frgdist}
&& \!\!\!\!\!\!\!\!\!\!\!\!\!\!\!\!\!\! \!\!\!\!\!\!\!\!\!\!\!\!\!\!\!\!\!\! \partial_t \Delta(k) =    \frac{2 t}{N-1} \frac{S_{N-1}}{(2 \pi)^N}
\int_{0}^{\infty} \rmd q \int_{0}^{\infty} \rmd p\, \Theta_{\Delta} (k,p,q)
\frac{B_{k,p,q}} {k^N p q}
\left[ \Delta(q) \Delta(p) -
 \Delta(q) \Delta(k) \right], \\
&&  \!\!\!\!\!\!\!\!\!\!\!\!\!\!\!\!\!\! \!\!\!\!\!\!\!\!\!\!\!\!\!\!\!\!\!\! B_{kpq} = \left[\frac{1}{4} (k+p+q) (p+q-k) (k+p-q) (k-p+q)\right]^{\frac{N-1}{2}}  \left[(k^2-q^2)  (p^2 - q^2) +
(N-2) k^2 p^2 \right], \nonumber  \\
&&  \!\!\!\!\!\!\!\!\!\!\!\!\!\!\!\!\!\! \!\!\!\!\!\!\!\!\!\!\!\!\!\!\!\!\!\!  \Theta_{\Delta} (k,p,q) =  \Theta
(k+p>q)\Theta (k+q>p)\Theta (p+q>k).
 \end{eqnarray}
The domain of integration is plotted in
figure \ref{f:domaine}. The distance geometry can be parameterized by
$p=\frac{k}{2} (s+t)$ and $q=\frac{k}{2} (s-t)$ with
\begin{eqnarray}
\!\!\!\!\!\!\!\!\!\!\!\!\!\!\! \int_0^\infty \rmd p \int_0^\infty \rmd q ~ \Theta_{\Delta} (k,p,q) f (k,p,q) = \frac{k^{2}}{2} \int_{1}^{\infty}\rmd s
\int_{-1}^{1}\rmd t f \big(k,\frac{k}{2} (s+t),\frac{k}{2} (s-t)\big).
\end{eqnarray}
It is immediate from (\ref{frgdist}) that $\int_0^\infty \rmd k  \,k^{N-1} \Delta(k)$ is conserved, using the symmetry under
the exchange $p \leftrightarrow k$. This implies energy conservation for all $N$. For $N=2$ one checks that  enstrophy is conserved, i.e.\  $\int_0^\infty \rmd k \,k^3 \Delta(k)$ can be brought to the form $\sim \int_{k,p,q} B_{k p q} (\frac{k}{pq} - \frac{p}{k q}) \Delta(p) \Delta(q) $ which leads to a factor of  $(k^2-p^2)(k^2-q^2)(p^2-q^2)$ times a symmetric function of $p,q$; hence it vanishes.

Violation of energy conservation necessitates a divergence in the integrals for large momenta so that the operations involved in the symmetrization, e.g.\ change of order in integration, are no more valid. For $N=3$ one sees that at $\zeta_{2}=1$, for fixed $s$ and $t$ there is
a logarithmic divergence $\int \rmd k/k$ at large $k$,  none for
$\zeta_{2}>1$, and a relevant one for $\zeta_{2}<1$.  The momentum
space-integrals are therefore no longer well-defined for $\zeta_{2}\le 1$.

\begin{figure}
\centerline{\fig{5cm}{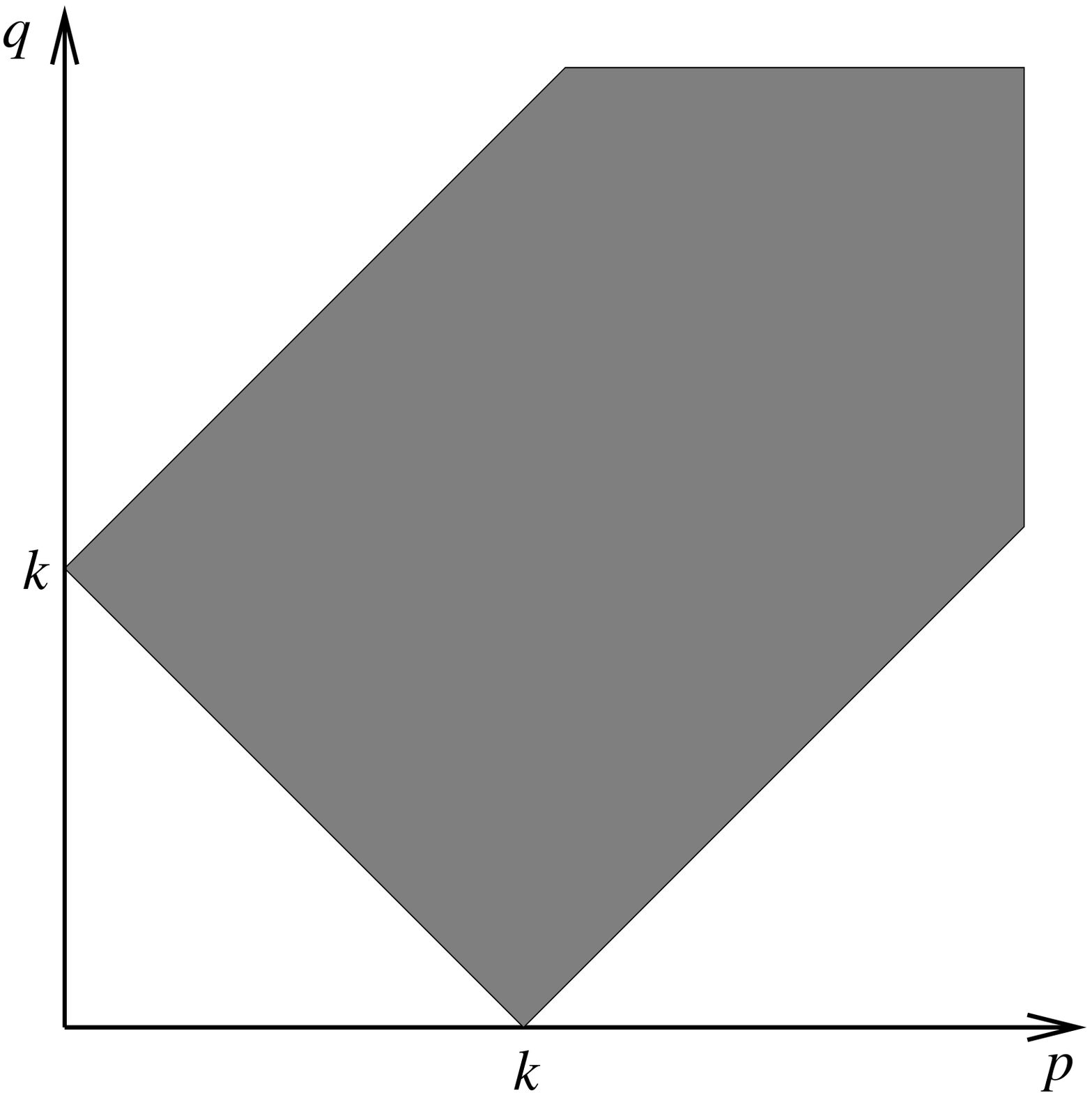}}
\caption{The domain of integration in equation (\ref{F.4bis}). }
\label{f:domaine}
\end{figure}

\subsection{Burgers}

One finds a similar expression for Burgers: 
\begin{eqnarray}\label{a4b}
\partial_t \Delta(k) &=&
\frac{S_{N-1}t}{4 (2 \pi)^N} k^{2-N}
\int_{0}^{\infty} \rmd q \int_{0}^{\infty} \rmd p\, \Theta_{\Delta} (k,p,q)  \frac{1}{pq} \, \left[\frac4{ (p{+}q{-}k) (k{+}p{-}q) (k{+}q{-}p) (k{+}p{+}q)} \right]^{\frac{3-N}2}\, \nonumber \\
&&~~~~~~~~~~~~~~~~~~~~ \times \bigg[ \frac{k^2}{2} (p^2+q^2-k^2)^2
\Delta(p)  \Delta(q) - p^2 (k^2+q^2-p^2)^2  \Delta(q) \Delta(k) \bigg].
\end{eqnarray}
\section{Short-distance expansion of $\Delta(u)$: Amplitudes $B$ and $C$ for
Burgers, Navier Stokes and surface quasi-geostrophic turbulence}
\label{s:amplitudes}
In this appendix, we calculate the necessary integrals for the short-distance
expansion of $\Delta (u)$  for Burgers, \ref{s:Ints-Burgers},
Navier-Stokes, \ref{s:Ints-NS} and quasi-geostrophic, \ref{sec-SQG-integral}.
For simplicity of notations, we use
\begin{equation}
b = N+\zeta_{2} \ .
\end{equation}

\subsection{Burgers}\label{s:Ints-Burgers}

To get the coefficient $C$ in (\ref{expansion}) we  need to compute
\begin{eqnarray}
\fl J(k) =\frac{1}{2} k^2  \int_\vq \frac{[\vq \cdot (\vk-\vq)]^2 }{q^2 (\vk-\vq)^2} G(q) G(k-q) = \frac{1}{2} k^2 \int_\vq \frac{(\frac{k^2}{4}-q^2)^2 }{(\frac{\vk}{2}+\vq)^2 (\frac{\vk}{2}-\vq)^2} G\Big(\Big|\frac{\vk}{2}+\vq\Big| \Big) G\Big(\Big|\frac{\vk}{2}-\vq \Big|\Big)
\end{eqnarray}
and set $G(k)=1/k^b$. We note generically $I$ an integral where such replacement
is performed, while $J$ denotes the same integral with IR cutoffs. Indeed,
the second integral in (\ref{expansion}) behaves as
$\sim m^{N-b} k^2 /(k^2+m^2)^{b/2}$ and contributes only to $B$ and $D$
(equivalent to the statement that $\int_\vq q^{-a}=0$ in dimensional regularization). While the calculation of $B$ and $D$ (for each integral) depends a priori on the IR details of $G(k)$, the coefficient $C$ can be obtained by the method of analytical continuation on the first integral only. This integral is both UV and IR convergent upon inserting $G(k)=1/k^b$ for $N/2 < b <N$. Its expression is then continued for $b>N$ to get $C$. The cancellation of $B$ between the two terms is easy to show. 
One has 
\begin{eqnarray}
  \fl I(k) &= & \frac{k^2}{2} \int_\vq \frac{(\frac{k^2}{4}-q^2)^2 }{|\frac{\vk}{2}+\vq|^{b+2} |\frac{\vk}{2}-\vq|^{b+2}}
  = \frac{k^2}{2} \frac1{\Gamma(b/2+1)^2} \int_{t_i>0} t_1^{b/2} t_2^{b/2} \partial^2_{v}\big|_{v=0} \int_\vq
e^{-(t_1+t_2) (\frac{k^2}{4}+q^2) + v  (\frac{k^2}{4}-q^2) - (t_1-t_2) \vk \cdot \vq } \nonumber  \\
 \fl &=& \frac{1}{2} k^2 \frac{(4 \pi)^{-N/2}}{\Gamma(b/2+1)^2} \int_{t_i>0} t_1^{b/2} t_2^{b/2} \partial^2_{v}\big|_{v=0}
(t_1+t_2+v)^{-N/2}
e^{ \frac{v^2-4 t_1 t_2}{t_1+t_2+v} \frac{k^2}{4}} \nn\\
\fl&  =&  \frac{1}{2} k^2 \frac{(4 \pi)^{-N/2}}{4 \Gamma(b/2+1)^2} \int_{t_i>0} t_1^{b/2} t_2^{b/2} (t_1+t_2)^{-4-\frac{N}{2}}  e^{- \frac{t_1 t_2}{t_1+t_2} k^2}
\nn\\
\fl&&\times \bigg\{ \Big[ N^2 (t_1 + t_2)^2 + 2 N (t_1 + t_2)\Big (t_2 +
t_1 (1 -
2 k^2 t_2)\Big)\Big]+2 k^2 \Big[t_1^3 - t_1 t_2^2 + t_2^3 - t_1^2 t_2 (1 - 2 k^2 t_2)\Big] \bigg\}.
\end{eqnarray}
Introducing $s=t_1+t_2$, $t_1=s u$, $t_2=s(1-u)$ with $\rmd t_1
\rmd t_2 = s\, \rmd u\, \rmd s$ one gets
\begin{eqnarray}
&&  \! \! \! \! \! \! \! \! \! \! \! \! \! \! \! \! \! \! \! \! \! \! \! \!\! \! \! \! \! \! \! \! \! \! \! \! \! \! \! \!
I(k) = \frac{1}{2} k^2 \frac{(4 \pi)^{-N/2}}{4 \Gamma(b/2+1)^2} \int_0^1 \rmd u  \int_{s>0}
s^{-1 +  b - \frac{N}{2}} (u (1-u))^{b/2}  \bigg\{
 \Big[N^2 + N (2 - 4 k^2 s (1 - u) u)\Big]\nn \\
&&  ~~~~~~~~~~~~~~~~~~~~~~~~~~~~~~~~~~~~~~~~~~~~~~~~+ 2 k^2 s \Big[1 - 2 (1 - u) u (2 - k^2 s (1 - u) u)\Big] \bigg\}
e^{- s u (1-u) k^2} \ .
\end{eqnarray}
The integration over $s$ can be performed, if $b>N/2, $ leading to
\begin{eqnarray}
&&  \! \! \! \! \! \! \! \! \! \! \! \! \! \! \! \! \! \! \! \! \! \! \! \!\! \! \! \! \! \! \! \! \! \! \! \! \! \! \! \!
I(k) = -  \frac{(4 \pi)^{-N/2} \Gamma( b - N/2)}{8 \Gamma(b/2+1)^2} k^{2+N-2 b} \int_0^1 \rmd u \;
[u (1-u)]^{\frac{N-b-2}{2}}
\Big[N - 4 b^2 (1 - u) u \nn\\
&& ~~~~~~~~~~~~~~~~~~~~~~~~~~~~~~~~~~- 4 N (1 + N) (1 - u) u  - 2 b (1 - 2 (1 + 2 N) (1 - u) u)\Big]\ .
\end{eqnarray}
This integral converges only for $b<N$, where it is\begin{eqnarray}
&&  \! \! \! \! \! \! \! \! \! \! \! \! \! \! \! \! \! \! \! \! \! \! \! \!\! \! \! \! \! \! \! \! \! \! \! \! \! \! \! \!
I(k) = C k^{2+N-2 b} = 2^{-2+b-2 N} \pi^{\frac{1-N}{2}} \frac{ \big[b^2 - 2 b (N-1) + N(N-1)\big]  \Gamma(b - \frac{N}{2}) \Gamma\big(\frac{1}{2} (N-b)\big)}{\Gamma(b/2+1)^2 \Gamma\big(\frac{1}{2} (1+N-b)\big)} k^{2+N-2 b}  \ .
 \label{Capp}
\end{eqnarray}
This identifies $C=C(b,N)$ for $b<N$. For $b>N$ the correct calculation requires regularization by a  mass $m^2$, and leads to $J(k) = B_1 k^{2-b} + C k^{2+N-2 b} + D k^{-b}$. While $B_1$ is cancelled by the second integral in (\ref{expansion}), the expression of $C$ remains equal to the analytical continuation of (\ref{Capp}).

\subsection{Navier-Stokes}
\label{s:Ints-NS}

The non-linear term in the FRG equation is the sum of two integrals,
\begin{eqnarray}
\fl  J(k) =  \frac{2}{(N-1)}
\int_\vq \frac{k^2 q^2 - (\vk \cdot \vq)^2}{k^2 q^2 (\vk-\vq)^2} \Big[(k^2-q^2)\Big ((\vk-\vq)^2 - q^2\Big) + (N-2)
k^2 (\vk-\vq)^2 \Big] G(q)\Big[G(k-q) - G(k)\Big]\ . \nonumber
\end{eqnarray}
If we  replace $G(p)$ by $p^{-b}$, we see that the first integral is both UV and IR convergent for $N/2 < b <N$, while the second is nowhere convergent. It is either UV divergent (for $b<N$) or IR divergent (for $b>N$)  at $q \approx 0$ (but not at $q \approx k$). It is thus convenient to split \(J(k)\) into two parts:
\begin{eqnarray}
 J(k) &=& \tilde J(k) + J_c(k),\\
J_c(k) &=&  2 \int_\vq \frac{k^2 q^2 - (\vk \cdot \vq)^2}{q^2} G(q)\Big[G(|\vk-\vq|) - G(k)\Big],  \\
 \tilde J(k) &=&  \frac{2}{(N-1)}
\int_\vq \frac{k^2 q^2 - (\vk \cdot \vq)^2}{k^2 (\vk-\vq)^2} \Big[q^2 - k^2 - (\vk-\vq)^2\Big]
 G(q)\Big[G(|\vk-\vq|) - G(k)\Big].
\end{eqnarray}
It is easy to see that the second integral in $J_c(k)$ contributes only to $B$ and $D$ but not to $C$, while $\tilde I(k)$, the same integral as $\tilde J(k), $ replacing $G(p)$ by $p^{-b}$, is now both UV and IR convergent for $N<b<N+2$. One thus has\begin{eqnarray}
\fl  ~~~~~~  \tilde I(k) &=&  - \frac{1}{2 k^2(N-1)} \int_{t_1,t_2>0} \bigg[ \frac{t_1^{b/2-1} t_2^{b/2}}{\Gamma(1+\frac{b}{2})\Gamma(\frac{b}{2})}  - \frac{t_1^{b/2-1} k^{-b}}{\Gamma(\frac{b}{2})} \bigg] B(t_1,t_2)
\int_\vq e^{- t_1 q^2 - t_2 (\vk-\vq)^2  } \\
\fl  B(t_1,t_2) &=&
\left[k^4 + 2 k^2 (\partial_{t_1} + \partial_{t_2}) + (\partial_{t_1} - \partial_{t_2})^2\right]
(-k^2 + \partial_{t_2} - \partial_{t_1})\ ,
\end{eqnarray}
using that $k^2 q^2-(\vk \cdot\ \vq)^2=-\frac{1}{4} \Big[k^4 - 2 k^2 (q^2 + (\vk-\vq)^2) + (q^2-(\vk-\vq)^2)^2\Big]$.
Using
\begin{eqnarray}\label{F12}
&& \int_\vq
e^{- t_1 q^2 - t_2 (\vk-\vq)^2  }= (4 \pi)^{-N/2} (t_1+t_2)^{-N/2} e^{- k^2 \frac{t_1 t_2}{t_1+t_2}}
\end{eqnarray}
we find\begin{eqnarray}
&&  \! \! \! \! \! \! \! \! \! \! \! \! \! \! \! \! \! \! \! \! \! \! \! \!\!  \tilde I(k) =
- \frac{2 k^2}{(4 \pi)^{N/2}} \int_{t_1,t_2>0} \bigg[ \frac{t_1^{b/2-1} t_2^{b/2}}{\Gamma(1+\frac{b}{2})\Gamma(\frac{b}{2})}  - \frac{t_1^{b/2-1} k^{-b}}{\Gamma(\frac{b}{2})} \bigg] t_1 (t_1+t_2)^{-2 - \frac{N}{2}} e^{- \frac{t_1 t_2}{t_1+t_2} k^2  }\ .
\end{eqnarray}
Introducing $s=t_1+t_2$, $t_1=s u$, $t_2=s(1-u)$ with $\rmd t_1 \rmd t_2 = s\, \rmd u\, \rmd s$ one gets
\begin{eqnarray}
&& \fl  \tilde I(k) = - \frac{2 k^2}{ (4 \pi)^{N/2} \Gamma(1+\frac{b}{2}) \Gamma(\frac{b}{2}) }
 \int_0^1 \rmd u  \int_{s>0} s^{-1+b - \frac{N}{2}} [u (1-u)]^{\frac{b}{2}} \Big[1 - s^{-b/2} (1-u)^{-b/2} \Gamma\Big(1+\frac{b}{2}\Big) k^{-b}\Big] e^{- k^2  s u(1-u)\ }\ .
\nonumber
\end{eqnarray}
The integral over $s$ can be performed in the first integral for $b>N/2$ and for $b>N$ in the second. This leads to
\begin{eqnarray}
&& \fl  \tilde I(k) = - \frac{2}{ (4 \pi)^{N/2} \Gamma(1+\frac{b}{2}) \Gamma(\frac{b}{2}) } \Gamma\Big(b-\frac{N}{2}\Big) k^{2-2b+N}
 \int_0^1 \rmd u  \,[u (1-u)]^{\frac{N-b}{2}}  \left[1  - u^{b/2} \frac{\Gamma(1+\frac{b}{2}) \Gamma(\frac{b-N}{2})}{\Gamma(b-\frac{N}{2})} \right] \ .
\end{eqnarray}
Both integrals are convergent for $b<N+2$ and one finds $ \tilde I(k) = (\tilde C_1 + \tilde C_2) k^{2-2b+N} $ with
\begin{eqnarray}
&& \tilde C_1 = - \frac{2^{b - N} \sqrt{\pi} b
  \Gamma\big(b - \frac{N}{2}\big) \Gamma\big(\frac{1}{2} (2 - b + N)\big)}{2 (4 \pi)^{N/2} \Gamma\big(1 + \frac{b}{2}\big)^2 \Gamma\big(\frac{1}{2}  (3 - b + N)\big)}, \\
  &&  \tilde C_2 =  \frac{ N \pi \Gamma\big(\frac{N}{2}\big)}{(4 \pi)^{N/2} \sin\big(\frac{1}{2} (b - N) \pi\big) \Gamma\big(\frac{b}{2}\big) \Gamma\big(2 - \frac{b}{2} + N\big)} \ .
\end{eqnarray}
We now compute the integral associated to $J_c(k)$
\begin{eqnarray}
\fl I_c(k) =  2 \int_\vq \frac{[k^2 q^2 - (\vk \cdot \vq)^2]}{q^2} q^{-b} \,|\vk-\vq|^{-b}  \nn\\
\fl \hphantom{ I_c(k)} =- \frac{1}{2}  \int_{t_1,t_2>0}  \frac{t_1^{b/2} t_2^{b/2-1}}{\Gamma(1+\frac{b}{2})\Gamma(\frac{b}{2})}  \bigg(k^4 + 2 k^2 (\partial_{t_1} + \partial_{t_2}) + (\partial_{t_1} - \partial_{t_2})^2\bigg)
\int_q e^{- t_1 q^2 - t_2 (\vk-\vq)^2  } \nn\\
\fl \hphantom{ I_c(k)} = \frac{(N-1) k^2}{ (4 \pi)^{N/2} \Gamma(1+\frac{b}{2}) \Gamma(\frac{b}{2}) }
 \int_{t_1,t_2>0} t_1^{b/2} t_2^{b/2-1} (t_1+t_2)^{-1-N/2} e^{-\frac{t_1 t_2}{t_1+t_2} k^2} \nn\\
\nn \fl \hphantom{ I_c(k)} =  \frac{(N-1) k^2}{ (4 \pi)^{N/2} \Gamma(1+\frac{b}{2}) \Gamma(\frac{b}{2}) }
 \int_0^1 \rmd u \int_{s>0} s^{-1+b-N/2} u^{\frac{b}{2}} (1-u)^{\frac{b}{2}-1} e^{- s u (1-u) k^2} \\
 \fl \hphantom{ I_c(k)} =  \frac{(N-1) \Gamma(b-\frac{N}{2})}{ (4 \pi)^{N/2} \Gamma(1+\frac{b}{2}) \Gamma(\frac{b}{2}) }  k^{2-2b+N}  \int_0^1 \rmd u\, u     [u(1-u)]^{-1+\frac{N-b}{2}}.
\nonumber
\end{eqnarray}
The integral over $s$ can be done for $b>N/2,$ and the one over $u$ for $b<N$.  It gives
\begin{eqnarray}
&& I_c(k) = C_c k^{2-2b+N}  \quad ,\qquad \quad C_c = \frac{2^{b - N}  (N-1) \sqrt{\pi}
  \Gamma\big(b - \frac{N}{2}\big) \Gamma\big(\frac{1}{2} (N-b)\big)}{(4 \pi)^{N/2} \Gamma\big(1 + \frac{b}{2}\big) \Gamma\big(\frac{b}{2}\big) \Gamma\big(\frac{1}{2} (1 - b + N)\big)}
\ .\end{eqnarray}
The total contribution is thus
\begin{eqnarray}
 J(k) &=& B k^{2-b} + C k^{2-2 b + N} + D k^{-b} + ... , \\
 ~~~C &=& \tilde C_1 + \tilde C_2 + C_c\ ,
\end{eqnarray}
where $ \tilde C_1 + C_c$ is the total contribution to $C$ from the first integral, and $C_2$ the contribution of the second integral. Of course one also shows $B=0,$ as a result of the  cancelations.

\subsection{Surface quasi-geostrophic turbulence}
\label{sec-SQG-integral}

\begin{figure}
\centerline{\Fig{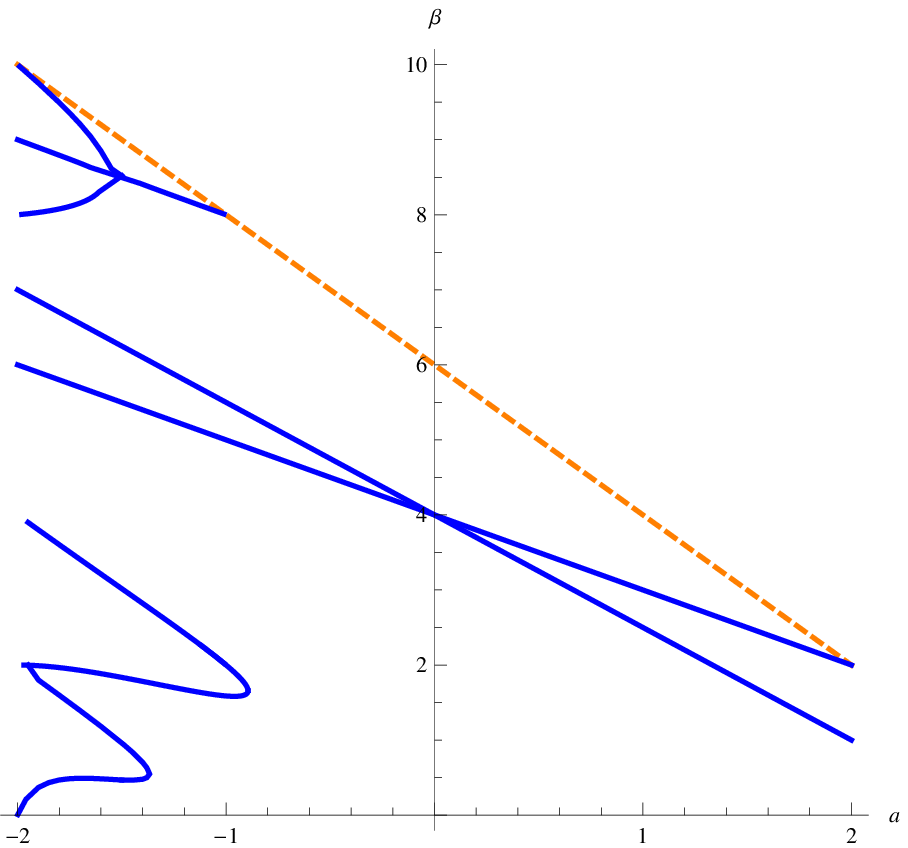}}
\caption{Blue solid curves: Locations in the $(a,\beta)$ plane, where
$F(a,\beta)$ given in equation~(\ref{eq-Fab}) vanishes. The two straight lines are
given by $\beta_1=1+\frac{3}{2}(2-a)$ and $\beta_2=4-a$.
Orange dashed line: $\beta=6-2a$.  }
\label{SQG-roots}
\end{figure}

Let us study the nonlinear term in the FRG equation for the SQG turbulence given by
\begin{eqnarray}
&&  J(k) =  \int \frac{d^2 \vq}{(2 \pi)^2} [\vq\times \vk]^2 (q^{-a}-p^{-a}) \tilde \Delta_{T}(q)
\big[ (q^{-a}-p^{-a}) \tilde \Delta_{T}(p)  - 2 (q^{-a}-k^{-a})
\tilde \Delta_{T}(k) \big], 
\end{eqnarray}
where $\vp=\vk-\vq$.
We now assume for the  correlator $\tilde \Delta_{T}(k)$ the form 
\begin{eqnarray}
&& \tilde \Delta_T(k) = \frac{1}{k^\beta}.
\end{eqnarray}
It is related to the velocity correlator $\Delta(k)$ by
\begin{eqnarray}
&& \beta = 4 - 2 a + \zeta_2, \ \quad , \quad
\tilde \Delta(k) = \frac{1}{k^{2+\zeta_2}}.
\end{eqnarray}
Keeping in mind that the case $a=2$ corresponds to the NS equation in $N = 2$,
we expect that $\beta$ is close to $2$ near $a=2$. We thus want to compute
\begin{eqnarray}
&&  I(k) :=  \int \frac{\rmd^2 \vq}{(2 \pi)^2} [\vq\times \vk]^2 (q^{-a}-p^{-a}) q^{-\beta}
\big[ (q^{-a}-p^{-a}) p^{-\beta}  - 2 (q^{-a}-k^{-a}) k^{-\beta} \big]. 
\end{eqnarray}
Using that 
\be[\vq\times \vk]^2=k^2 q^2-(\vk \cdot \vq)^2=-\frac{1}{4}%
\big[k^4 - 2 k^2 (q^2 + (\vk-\vq)^2) + (q^2-(\vk-\vq)^2)^2\big]
\ee we obtain
\begin{eqnarray}
&&  \! \! \! \! \! \! \! \! \! \! \! \! \! \! \! \! \! \! \! \! \! \! \! \!\!
I(k) =  - \frac{1}{4} \int_{t_1,t_2>0}
\Bigg[ \frac{  t_1^{a+\beta/2-1} t_2^{\beta/2-1} + t_1^{\beta/2-1} t_2^{a+\beta/2-1}  }{\Gamma(a+\frac{\beta}{2})\Gamma(\frac{\beta}{2})}  - 2 \frac{t_1^{a/2+\beta/2-1} t_2^{a/2+\beta/2-1} }{\Gamma(a/2+\frac{\beta}{2})^2}
\\
&& ~~~~~~~~~~~~~~- 2 k^{-\beta} \bigg( \frac{t_1^{a+\beta/2-1}}{\Gamma(a+\beta/2)} - k^{-a} \frac{t_1^{a/2+\beta/2-1}}{\Gamma(a/2+\beta/2)} \bigg) (- \partial_{t_2}) + 2 k^{-\beta}  \frac{t_1^{a/2+\beta/2-1} t_2^{a/2-1}}{\Gamma(a/2+\beta/2) \Gamma(a/2)} \nn\\
&& ~~~~~~~~~~~~~~- 2 k^{-a-\beta}  \frac{t_1^{\beta/2-1} t_2^{a/2-1}}{\Gamma(\beta/2) \Gamma(a/2)}
\Bigg] \tilde B(t_1,t_2)
\int_\vq e^{- t_1 q^2 - t_2 ( \vk-\vq)^2  }, \nonumber
\end{eqnarray}
where $\tilde B(t_1,t_2) =
k^4 + 2 k^2 (\partial_{t_1} + \partial_{t_2}) + (\partial_{t_1} - \partial_{t_2})^2$.
Using equation (\ref{F12}) for $N=2$
we find
\begin{eqnarray}
&&  I(k) = F(a,\beta) k^{-2 (a+ \beta -3)},  \\
&& F(a,{\beta}) = \frac{1}{32 \pi} \Bigg[
\Gamma (a+{\beta}-2) \left(\frac{8 \Gamma
   \left(2-\frac{{\beta}}{2}\right) \Gamma
   \left(-a-\frac{{\beta}}{2}+2\right)}{\Gamma
   \left(\frac{{\beta}}{2}\right) \Gamma (-a-{\beta}+4)
   \Gamma
   \left(a+\frac{{\beta}}{2}\right)}-\frac{\sqrt{\pi } 2^{a+{\beta}} \Gamma
   \left(-\frac{a}{2}-\frac{{\beta}}{2}+2\right)}{\Gamma \left(\frac{1}{2} (-a-{\beta}+5)\right)
   \Gamma
   \left(\frac{a+{\beta}}{2}\right)^2}\right) \nonumber \\
   && ~~~~~~~~~~~~~~~~~~~~+ \frac{8 \Gamma \left(2-\frac{a}{2}\right)
   \left(\frac{\Gamma
   \left(-\frac{a}{2}-\frac{{\beta}}{2}+2\right)
   \Gamma
   \left(a+\frac{{\beta}}{2}-2\right)}{\Gamma
   \left(-a-\frac{{\beta}}{2}+4\right) \Gamma
   \left(\frac{a+{\beta}}{2}\right)}+\frac{\Gamma
   \left(2-\frac{{\beta}}{2}\right) \Gamma
   \left(\frac{1}{2} (a+{\beta}-6)\right)}{\Gamma
   \left(\frac{{\beta}}{2}\right) \Gamma
   \left(-\frac{a}{2}-\frac{{\beta}}{2}+3\right)}\right)}{\Gamma
   \left(\frac{a}{2}\right)} \Bigg]. \label{eq-Fab}
\end{eqnarray}
One checks that for $a=2$ one recovers $C(N=2,\zeta_2)=F(a=2,\beta=\zeta_2)$
where $C(N,\zeta_2)$ is given by equation (\ref{Ctext}).
As we already know this means that the limits $\beta \to 2$ and $a \to 2$
are not exchangeable without an IR cutoff. Indeed, we have
$\lim_{\beta \to 2} F(a=2,\beta)=-1/(4 \pi)$
and not $-1/(8 \pi)$, which we expect in the presence of an IR cutoff.
However, we find that if one keeps $a \neq 2$ infinitesimally close to $2$,
and then takes the limit $\beta \to 2,$ one does find $-1/(8 \pi)$.
This means that one needs to keep $a>2$, and that  $\lim_{a \to 2} F(a,\beta=2)=-1/(8 \pi)$ is the correct limit.

This result leaves two options for the large-$k$  behavior of possible
fixed points parameterized by $\beta$:

(i) $\beta=6-2a$ leads to the nonlinear term in the FRG equation which
for large $k$ asymptotics is also $\sim F k^{-\beta}$. Thus, it
can  be  balanced  by  the rescaling terms and may have a self consistent
solution  even  if  $F$  is  not  zero. The tail is then given by $\Delta_T(k) \sim 1/k^{6-2a}$.
In this case the tail $\Delta(k) \sim 1/k^4$ for the velocity remains
independent of  $a$.

(ii) there is a value of $\beta < 6 - 2 a$ where $F(a,\beta)$ vanishes.
 All possible values are shown on figure~\ref{SQG-roots} for $-2 \le a \le 2$.
 For instance we find that   $F(a,\beta)$ vanishes for 
 $\beta_1=1+\frac{3}{2}(2-a)$ and $\beta_2=4-a$, which cross at $a=0$. However, there are other values.

If we ask that the function $\zeta_2(a)$ for the physical fixed point is continuous, it follows from our analysis for the 2D NS equation (where $\zeta_2(a=2)=2$)
that likely values are $\zeta_2(a)=2$ or $\zeta_2(a)=a$ for $a>0$. However more work is clearly called for, in view of these
results, to study possible fixed points as a function of $a$.

\section{Small-$k$ expansion for Navier-Stokes in dimension $N=2$}
\label{sec:app-small-k}
In this appendix, we calculate the small-$k$ expansion of the nonlinear term
in the FRG equation for the 2-dimensional   decaying Navier-Stokes turbulence. This will
allow us to find in a self-consistent way the small-$k$ behavior of the
FP solution $\tilde{\Delta}^*(k)$.
To this aim, consider the non-linear contribution to the
flow-equation in the distance geometry representation given by
equation~(\ref{F.13bis}). It is useful to symmetrize it in $t \to -t$:
\begin{eqnarray}\label{F.14bis}
\fl   \delta \tilde \Delta(k) =
\frac{k^4}{4 \pi^2} \int_{1}^{\infty}\rmd s
\int_{0}^{1}\rmd t \frac{s t}{s^2-t^2} \sqrt{(s^2-1)(1-t^2)} \\
\fl
\hphantom{ \delta \tilde \Delta(k) =}
\times \left\{\! \left[ \Big((s-t)^2 - 4\Big)  \tilde \Delta\big({\textstyle \frac{k}{2}} (s{-}t)\big) -  \Big((s+t)^2 {-} 4\Big)  \tilde \Delta\big({\textstyle\frac{k}{2}} (s{+}t)\big) \right]\!
 \tilde \Delta(k) + 4 { s t} \tilde \Delta\big({\textstyle \frac{k}{2}} (s{-}t)\big) \tilde \Delta\big({\textstyle\frac{k}{2}} (s{+}t)\big)  \!\right\}.\! \nonumber
 \end{eqnarray}
We now want to expand equation~(\ref{F.14bis}) in small $k$ for an arbitrary function
$\tilde{\Delta}(k)$. In Sec.~\ref{2d-isotrop-fp-search} we already discussed that
the expected behavior
of the fixed-point solution $\tilde{\Delta}^*(k)$ at small $k$ is
$\tilde{\Delta}(k) \sim k^{n-1}$  with $n=3$.
Let us for the moment consider a more general class of functions with a
finite $\tilde \Delta(k=0)$ and $\tilde \Delta'(k=0)$.
 It is not possible to expand the integrand of (\ref{F.14bis}) in small $k$,
since this gives integrals diverging at large $s$.
Instead, one can rescale $s$, by defining $s=2 q/k$, and only then expand in  $k$.
This allows one to integrate term by term over $t \in [0,1]$,
\begin{eqnarray}
&& \fl \delta \tilde \Delta(k) = - \frac{k^2}{8 \pi} \int_{k/2}^\infty \rmd q
 \, \Big[ q^2 \tilde \Delta'(q) \tilde \Delta(0) + 2 q \tilde \Delta(q)\big
 (\tilde \Delta(0) - \tilde \Delta(q\big)\Big]  - \frac{k^3}{8 \pi} \tilde \Delta'(0)
 \int_{k/2}^\infty \rmd q\, \Big[2 q \tilde \Delta(q) + q^2 \Delta'(q)\Big]
 + O(k^4) \nonumber \\
&& \fl \hphantom{\delta \tilde \Delta(k)}=  \frac{k^2}{4 \pi}
\int_{0}^\infty \rmd q\,  q  \tilde \Delta(q)^2 + O(k^4).  \label{smallk}
 \end{eqnarray}
We have used integrations
by parts and the large-$k$ behavior (\ref{2dlarge}) which suggests that
$\lim_{k \to \infty} k^2 \tilde \Delta(k)=0$.
Note that the expansion of $\tilde{\Delta}(q)$ in the integrands of the first line
of equation~(\ref{smallk})  can not produce terms of order $k^4$.
Expanding (\ref{F.14bis}) further to order $k^4$, we find
\begin{eqnarray}
\delta \tilde \Delta(k) = ...+\frac{k^4}{384 \pi}\int_{k/2}^\infty &\rmd q&
\Big\{6 \tilde \Delta '(q)\Big[7 \tilde \Delta (0)-4 q^2 \tilde \Delta ''(0)\Big]
-12 q \tilde \Delta '(q)^2\nonumber\\
  && -q\Big[12 \tilde \Delta (q)
  \Big(4 \tilde \Delta ''(0)-\tilde \Delta ''(q)\Big) +\tilde \Delta (0)
  \Big(q \tilde \Delta'''(q)+6 \tilde \Delta
   ''(q)\Big)\Big] \Big\}.
\end{eqnarray}
Together with (\ref{smallk}),  this yields 
\begin{equation}\label{F.18}
\delta \tilde \Delta(k) = \frac{k^2}{4 \pi}  \int_{0}^\infty \rmd q\,  q
\tilde \Delta(q)^2 -\frac{k^4}{16 \pi}  \int_0^\infty\rmd q\, q \tilde \Delta'(r)^2
-\frac{5 k^4}{ 48 \pi} \tilde \Delta(q=0)^2 +O(k^5).
\end{equation}
Note that at least  the  first few terms of the expansion (\ref{F.18}) depend mainly on
the integral properties of $\tilde \Delta(k)$ and not on its small-$k$ expansion.

\section{Asymptotic large-$k$ behavior for Navier-Stokes in $N=2$}

\label{s:NS-D=2-large-k-Andrei}
In this appendix, we calculate the asymptotic large-$k$ behavior of the non-linear term in the flow-equation for 2D Navier Stokes. We start from the rescaled dimensionless version of (\ref{NSFRG}), setting
\begin{equation}\label{la;ksjdf}
\tilde\Delta(k) \to  \frac{1}{(k^{2}+m^{2})^{2}}.
\end{equation}
Thus we need to compute for $N=2$ the  convergent integral
\begin{eqnarray} \label{integratecrazy3}
\fl   { \delta \tilde\Delta(k)}=  2
\int \frac{\rmd^2 \vq}{(2 \pi)^2} \frac{k^2 q^2 - (\vk \cdot \vq)^2}{k^2 q^2 (\vk-\vq)^2} \left[(k^2-q^2)\Big ((\vk-\vq)^2 - q^2\Big) \right] \frac{1}{(q^2+m^2)^2} \left[\frac{1}{((\vk-\vq)^2+m^2)^2} - \frac{1}{(k^2+m^2)^2} \right]. \
\end{eqnarray}
In polar coordinates
this integral reads
\begin{eqnarray}
 { \delta \tilde\Delta(k)} &=&  \frac{2}{(2 \pi)^2}
\int\limits_{0}^{\infty} q\, \rmd q \int\limits_{0}^{2\pi}\rmd \phi\,
\frac{1-\cos ^2(\phi )}{k^2-2 k q \cos (\phi
   )+q^2 }\left(k^2-q^2\right) \left[k^2-2 k q \cos (\phi )\right]
\frac{1 }{ \left(m^2+q^2\right)^2 } \nonumber\\
&& \qquad\times  \left[\frac{1}{\left(k^2-2
   k q \cos (\phi
   )+m^2+q^2\right)^2}-\frac{1}{\left(k^2+m^2\right)^2}\right].
\end{eqnarray}
After rescaling $q\to qk$ and $m\to m_1 k$ it becomes
\begin{eqnarray}
{ \delta \tilde\Delta(k)} &=&  \frac{2}{k^4 (2 \pi)^2}
\int\limits_{0}^{\infty} q\, \rmd q \int\limits_{0}^{2\pi}\rmd \phi\,
\frac{1-\cos ^2(\phi )}{1-2  q \cos (\phi
   )+q^2 }\left(1-q^2\right) \left[1-2  q \cos (\phi )\right]
\frac{1 }{ \left(m_1^2+q^2\right)^2 } \nonumber\\
&&\qquad \times \left[\frac{1}{\left(1-2
    q \cos (\phi
   )+m_1^2+q^2\right)^2}-\frac{1}{\left(1+m_1^2\right)^2}\right].
\end{eqnarray}
Changing variables to $t$ in such a way that
$ \cos (\phi )\to \frac{1-t^2}{t^2+1}$ and $\sin (\phi )\to \frac{2 t}{t^2+1}$
with Jacobian $\frac{4}{t^2+1}$ we arrive at
\begin{eqnarray}
{ \delta \tilde\Delta(k)} &=&  \frac{8}{\pi ^2 k^4}
\int\limits_{0}^{\infty} q^2 \rmd q \int\limits_{0}^{\infty}t^2 \rmd t
 \frac{  \left(q^2-1\right)  \left(q t^2+q+2 t^2-2\right)
   \left(2 q \left(t^2-1\right)+t^2+1\right) }
   {
   (m_1^2+q^2)^2 \left[ q^2 \left(t^2+1\right)+2 q
   (t^2-1)+t^2+1\right] } \nonumber \\
&&\qquad\times \frac{
    \left(t^2+1\right) (2 m_1^2    +q^2+ 2)  +2 q
   \left(t^2-1\right)}
   { \left(m_1^2+1\right)^2 \left(t^2+1\right)^3
   \left[ \left(t^2+1\right)(m_1^2 +1 +q^2) +2 q
   \left(t^2-1\right)\right]^2}.
\end{eqnarray}
The integral over $t$ has to be taken independently for $0<q<1$ and $1<q<\infty$
so that $\delta \tilde\Delta(k)=\delta_{1} \tilde\Delta(k)+\delta_{2} \tilde\Delta(k)$. Introducing
$A=\sqrt{m_1^2+(q-1)^2}$ and  $B=\sqrt{m_1^2+(q+1)^2}$ one can write
these integrals in the following form
\begin{eqnarray}
  \delta_{1} \tilde\Delta(k) &=& \frac{1}{\pi k^4} \int\limits_{0}^{1}  \rmd q\,
 \frac{(B-A)^3 \left[\left(A^2-B^2\right)^2-16\right]
 \left[A^4-2 A^2 \left(B^2+4\right)+B^4-8
   B^2\right]^{-2}}{  A B
    \left[(A-B)^2-4\right]^2 (A+B) \left(A^2+B^2-2\right)^2
    }
\nonumber    \\
&&\qquad \times
 \bigg[ A^8-4 A^7 B+2 A^6 \left(2 B^2-9\right)+4 A^5 B
   \left(B^2+7\right)-2 A^4 \left(5 B^4+7 B^2-48\right)
\nonumber    \\
&&\qquad\quad~ +2 A^2    \left(2 B^6-7 B^4+32 B^2-64\right)-4 A
   B \left(B^6-7 B^4+32\right) \nonumber    \\
&&\qquad\quad~ +4 A^3 B^3    \left(B^2+2\right) +B^2 \left(B^2-8\right)^2
   \left(B^2-2\right)    \bigg]
\end{eqnarray}
and
\begin{eqnarray}
\delta_{2} \tilde\Delta(k)  &=& \frac{1}{\pi k^4} \int\limits_{1}^{\infty}    \rmd q\,  \frac{(B-A) \left[(A+B)^2-2\right]
   \left[\left(A^2-B^2\right)^2-16\right]}{  A B  (A+B-2)^2
   (A+B) (A+B+2)^2 \left(A^2+B^2-2\right)^2}.
\end{eqnarray}
Integration over $q$ and combining both terms gives
\begin{equation}
\delta \tilde\Delta(k)=-\frac{1}{8\pi k^4} f(m/k)
\end{equation}
with
\begin{eqnarray}
f(x)&=&\frac{1}{x^4 \left(x^2+1\right)^2} \bigg\{ \left(6 x^4+8
   x^2+2\right) \log \left(x^2+1\right)
   +2 \left(2 x^4+x^2\right)\nonumber \\
&& + \sqrt{4 x^2+1} \left(x^2+1\right)^2 \bigg[\log \left(\sqrt{4
   x^2+1}-1\right)
   -\log \left(\left(x^2+1\right) \left(\sqrt{4
   x^2+1}+3\right)-2\right)\bigg] \nonumber \\
&&
   + \left[\left(\sqrt{4
   x^2+1}-6\right) x^2+\sqrt{4 x^2+1}-2\right] \left(x^2+1\right)
   \log \left(x^2\right) \bigg\}.
\end{eqnarray}
Expanding in  small $x$, i.e.\ large $k/m$,  we obtain
\begin{equation}
f(x)=1+x^2 \left(8 \log (x)-\frac{2}{3}\right)- x^4 \left(32 \log
   (x)+\frac{53}{6}\right)+O\left(x^5\right).
\end{equation}
Note that only the leading term is universal, while the higher
ones depend on the regularization by $m$ introduced in
equation~(\ref{la;ksjdf}). This implies the leading-order term
$-\frac{A^2}{8\pi k^4}  $ given in equation~(\ref{76}).

\section{Numerical solution for the fixed point in dimension $N=2$}\label{a6c}

\begin{figure}[t]
\centerline{\Fig{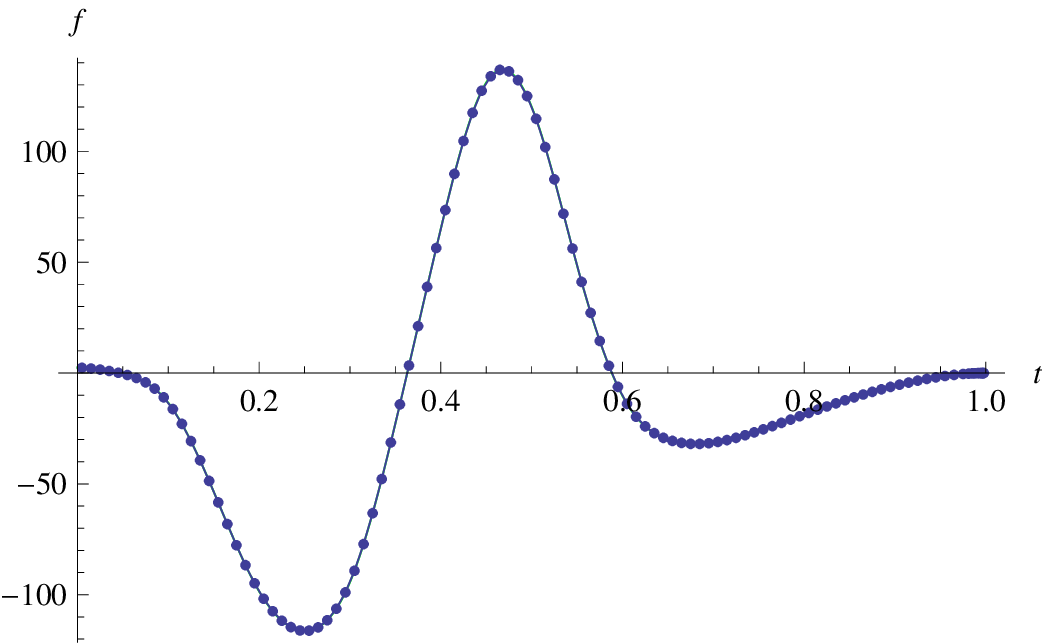} \fig{8.5cm}{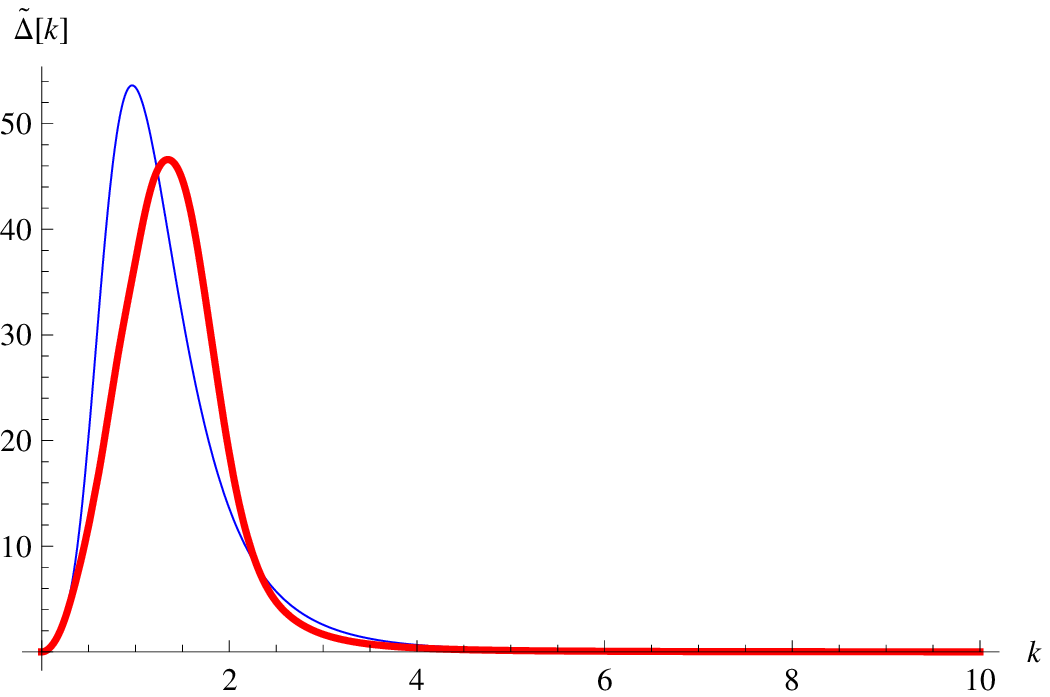}}
\caption{Left: Check for the precision of the fit for $f (t)$
 defined in Eq.\
(\ref{F.49}). Right:
The guessed fixed-point $\tilde \Delta_{\mathrm{guess}} (k)$ (blue,
thin line), and the numerical solution $\tilde \Delta (k) $ defined in Eq.~(\ref{70}) (red, fat line).
See figure \ref{f:2D-E(k)} for a log-log plot of $\tilde \Delta (k) $.}
\label{f:fit-check}
\end{figure}
\begin{figure}[t]
\centerline{\Fig{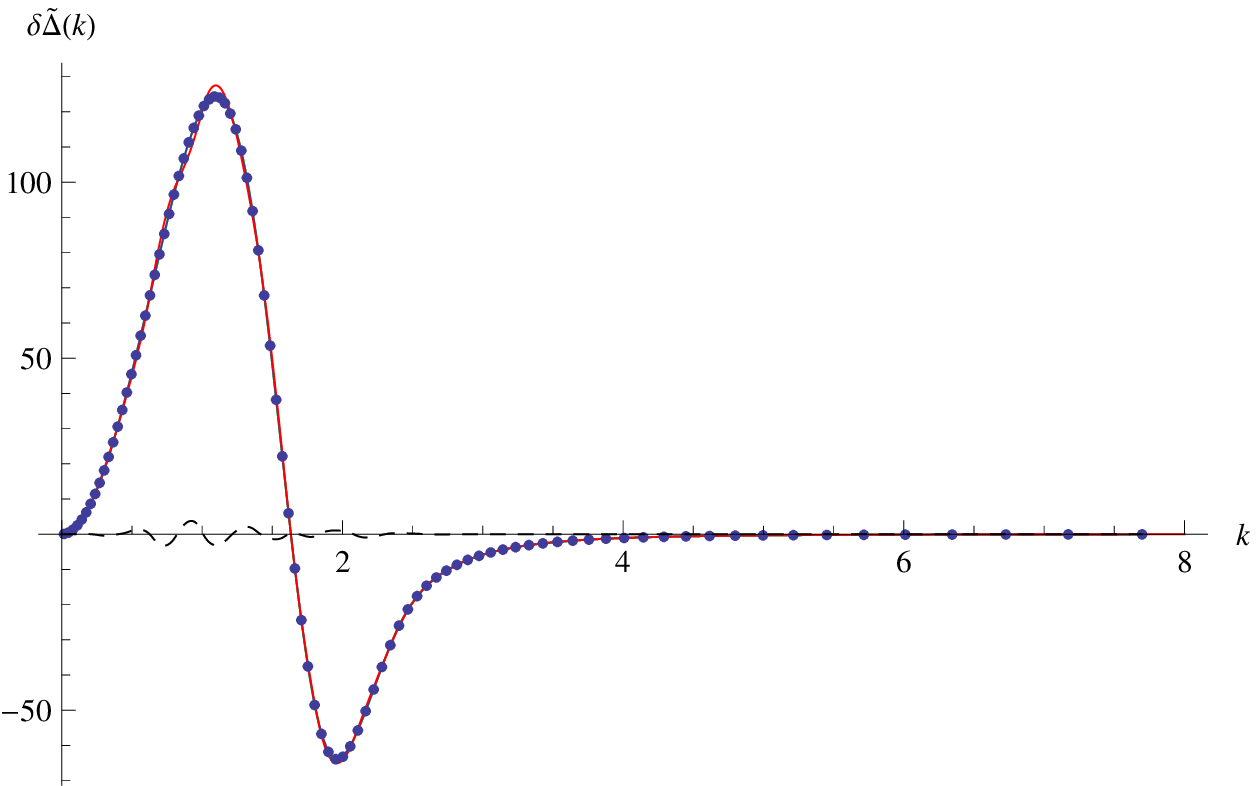}\Fig{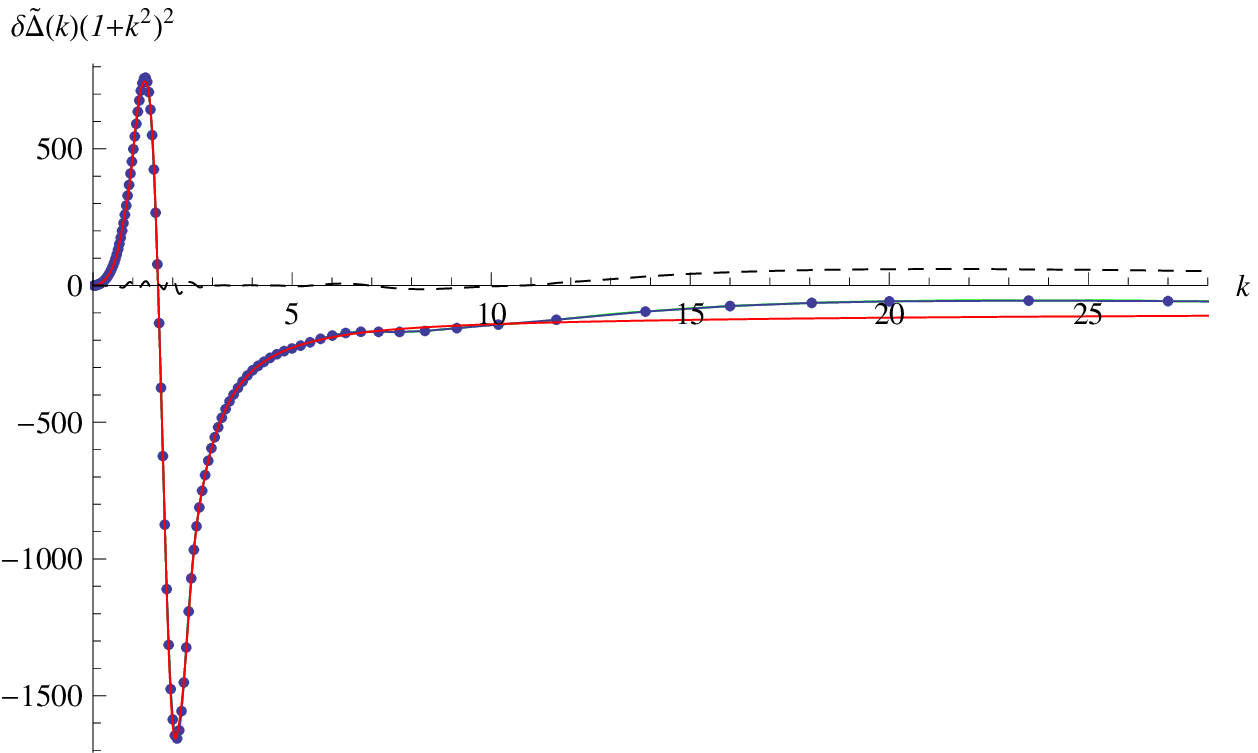}}
\centerline{\Fig{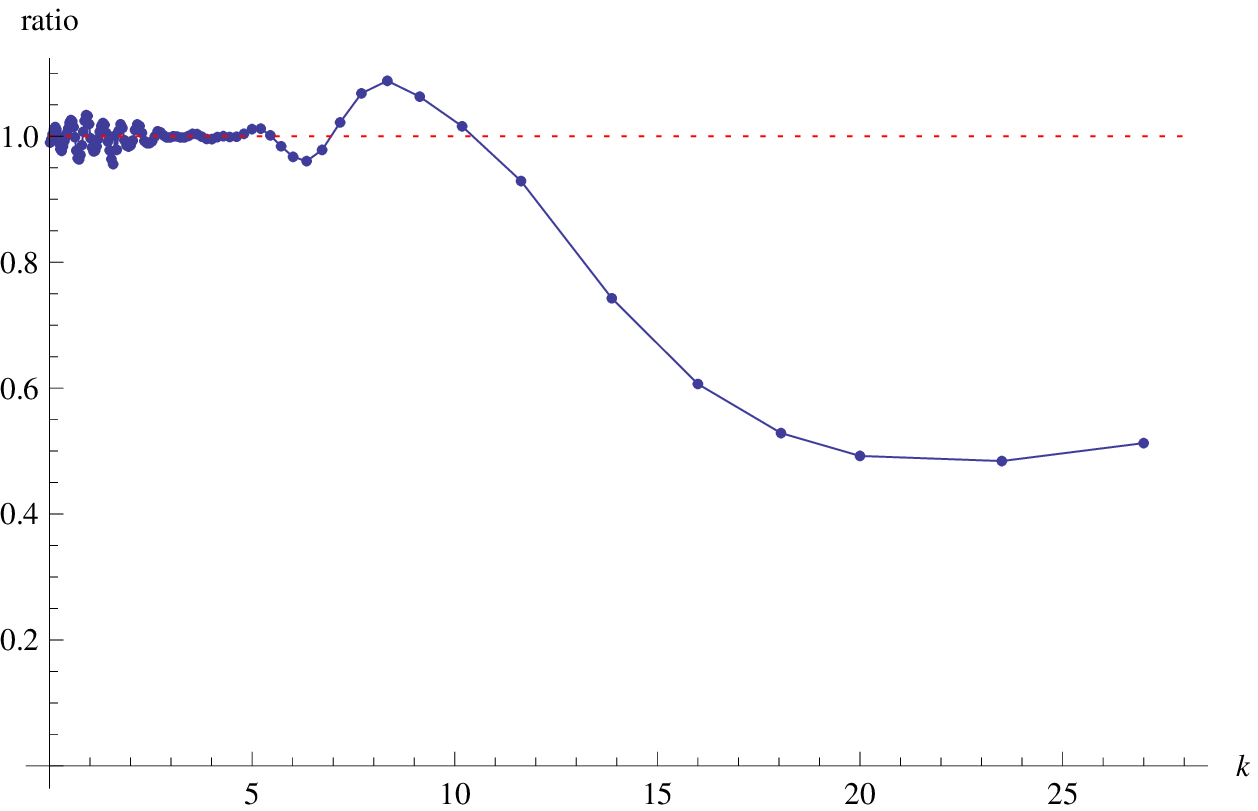}}
\caption{Top left: $ \delta \tilde \Delta (k)$ (green solid
line with superimposed blue dots),
$2 \tilde \Delta(k) +
k \tilde \Delta'(k) $ (solid red line) and $
\partial_{\ell}\tilde \Delta (k)$ (dashed black line). Top right: The
same 3 terms, multiplied by $(1+k^{2})^{2}$.
Bottom: plot of the ratio $\frac{\delta \tilde \Delta(k)}{2 \tilde \Delta(k) +
k \tilde \Delta'(k)} $ which must be 1 at the fixed point.
}\label{fig:precision}
\end{figure}

We wish to integrate equations~(\ref{51}) and (\ref{F.13bis}) numerically,
using $\zeta =2$. If the
fixed point is attractive,  $\tilde \Delta (k)$ will converge against it.
Numerically, the problem is hard for several reasons: \begin{enumerate}
\item convergence of the integral (\ref{F.13bis}) for large $k$ is slow
and imprecise.
\item one needs high precision for $\tilde \Delta (k)$, which respects
the conditions (\ref{2dsmall}) and (\ref{eqA2})  at small $k$, and the asymptotic form
(\ref{76}) for large $k$.
\item the r.h.s.\ of equation~(\ref{51}) must be  calculated numerically at
many different points; we will use 108 points (see table~\ref{tab:Deltak}).
\item \label{num-inst} one easily runs into numerical instabilities, when using a
spline-interpolation through all points, or a polynomial fit of degree
$\approx 20$, which is necessary to represent faithfully the
data-points.
\end{enumerate}
In order to circumvent these problems, we do the following: Write
the ansatz (\ref{70})-(\ref{Deltaguess}).
The  guess in equation (\ref{Deltaguess})  was obtained by {\em (i)} imposing the correct
asymptotic forms, {\em (ii)} trying to optimize  consistency
relations as (\ref{F.18}).  Since we have no small-$k$ expansion with
rapidly decaying coefficients, the final criterion {\em (iii)} for
(\ref{Deltaguess}) was a  r.h.s.\ in the
flow-equation which is small in the intermediate regime of finite
$k$.

Our information is stored in   $\tilde \Delta_{\mathrm{cor}} (k)$,
spaced as the first column in table \ref{tab:Deltak}.
It is updated via
\begin{equation}\label{update}
\tilde
 \Delta_{\mathrm{cor}} (k) \to  \tilde \Delta_{\mathrm{cor}} (k) + \kappa \, t
\partial_{t} \tilde \Delta (k)\ ,
\end{equation}
at the given values of $k$. The quantity \(
\partial_{t} \tilde \Delta (k) \) is given by equations~(\ref{51}) and (\ref{F.13bis}).
Note that while we use the distance geometry formula (\ref{F.13bis}) for the
correction since it is numerically more efficient, we have checked that the result
is the same when replacing equation (\ref{F.13bis}) by equations (\ref{52}) and
(\ref{53}). The step-size $\kappa$ is finally reduced to $\kappa =0.005$.

To circumvent the numerical problems mentioned under (\ref{num-inst})
above when obtaining the rescaling term in (\ref{51}) and the numerical
integral in (\ref{F.13bis}),  we approximate $\tilde
\Delta_{\mathrm{cor}} (k)$ by the smooth function
\begin{eqnarray}\label{F.49}
\tilde \Delta_{\mathrm{cor}} (k) = \frac{f \big(t (k)\big)}{m (k)}\\
t (k):=\frac{k}{\sqrt{10+k^{2}}}  \\
m (k):= \frac{\left(k^2+1\right)^{5/2}}{k^2}
\end{eqnarray}
The function $f (t)$ is defined on $[0,1]$, bounded (of order 100), and
converges to 0 for $t$ to 0 or 1, see figure \ref{f:fit-check}. In order to produce a smooth fit for
$f (t)$, we
use the {\em best} (i.e.\ least variance) cubic spline with $o$
equidistant points and  \(f(0)=f(1)=f'(0)=f'(1)=0\). $o$ is
initially chosen to be 13, and then increased up to 20. Using splines
of this relatively low order (compared with the number of data points)
effectively filters out numerical noise. (Note: A polynomial fit
of the same order is not adequate, but generates a dynamic
instability. To further increase precision, a spline with variable
switch-points could be used. $o=20$ is our upper limit due to RAM
problems in the
implementation of the integration routines, which split the integrals in
pieces for each spline part.) Before
updating $ \tilde  \Delta_{\mathrm{cor}} (k)$ via (\ref{update}), the
table of stored values for $ \tilde  \Delta_{\mathrm{cor}} (k)$ is
replaced by the approximation (\ref{F.49}). This is necessary for
consistency.
On the left of figure \ref{f:fit-check}, we show our final result of the function $f (t)$, and the points
from which it is constructed. One sees that no numerical artefact is
present.

Our final result  for $\tilde \Delta (k)$ is shown in a  log-log plot
on figure~\ref{f:2D-E(k)}. A linear plot is presented on
the right of  figure \ref{f:fit-check} (fat  red line), as well as our
initial guess $\tilde \Delta_{\mathrm{guess}} (k)$. Note that since
there exists a redundant mode (the choice of \(k_{\rm max}\) at which $E(k) = k\tilde  \Delta (k) $ is maximal, one could get these two
curves closer. Numerical values for  $\tilde \Delta (k)$ are
given in table \ref{tab:Deltak}.  The  precision can be inferred from
figure \ref{fig:precision}. It should be few percent (relative
precision) for $k<5$, but the precision decreases for larger values of $k$. Note however
that we know the exact asymptotic form $\tilde \Delta (k) = 16 \pi/k^{4}$.

\begin{table}\label{xx2}
\begin{tabular}{|c|c|}\hline
$k$ & $\tilde  \Delta (k)$\\
\hline
 0 & 0 \\
 0.0158116 & 0.0131526 \\
 0.0474395 & 0.117591 \\
 0.0790817 & 0.323878 \\
 0.110748 & 0.629261 \\
 0.142447 & 1.03318 \\
 0.174189 & 1.53977 \\
 0.205984 & 2.15575 \\
 0.237841 & 2.8848 \\
 0.26977 & 3.72542 \\
 0.301781 & 4.66717 \\
 0.333885 & 5.68728 \\
 0.366091 & 6.76818 \\
 0.39841 & 7.91167 \\
 0.430852 & 9.12235 \\
 0.463428 & 10.4041 \\
 0.496149 & 11.7583 \\
 0.529027 & 13.1873 \\
 0.562072 & 14.6963 \\
 0.595297 & 16.2907 \\
 0.628713 & 17.9758 \\
 0.662334 & 19.758 \\
 0.69617 & 21.6242 \\
 0.730237 & 23.5339 \\
 0.764546 & 25.447 \\
 0.799113 & 27.3277 \\
 0.83395 & 29.1463 \\
\hline
\end{tabular}
\begin{tabular}{|c|c|}\hline
$k$ & $\tilde  \Delta (k)$\\
\hline
 0.869074 & 30.8943 \\
 0.9045 & 32.5893 \\
 0.940243 & 34.2548 \\
 0.976321 & 35.9175 \\
 1.01275 & 37.6048 \\
 1.04955 & 39.3028 \\
 1.08673 & 40.9413 \\
 1.12433 & 42.4538 \\
 1.16235 & 43.782 \\
 1.20082 & 44.876 \\
 1.23977 & 45.7066 \\
 1.2792 & 46.2688 \\
 1.31916 & 46.5628 \\
 1.35967 & 46.5915 \\
 1.40074 & 46.3608 \\
 1.44242 & 45.8577 \\
 1.48473 & 45.0426 \\
 1.5277 & 43.8845 \\
 1.57137 & 42.3626 \\
 1.61578 & 40.4659 \\
 1.66095 & 38.2213 \\
 1.70694 & 35.7034 \\
 1.75378 & 32.9836 \\
 1.80152 & 30.1263 \\
 1.85021 & 27.1889 \\
 1.8999 & 24.2391 \\
 1.95064 & 21.362 \\
\hline
\end{tabular}
\begin{tabular}{|c|c|}\hline
$k$ & $\tilde  \Delta (k)$\\
\hline
 2.0025 & 18.629 \\
 2.05554 & 16.0965 \\
 2.10983 & 13.8072 \\
 2.16544 & 11.7798 \\
 2.22246 & 10.005 \\
 2.28096 & 8.46986 \\
 2.34104 & 7.15983 \\
 2.40281 & 6.05898 \\
 2.46637 & 5.14416 \\
 2.53185 & 4.38304 \\
 2.59937 & 3.74686 \\
 2.66909 & 3.21119 \\
 2.74115 & 2.7555 \\
 2.81574 & 2.36475 \\
 2.89304 & 2.02966 \\
 2.97328 & 1.74234 \\
 3.05668 & 1.49586 \\
 3.14352 & 1.28418 \\
 3.23409 & 1.10221 \\
 3.32872 & 0.945681 \\
 3.42779 & 0.810933 \\
 3.53174 & 0.694791 \\
 3.64104 & 0.59454 \\
 3.75626 & 0.507916 \\
 3.87803 & 0.433078 \\
 4.00711 & 0.368428 \\
 4.14436 & 0.312578 \\
\hline
\end{tabular}
\begin{tabular}{|c|c|}\hline
$k$ & $\tilde  \Delta (k)$\\
\hline
 4.29082 & 0.264334 \\
 4.44768 & 0.222674 \\
 4.6164 & 0.186731 \\
 4.79873 & 0.15576 \\
 4.99681 & 0.12912 \\
 5.21327 & 0.106266 \\
 5.45142 & 0.0867232 \\
 5.71548 & 0.0700738 \\
 6.01089 & 0.0559517 \\
 6.34489 & 0.044037 \\
 6.72728 & 0.0340508 \\
 7.17178 & 0.0257567 \\
 7.69832 & 0.0189527 \\
 8.33709 & 0.0134578 \\
 9.13673 & 0.00910839 \\
 10.1818 & 0.00575792 \\
 11.6362 & 0.00328225 \\
 13.8756 & 0.00157192 \\
 16 & 0.000868729 \\
 18.0514 & 0.00052673 \\
 20 & 0.000344759 \\
 23.5 & 0.000177434 \\
 27 & 0.000100406 \\
 31.5 & 0.0000534954 \\
 35.5 & 0.0000328852 \\
 41 & 0.0000183294 \\
 48 & 9.68650 \mbox{$\times 10^{-6}$}\\
\hline
\end{tabular}
\caption{Numerical fixed point of equations (\ref{51}), (\ref{F.13bis})
for $\tilde  \Delta (k)$.}
\label{tab:Deltak}
\end{table}

\section{Navier-Stokes equation in the limit of large $N$}
\label{a:K}
In this appendix we consider the
FRG flow for the Navier-Stokes equation  (\ref{FRG-real-space}) in the limit of
large $N$.
We start from the  flow equation in real space~(\ref{FRG-real-space}) to derive an equation
for $\nabla_{\vu}^2  \Delta(u)$:
\begin{eqnarray}\label{FRG-real-space-N} &&
 t \partial_{t}\nabla_{\vu}^2   \tilde \Delta(u) =(2-\zeta)\,\nabla_{\vu}^2   \tilde   \Delta(u)
+ \frac \zeta2 \vu \partial_\vu \big[\nabla_{\vu}^2    \tilde \Delta(u)\big ]+ \nabla_{\vu}^2    \delta \tilde\Delta_{\rm L} (u)
+ \nabla_{\vu}^2    \delta\tilde \Delta_{\rm NL} (u).
\end{eqnarray}
The  Laplacian is  $\nabla_{\vu}^2=N \rmd/\rmd y + 2 y\, \rm d^2/dy^2$;
rewriting \(\Delta(u)\) in terms of $r(y),$ given by equation~(\ref{220}), we obtain
\begin{eqnarray}
 \nabla_{\vu}^2 \Delta(u)=-\bigg[ N\,\left( 2 + N \right) \,r''(y) +
  4\,y\,\left( \left( 2 + N \right) \,r'''(y) +
     y\,r^{(4)}(y) \right)\bigg].
\end{eqnarray}
The Laplacian of
$\nabla_{\vu}^2 \delta \tilde\Delta_{\rm L} (u)$ is local and has been
expressed in terms of $r(y)$ in equation~(\ref{99-2}).
The nonlocal part $\delta \tilde\Delta_{\rm NL} (u)$ is given by equation~(\ref{53-2})
with the kernel~(\ref{a44-2}). 
Taking into account that for
a large-$N$ expansion in real space,   both $p^2$ and $k^2 \sim N$,
we can expand the kernel in $1/N$ as follows
\begin{eqnarray} 
 && A(p,k) = 2k^4 \left(1 - \frac{1}{N}\right) - \frac{4}{N} \frac{k^4 p^2}{k^2 + p^2} +
O\!\left(\frac{1}{N^2} \right)\ . \label{Anew-1}
\end{eqnarray}
The first term in equation~(\ref{Anew-1}) can be written
as a derivative, thus  becomes also local
\begin{eqnarray}
&&\fl  \nabla_{\vu}^2 \delta_{\rm }{ \tilde \Delta_{\rm NL}^{(1)} (u)} =
         2 \left( N -1\right) \,r'(0)
     \left[ \left( 4 + N \right)
        \left( \left( 2 + N \right)
           \left( N\,r'''(y)
           + 6\,y\,r^{(4)}(y) \right)  +
          12\,y^2\,r^{(5)}(y) \right)  + 8\,y^3\,r^{(6)}(y) \right].
        \ \ \ \ \label{990}
\end{eqnarray}
Note that apart from the factor of  $r'(0)$, it has the structure of the last term in equation~(\ref{99-2}).
Let us now evaluate the second term in equation~(\ref{Anew-1}) to leading order at
large $N$,
\begin{eqnarray}
   \nabla_{\vu}^2 \delta\tilde \Delta_{\rm NL}^{ (2)} (u)
  &=& -\frac{{ 4}}{N}
\int_{k,p} e^{-i k u} k^6 \tilde R(k) \frac{1}{k^2 +
p^2} p^4 \tilde R(p)\nn\\
 &=& -\frac{{ 4}}{N} \int_0^\infty \rmd t \int_{k,p} e^{-i k u} k^6 \tilde R(k) p^4
\tilde R(p) e^{-t k^2 - t p^2}\ .
\end{eqnarray}
Since $\int_{k} e^{-i k u} \tilde R(k) = \tilde R(u) = r(y),$
we have in the large-$N$ limit
\begin{eqnarray}
&& \int_k e^{-i k u} e^{- t k^2} {\tilde  R(k)}= r(y + N t)\ , \label{179}
\end{eqnarray}
as can be checked by Taylor expanding and using that $-k^2=\nabla_{\vu}^2=N\rmd/\rmd y$
to leading order. Hence
\begin{eqnarray}
 && { \nabla_{\vu}^2 \delta\tilde \Delta_{\rm NL}^{(2)}(u)}  =
 \frac{{ 4}}{N} N^5 \int_0^\infty \rmd t\,
r'''(y + N t)
r''(N t)  =  \frac{{ 4}}{N} N^4 \int_0^\infty \rmd z\, r'''(y + z) r''(z). \label{181}
\end{eqnarray}
We  now put all pieces together, setting  $\zeta= \zeta_{0} + \zeta_{-1}/N+ ...$
Expanding to order  $1/N$ we obtain equation (\ref{K8-bis}).

\section*{References}


\section*{Contents}

\contentsline {section}{\numberline {1}Introduction}{2}
\contentsline {section}{\numberline {2}Model and notations}{3}
\contentsline {section}{\numberline {3}Known results and phenomenology}{4}
\contentsline {subsection}{\numberline {3.1}Decaying Burgers}{4}
\contentsline {subsection}{\numberline {3.2}Decaying Navier Stokes}{5}
\contentsline {section}{\numberline {4}FRG equations}{6}
\contentsline {subsection}{\numberline {4.1}Loop expansion: General strategy}{6}
\contentsline {subsection}{\numberline {4.2}FRG equation for Burgers}{7}
\contentsline {subsection}{\numberline {4.3}FRG equation for Navier-Stokes in momentum space}{8}
\contentsline {subsection}{\numberline {4.4}FRG equation for Navier-Stokes in real space: isotropic turbulence}{9}
\contentsline {subsection}{\numberline {4.5}Energy conservation and energy anomaly.}{9}
\contentsline {subsection}{\numberline {4.6}Enstrophy conservation and enstrophy anomaly in dimension 2.}{10}
\contentsline {section}{\numberline {5}Short-distance analysis, cusp or no cusp?}{11}
\contentsline {subsection}{\numberline {5.1}Burgers}{11}
\contentsline {subsection}{\numberline {5.2}Navier-Stokes}{12}
\contentsline {section}{\numberline {6}Two-dimensional decaying turbulence ($N=2$)}{12}
\contentsline {subsection}{\numberline {6.1}Basic properties}{12}
\contentsline {subsection}{\numberline {6.2}Isotropic turbulence}{13}
\contentsline {subsubsection}{\numberline {6.2.1}FRG equations.}{13}
\contentsline {subsubsection}{\numberline {6.2.2}Searching for an isotropic fixed point}{14}
\contentsline {subsubsection}{\numberline {6.2.3}Small-$k$ expansion.}{14}
\contentsline {subsubsection}{\numberline {6.2.4}Large-$k$ expansion.}{15}
\contentsline {subsubsection}{\numberline {6.2.5}Numerical solution}{15}
\contentsline {subsubsection}{\numberline {6.2.6}Physical interpretation of the solution}{16}
\contentsline {subsection}{\numberline {6.3}Periodic 2D-turbulence}{16}
\contentsline {section}{\numberline {7}Analysis of the FRG equation in three dimensions ($N=3$), and in large dimensions ($N\to \infty$)}{18}
\contentsline {section}{\numberline {8}Decaying surface quasi-geostrophic turbulence}{19}
\contentsline {section}{\numberline {9}Conclusions and Perspectives}{20}
\contentsline {section}{\numberline {A}1-loop FRG equation in real space}{20}
\contentsline {section}{\numberline {B}1-loop FRG equation in Fourier space}{21}
\contentsline {section}{\numberline {C}FRG equation for $N=3$ periodic flows}{23}
\contentsline {section}{\numberline {D}Generating functional approach and diagrammatics}{23}
\contentsline {section}{\numberline {E}Distance geometry for FRG equation}{26}
\contentsline {subsection}{\numberline {E.1}Navier-Stokes}{26}
\contentsline {subsection}{\numberline {E.2}Burgers}{27}
\contentsline {section}{\numberline {F}Short-distance expansion of $\Delta (u)$: Amplitudes $B$ and $C$ for Burgers, Navier Stokes and surface quasi-geostrophic turbulence}{27}
\contentsline {subsection}{\numberline {F.1}Burgers}{27}
\contentsline {subsection}{\numberline {F.2}Navier-Stokes}{28}
\contentsline {subsection}{\numberline {F.3}Surface quasi-geostrophic turbulence}{29}
\contentsline {section}{\numberline {G}Small-$k$ expansion for Navier-Stokes in dimension $N=2$}{31}
\contentsline {section}{\numberline {H}Asymptotic large-$k$ behavior for Navier-Stokes in $N=2$}{31}
\contentsline {section}{\numberline {I}Numerical solution for the fixed point in dimension $N=2$}{33}
\contentsline {section}{\numberline {J}Navier-Stokes equation in the limit of large $N$}{34}

\end{document}